\pdfoutput=1

\documentclass[12pt,a4paper]{article}

\usepackage[british]{babel}

\usepackage{ifthen} 
\newboolean{pdflatex}
\setboolean{pdflatex}{true} 

\newboolean{articletitles}
\setboolean{articletitles}{true} 

\newboolean{uprightparticles}
\setboolean{uprightparticles}{true} 

\newboolean{inbibliography}
\setboolean{inbibliography}{false} 

\usepackage{multirow}
\usepackage{rotating}
\usepackage{mathrsfs}

\usepackage[normalem]{ulem}

\usepackage{microtype}
\usepackage{lineno}  
\usepackage{xspace} 
\usepackage{caption} 

\usepackage{graphicx}  
\usepackage{color}
\usepackage{colortbl}
\graphicspath{{./figs/}} 

\usepackage{amsmath} 
\usepackage{amssymb}
\usepackage{amsfonts}
\usepackage{upgreek} 

\newcommand*\patchAmsMathEnvironmentForLineno[1]{%
\expandafter\let\csname old#1\expandafter\endcsname\csname #1\endcsname
\expandafter\let\csname oldend#1\expandafter\endcsname\csname
end#1\endcsname
 \renewenvironment{#1}%
   {\linenomath\csname old#1\endcsname}%
   {\csname oldend#1\endcsname\endlinenomath}%
}
\newcommand*\patchBothAmsMathEnvironmentsForLineno[1]{%
  \patchAmsMathEnvironmentForLineno{#1}%
  \patchAmsMathEnvironmentForLineno{#1*}%
}
\AtBeginDocument{%
\patchBothAmsMathEnvironmentsForLineno{equation}%
\patchBothAmsMathEnvironmentsForLineno{align}%
\patchBothAmsMathEnvironmentsForLineno{flalign}%
\patchBothAmsMathEnvironmentsForLineno{alignat}%
\patchBothAmsMathEnvironmentsForLineno{gather}%
\patchBothAmsMathEnvironmentsForLineno{multline}%
\patchBothAmsMathEnvironmentsForLineno{eqnarray}%
}

\usepackage{hyperref}    
\usepackage[all]{hypcap} 

\def\lhcb {\mbox{LHCb}\xspace}

\def\MagUp {\mbox{\em Mag\kern -0.05em Up}\xspace}

\ifthenelse{\boolean{uprightparticles}}%
{

 \def\Peta        {\ensuremath{\upeta}\xspace}

 \def\Pmu         {\ensuremath{\upmu}\xspace}

 \def\Ppi         {\ensuremath{\uppi}\xspace}

 \def\Pphi        {\ensuremath{\upphi}\xspace}

 \def\Ppsi        {\ensuremath{\uppsi}\xspace}

 \def\PDelta      {\ensuremath{\Delta}\xspace}                 
 \def\PXi      {\ensuremath{\Xi}\xspace}                 
 \def\PLambda      {\ensuremath{\Lambda}\xspace}                 
 \def\PSigma      {\ensuremath{\Sigma}\xspace}                 
 \def\POmega      {\ensuremath{\Omega}\xspace}                 
 \def\PUpsilon      {\ensuremath{\Upsilon}\xspace}                 
 
 \def\PB      {\ensuremath{\mathrm{B}}\xspace}                 
                  
 \def\PD      {\ensuremath{\mathrm{D}}\xspace}

 \def\PJ      {\ensuremath{\mathrm{J}}\xspace}                 
 \def\PK      {\ensuremath{\mathrm{K}}\xspace}

 \def\Pb      {\ensuremath{\mathrm{b}}\xspace}                 
 \def\Pc      {\ensuremath{\mathrm{c}}\xspace}                 
                  
 \def\Pe      {\ensuremath{\mathrm{e}}\xspace}

 \def\Pi      {\ensuremath{\mathrm{i}}\xspace}

 \def\Pp      {\ensuremath{\mathrm{p}}\xspace}

 \def\Ps      {\ensuremath{\mathrm{s}}\xspace}

}
{

 \def\Peta        {\ensuremath{\eta}\xspace}

 \def\Pmu         {\ensuremath{\mu}\xspace}

 \def\Ppi         {\ensuremath{\pi}\xspace}

 \def\Pphi        {\ensuremath{\phi}\xspace}

 \def\Ppsi        {\ensuremath{\psi}\xspace}                 
                  
 \mathchardef\PDelta="7101
 \mathchardef\PXi="7104
 \mathchardef\PLambda="7103
 \mathchardef\PSigma="7106
 \mathchardef\POmega="710A
 \mathchardef\PUpsilon="7107
                  
 \def\PB      {\ensuremath{B}\xspace}                 
                  
 \def\PD      {\ensuremath{D}\xspace}

 \def\PJ      {\ensuremath{J}\xspace}                 
 \def\PK      {\ensuremath{K}\xspace}

 \def\Pb      {\ensuremath{b}\xspace}                 
 \def\Pc      {\ensuremath{c}\xspace}                 
                  
 \def\Pe      {\ensuremath{e}\xspace}

 \def\Pi      {\ensuremath{i}\xspace}

 \def\Pp      {\ensuremath{p}\xspace}

 \def\Ps      {\ensuremath{s}\xspace}

}

\makeatletter
\ifcase \@ptsize \relax
  \newcommand{\miniscule}{\@setfontsize\miniscule{4}{5}}
\or
  \newcommand{\miniscule}{\@setfontsize\miniscule{5}{6}}
\or
  \newcommand{\miniscule}{\@setfontsize\miniscule{5}{6}}
\fi
\makeatother

\DeclareRobustCommand{\optbar}[1]{\shortstack{{\miniscule (\rule[.5ex]{1.25em}{.18mm})}
  \\ [-.7ex] $#1$}}

\def\epem       {{\ensuremath{\Pe^+\Pe^-}}\xspace}

\def\mup        {{\ensuremath{\Pmu^+}}\xspace}
\def\mun        {{\ensuremath{\Pmu^-}}\xspace} 
\def\mumu       {{\ensuremath{\Pmu^+\Pmu^-}}\xspace}

\def\squark    {{\ensuremath{\Ps}}\xspace}

\def\cquark    {{\ensuremath{\Pc}}\xspace}
\def\cquarkbar {{\ensuremath{\overline \cquark}}\xspace}

\def\bquark    {{\ensuremath{\Pb}}\xspace}
\def\bquarkbar {{\ensuremath{\overline \bquark}}\xspace}

\def\pion   {{\ensuremath{\Ppi}}\xspace}

\def\pip    {{\ensuremath{\pion^+}}\xspace}
\def\pim    {{\ensuremath{\pion^-}}\xspace}

\def\kaon    {{\ensuremath{\PK}}\xspace}
  \def\Kbar    {{\kern 0.2em\overline{\kern -0.2em \PK}{}}\xspace}

\def\KorKbar    {\kern 0.18em\optbar{\kern -0.18em K}{}\xspace}

\def\Kp      {{\ensuremath{\kaon^+}}\xspace}
\def\Km      {{\ensuremath{\kaon^-}}\xspace}

  \def\Dbar    {{\kern 0.2em\overline{\kern -0.2em \PD}{}}\xspace}
\def\D       {{\ensuremath{\PD}}\xspace}

\def\DorDbar    {\kern 0.18em\optbar{\kern -0.18em D}{}\xspace}
\def\Dz      {{\ensuremath{\D^0}}\xspace}

\def\Dp      {{\ensuremath{\D^+}}\xspace}

\def\Dstarp  {{\ensuremath{\D^{*+}}}\xspace}

\def\Ds      {{\ensuremath{\D^+_\squark}}\xspace}

\def\B       {{\ensuremath{\PB}}\xspace}
\def\Bbar    {{\ensuremath{\kern 0.18em\overline{\kern -0.18em \PB}{}}}\xspace}

\def\BorBbar    {\kern 0.18em\optbar{\kern -0.18em B}{}\xspace}

\def\Bc      {{\ensuremath{\B_\cquark^+}}\xspace}

\def\jpsi     {{\ensuremath{{\PJ\mskip -3mu/\mskip -2mu\Ppsi\mskip 2mu}}}\xspace}

\def\proton      {{\ensuremath{\Pp}}\xspace}
\def\antiproton  {{\ensuremath{\overline \proton}}\xspace}

\def\Lz          {{\ensuremath{\PLambda}}\xspace}
\def\Lbar        {{\ensuremath{\kern 0.1em\overline{\kern -0.1em\PLambda}}}\xspace}
\def\LorLbar    {\kern 0.18em\optbar{\kern -0.18em \PLambda}{}\xspace}

\def\Lc      {{\ensuremath{\Lz^+_\cquark}}\xspace}

\def\to                 {\ensuremath{\rightarrow}\xspace}

\def\AT#1     {\ensuremath{A_{\mathrm{T}}^{#1}}\xspace}           

\def\C#1      {\ensuremath{\mathcal{C}_{#1}}\xspace}                       
\def\Cp#1     {\ensuremath{\mathcal{C}_{#1}^{'}}\xspace}                    
\def\Ceff#1   {\ensuremath{\mathcal{C}_{#1}^{\mathrm{(eff)}}}\xspace}        
\def\Cpeff#1  {\ensuremath{\mathcal{C}_{#1}^{'\mathrm{(eff)}}}\xspace}       
\def\Ope#1    {\ensuremath{\mathcal{O}_{#1}}\xspace}                       
\def\Opep#1   {\ensuremath{\mathcal{O}_{#1}^{'}}\xspace}                    

\newcommand{\tev}{\ensuremath{\mathrm{\,Te\kern -0.1em V}}\xspace}
\newcommand{\gev}{\ensuremath{\mathrm{\,Ge\kern -0.1em V}}\xspace}
\newcommand{\mev}{\ensuremath{\mathrm{\,Me\kern -0.1em V}}\xspace}
\newcommand{\kev}{\ensuremath{\mathrm{\,ke\kern -0.1em V}}\xspace}
\newcommand{\ev}{\ensuremath{\mathrm{\,e\kern -0.1em V}}\xspace}
\newcommand{\gevc}{\ensuremath{{\mathrm{\,Ge\kern -0.1em V\!/}c}}\xspace}
\newcommand{\mevc}{\ensuremath{{\mathrm{\,Me\kern -0.1em V\!/}c}}\xspace}
\newcommand{\gevcc}{\ensuremath{{\mathrm{\,Ge\kern -0.1em V\!/}c^2}}\xspace}
\newcommand{\gevgevcccc}{\ensuremath{{\mathrm{\,Ge\kern -0.1em V^2\!/}c^4}}\xspace}
\newcommand{\mevcc}{\ensuremath{{\mathrm{\,Me\kern -0.1em V\!/}c^2}}\xspace}

\def\mum  {\ensuremath{{\,\upmu\rm m}}\xspace}

\def\mbarn{\ensuremath{\rm \,mb}\xspace}

\def\pb {\ensuremath{\rm \,pb}\xspace}

\def\invfb   {\ensuremath{\mbox{\,fb}^{-1}}\xspace}

\newcommand{\stat}{\ensuremath{\mathrm{\,(stat)}}\xspace}
\newcommand{\syst}{\ensuremath{\mathrm{\,(syst)}}\xspace}

\newcommand{\chisq}{\ensuremath{\chi^2}\xspace}

\def\deriv {\ensuremath{\mathrm{d}}}

\def\gsim{{~\raise.15em\hbox{$>$}\kern-.85em
          \lower.35em\hbox{$\sim$}~}\xspace}
\def\lsim{{~\raise.15em\hbox{$<$}\kern-.85em
          \lower.35em\hbox{$\sim$}~}\xspace}

\def\sPlot{\mbox{\em sPlot}\xspace}

\def\sqs   {\ensuremath{\protect\sqrt{s}}\xspace}

\def\ptot       {\mbox{$p$}\xspace}
\def\pt         {\mbox{$p_{\rm T}$}\xspace}

\def\evtgen     {\mbox{\textsc{EvtGen}}\xspace}

\def\geant      {\mbox{\textsc{Geant4}}\xspace}

\def\photos     {\mbox{\textsc{Photos}}\xspace}

\def\tell1  {TELL1\xspace}
\def\ukl1   {UKL1\xspace}

\newcommand{\ie}{\mbox{\itshape i.e.}\xspace}

\newcommand{\ups}    {\PUpsilon}

\newcommand{\YoneS}  {\ensuremath{\ups\mathrm{(1S)}}\xspace}
\newcommand{\YtwoS}  {\ensuremath{\ups\mathrm{(2S)}}\xspace}
\newcommand{\YthreeS}{\ensuremath{\ups\mathrm{(3S)}}\xspace}
\newcommand{\YfourS} {\ensuremath{\ups\mathrm{(4S)}}\xspace}

\newcommand{\Charm}   {\ensuremath{\mathsf{C}}\xspace}

\newcommand{\ptC}     {\ensuremath{p^{\Charm}_{\mathrm{T}}}\xspace}
\newcommand{\yC}      {\ensuremath{y^{\Charm}}\xspace}
\newcommand{\pty}     {\ensuremath{p^{\ups}_{\mathrm{T}}}\xspace}
\newcommand{\yy}      {\ensuremath{y^{\ups}}\xspace}
\newcommand{\kT}      {\ensuremath{k_{\mathrm{T}}}\xspace}

\usepackage{cite} 
\usepackage{mciteplus}

\begin{document}

\renewcommand{\thefootnote}{\fnsymbol{footnote}}
\setcounter{footnote}{1}


\begin{titlepage}
\pagenumbering{roman}

\vspace*{-1.5cm}
\centerline{\large EUROPEAN ORGANIZATION FOR NUCLEAR RESEARCH (CERN)}
\vspace*{1.5cm}
\noindent
\begin{tabular*}{\linewidth}{lc@{\extracolsep{\fill}}r@{\extracolsep{0pt}}}
\ifthenelse{\boolean{pdflatex}}
{\vspace*{-2.7cm}\mbox{\!\!\!\includegraphics[width=.14\textwidth]{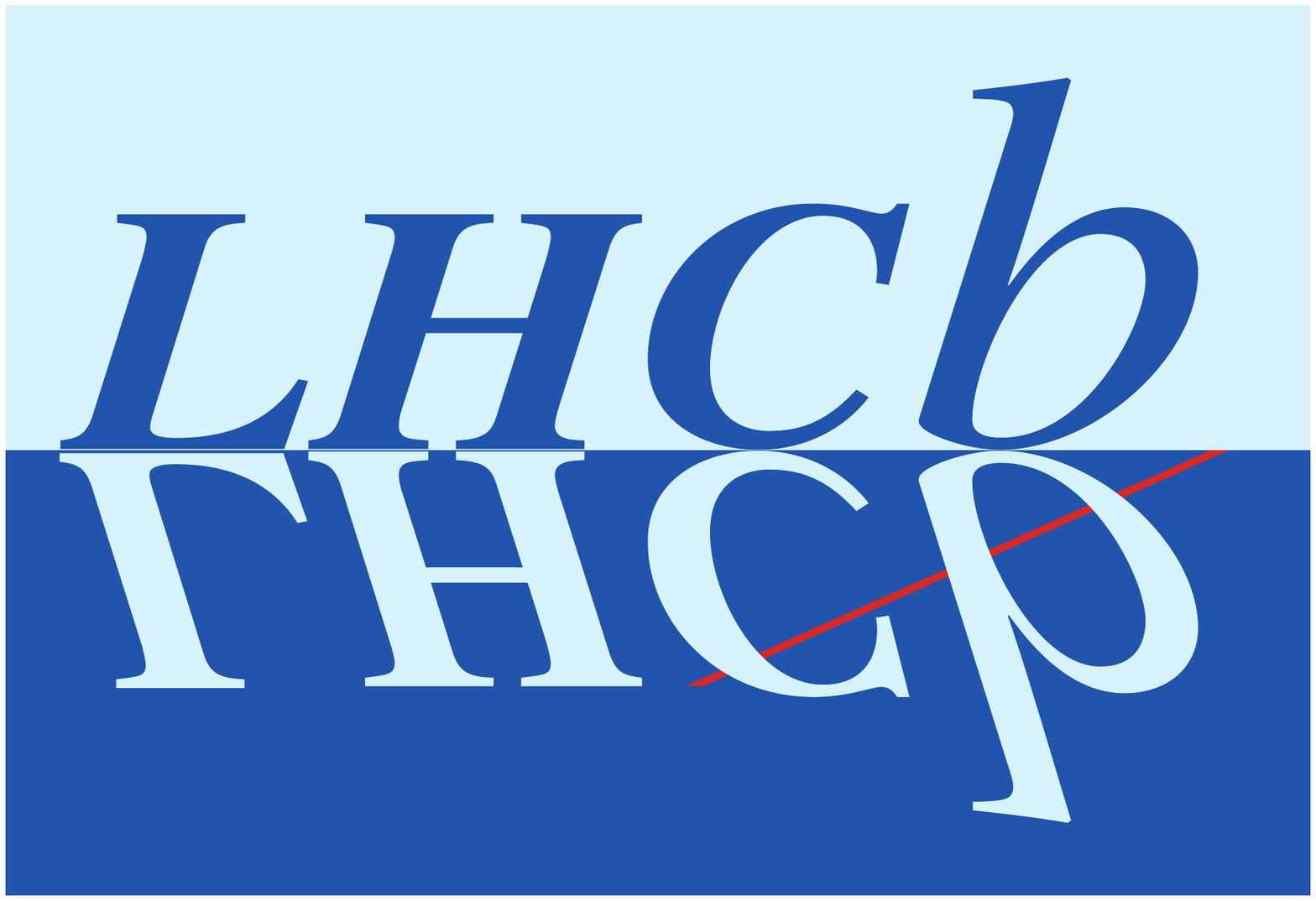}} & &}%
{\vspace*{-1.2cm}\mbox{\!\!\!\includegraphics[width=.12\textwidth]{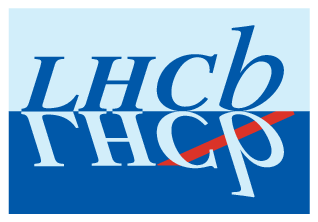}} & &}%
\\
 & & CERN-PH-EP-2015-279 \\  
 & & LHCb-PAPER-2015-046 \\  
 & & 20th October 2015 \\ 
 & & \\
\end{tabular*}

\vspace*{2.5cm}

{\bf\boldmath\huge
  \begin{center}
    Production of
    associated \ups~and open charm hadrons in 
    $\proton\proton$~collisions at \mbox{$\sqs=7$} and~\mbox{$8\tev$}
    via double parton scattering
  \end{center}
}

\vspace*{0.25cm}

\begin{center}
The LHCb collaboration\footnote{Authors are listed at the end of this paper.}
\end{center}

\vspace{\fill}

\begin{abstract}
  \noindent
  Associated production of bottomonia and open charm hadrons in
  $\proton\proton$~collisions at~$\sqs=7$~and~8\tev is observed 
  using data corresponding to an~integrated
  luminosity of 3\invfb accumulated with the~LHCb~detector.
  The~observation of five combinations, 
  $\YoneS\Dz$,
  $\YtwoS\Dz$,
  $\YoneS\Dp$,
  $\YtwoS\Dp$ and $\YoneS\Ds$, is reported.
  Production cross-sections
  are measured for \YoneS\Dz and \YoneS\Dp~pairs in
  the~forward region.
  The~measured cross-sections and the~differential distributions
  indicate the~dominance of double parton scattering as 
  the~main production mechanism.
\end{abstract}

\vspace*{0.5cm}

\begin{center}
  Published in JHEP 07\,(2016)\,052
\end{center}

\vspace{\fill}

{\footnotesize 
\centerline{\copyright~CERN on behalf of the \lhcb collaboration, licence \href{http://creativecommons.org/licenses/by/4.0/}{CC-BY-4.0}.}}
\vspace*{2mm}

\end{titlepage}


\newpage
\setcounter{page}{2}
\mbox{~}
%
%
%
%

\cleardoublepage


\renewcommand{\thefootnote}{\arabic{footnote}}
\setcounter{footnote}{0}



\pagestyle{plain} 
\setcounter{page}{1}
\pagenumbering{arabic}


%

\section{Introduction}\label{sec:intro}

Production of multiple heavy quark pairs in
high\nobreakdash-energy hadron collisions
was first observed in 1982 
by the~NA3~collaboration 
in the~channels 
\mbox{$\pim(\proton)\,{\mathrm{nucleon}}\to\jpsi\jpsi+\mathrm{X}$}~\mbox{\cite{Badier:1982ae,Badier:1985ri}}.
Soon after, evidence for the~associated production of  
four open charm particles in pion\nobreakdash-nucleon reactions 
was obtained by the~WA75~collaboration~\cite{Aoki:1986uq}.
A~measurement of \jpsi~pair production in 
proton\nobreakdash-proton\,\mbox{($\proton\proton$)} 
collisions at $\sqs=7\tev$~\cite{LHCb-PAPER-2011-013}
has been made by the~LHCb collaboration in 2011.
This measurement appears to be in good agreement 
with two models within
the~single parton scattering\,(SPS) mechanism, 
namely non\nobreakdash-relativistic quantum 
chromodynamics\,(NRQCD)~calculations~\cite{Berezhnoy:2011xy}
and $k_{\mathrm{T}}$\nobreakdash-\hspace{0pt}factorization~\cite{Baranov:2011zz}.
However the~obtained result also agrees with
predictions~\cite{Kom:2011bd} 
of the~double parton 
scattering\,(DPS)~mechanism~\mbox{\cite{Paver:1982yp,Diehl:2011tt,Diehl:2011yj,Bansal:2014paa,Szczurek:2015vha}}.

The~production of \jpsi~pairs 
has also been observed  by the~D0~\cite{Abazov:2014qba} 
and CMS~\cite{Khachatryan:2014iia} collaborations.
A~large double charm production 
cross\nobreakdash-section 
involving open charm in $\proton\proton$~collisions 
at $\sqs=7\tev$ has been observed by the~\lhcb 
collaboration~\cite{LHCb-PAPER-2012-003}.
The~measured cross-sections exceed the~SPS expectations
significantly~\cite{Berezhnoy:1998aa,Baranov:2006dh,Lansberg:2008gk,Maciula:2014xba,Maciula:2014oya}
and agree with the~DPS estimates.
A~study of differential distributions supports a~large role
for the~DPS~mechanism in
multiple production of heavy quarks.

The~study of \mbox{$(\bquark\bquarkbar)(\cquark\cquarkbar)$}~production 
in hadronic collisions started with the~observation 
of \Bc~mesons in $\proton\antiproton$~collisions by the~CDF collaboration~\cite{Abe:1998wi}.
A~detailed study of \Bc~production spectra in 
$\proton\proton$~collisions by the~LHCb collaboration~\cite{LHCb-PAPER-2014-050}
showed good agreement with leading-order 
\mbox{NRQCD}~calculations~\cite{Berezhnoy:1994ba,Kolodziej:1995nv,Chang:1994aw}
including the~SPS~contribution only.

The~leading-order 
\mbox{NRQCD}~calculations using 
the~same matrix element as in Ref.~\cite{Berezhnoy:1994ba},
applied to another class of
\mbox{$(\bquark\bquarkbar)(\cquark\cquarkbar)$}~production, namely  
associated production of bottomonia and open charm hadrons
in the~forward region, defined in terms of the~rapidity $y$ as $2<y<4.5$,
predict~\cite{Berezhnoy:2015jga}
\begin{equation}
  R_{\mathrm{SPS}} =  
  \dfrac{\upsigma^{\ups\cquark\cquarkbar}}{\upsigma^{\ups}} 
  = (0.2-0.6)\,\%\,,\label{eq:r}
\end{equation}
where 
$\upsigma^{\ups\cquark\cquarkbar}$ denotes the~production cross-section 
for associated production of $\ups\cquark\cquarkbar$-pair and 
$\upsigma^{\ups}$ denotes the~inclusive production 
cross-section of \ups~mesons.
A~slightly smaller value of $R_{\mathrm{SPS}}$ is 
    obtained through the~\kT-factorization 
    approach~\cite{
       Gribov:1984tu,*Levin:1990gg,
       Baranov:2002cf,
       Andersson:2002cf,*Andersen:2003xj,*Andersen:2006pg,
       Baranov:2006dh,Baranov:2006rz,Baranov:2012fb}
    using the~transverse momentum dependent gluon density 
    from Refs.~\cite{Jung:2012hy,Hautmann:2013tba,Jung:2014vaa},
    \begin{equation}
      R_{\mathrm{SPS}} =  
      \dfrac{\upsigma^{\ups\cquark\cquarkbar}}{\upsigma^{\ups}} 
      = (0.1-0.3)\,\%\,.\label{eq:rkt}
    \end{equation}

Within the~DPS mechanism, the~\ups~meson and
$\cquark\cquarkbar$-pair are produced
independently in different partonic
interactions. Neglecting
the~parton correlations in the~proton, 
the~contribution of this mechanism 
is estimated according to the~formula~\cite{Calucci:1997ii,Calucci:1999yz,DelFabbro:2000ds}
\begin{equation}
  \upsigma^{\ups\cquark\cquarkbar} = 
  \dfrac{ \upsigma^{\ups}\times \upsigma^{\cquark\cquarkbar}}
        {\upsigma_{\mathrm{eff}}}\,, \label{eq:pocket}
\end{equation}
where $\upsigma^{\cquark\cquarkbar}$~and 
$\upsigma^{\ups}$~are the~inclusive charm and \ups~cross-sections, 
and $\upsigma_{\mathrm{eff}}$ 
is an~effective cross\nobreakdash-section, 
which provides the~proper normalization
of the~DPS cross\nobreakdash-section estimate.
The~latter is related to 
the~transverse overlap function 
between partons in the~proton.
Equation~\eqref{eq:pocket} can be used to calculate
the~ratio $R_{\mathrm{DPS}}$ as 
\begin{equation}
  R_{\mathrm{DPS}} = \dfrac{\upsigma^{\ups\cquark\cquarkbar}}{\upsigma^{\ups}}  = 
  \dfrac{\upsigma^{\cquark\cquarkbar}}{\upsigma_{\mathrm{eff}}}. \label{eq:dpsr}
\end{equation}
Using  the~measured production cross\nobreakdash-section 
for inclusive charm in $\proton\proton$~collisions 
at the~centre\nobreakdash-of\nobreakdash-mass energy \mbox{$7\tev$}~\cite{LHCb-PAPER-2012-041} 
in the~forward region
and \mbox{$\upsigma_{\mathrm{eff}}\sim14.5\mbarn$~\cite{Abe:1997bp,Abe:1997xk}}, 
one obtains~\mbox{$R_{\mathrm{DPS}} \sim 10\%$}, 
which is significantly
larger than $R_{\mathrm{SPS}}$~from
Eq.~\eqref{eq:r}.
The~production cross\nobreakdash-sections for 
$\YoneS\Dz$~and $\YoneS\Dp$~at $\sqs=7\tev$  are calculated using 
the~measured prompt charm production cross\nobreakdash-section 
from Ref.~\cite{LHCb-PAPER-2012-041} and 
the~\YoneS~cross\nobreakdash-section 
from~Ref.~\cite{LHCb-PAPER-2015-045}.
In~the~LHCb kinematic region, covering 
transverse momenta~\pt and 
rapidity~$y$ of \YoneS and $\D^{0,+}$~mesons of 
\mbox{$p_{\mathrm{T}}(\YoneS)<15\gevc$},
\mbox{$1<p_{\mathrm{T}}(\D^{0,+})<20\gevc$},
\mbox{$2.0<y(\YoneS)<4.5$} and 
\mbox{$2.0<y(\D^{0,+})<4.5$},
the~expected production cross\nobreakdash-sections are
\begin{subequations}
  \begin{eqnarray}
    \mathscr{B}_{\mumu}\times\left.\upsigma^{\YoneS\Dz}_{\sqs=7\tev}\right|_{\mathrm{DPS}} 
     & = &  206 \pm 17 \pb, \\ 
    \mathscr{B}_{\mumu}\times\left.\upsigma^{\YoneS\Dp}_{\sqs=7\tev}\right|_{\mathrm{DPS}}
    & = & \phantom{0}  86 \pm 10 \pb, 
  \end{eqnarray}\label{eq:dpsth}\end{subequations}
where $\mathscr{B}_{\mumu}$ is the~branching fraction of $\YoneS\to\mumu$~\cite{PDG2014},
\mbox{$\upsigma_{\mathrm{eff}}=14.5\mbarn$}~is used with no~associated uncertainty 
included~\cite{Abe:1997bp,Abe:1997xk}.
The~basic DPS~formula, Eq.~\eqref{eq:pocket},  
leads to the~following predictions
for the~ratios of production cross-sections
$R^{\Dz/\Dp}$ and
$R_{\Charm}^{\YtwoS/\YoneS}$ 
\begin{subequations}
  \begin{gather}
    R^{\Dz/\Dp}            =  
    \dfrac{ \upsigma^{\ups\Dz}}{ \upsigma^{\ups\Dp}}
    =   \dfrac{ \upsigma^{\Dz}}{ \upsigma^{\Dp}}  =  2.41\pm0.18\,, \label{eq:rdd}\\
    R_{\Charm}^{\YtwoS/\YoneS}   =   
    \mathscr{B}_{2/1} 
    \dfrac {\upsigma^{ \YtwoS \Dz}  } 
    {\upsigma^{ \YoneS \Dz } } =
    \mathscr{B}_{2/1} 
    \dfrac {\upsigma^{ \YtwoS \Dp} } 
    {\upsigma^{ \YoneS \Dp } }   =  
    \mathscr{B}_{2/1}  
    \dfrac {\upsigma^{ \YtwoS} } 
    {\upsigma^{ \YoneS}}  =   0.249 \pm 0.033\,, \label{eq:r21}
  \end{gather}\label{eq:dpsth2}\end{subequations}
where
$\upsigma^{\Dz}$,
$\upsigma^{\Dp}$ and
$\upsigma^{\ups}$ stand for the~measured production cross\nobreakdash-sections of
\Dz, \Dp~and \ups~mesons~\cite{LHCb-PAPER-2012-041,LHCb-PAPER-2015-045},
and $\mathscr{B}_{2/1}$~is the~ratio of dimuon branching fractions
of \YtwoS and \YoneS~mesons.

Here we report the~first observation of 
associated production of bottomonia and open charm hadrons. 
The~production cross-sections
and the~differential distributions
are measured.
The~latter provide crucial information for understanding
the~production mechanism. 
The~analysis is performed using the~Run~1 data~set
recorded by the~LHCb detector,
consisting of 1\invfb of integrated luminosity 
accumulated at~\mbox{$\sqs=7\tev$} 
and \mbox{$2\invfb$}~accumulated at~\mbox{$8\tev$}.

\section{Detector and data sample}
\label{sec:lhcb}

The \lhcb detector~\cite{Alves:2008zz,LHCb-DP-2014-002} is a single-arm forward
spectrometer covering the \mbox{pseudorapidity} range $2<\Peta <5$,
designed for the study of particles containing \bquark or \cquark
quarks. The~detector includes a high-precision tracking system
consisting of a silicon-strip vertex detector surrounding the~$\proton\proton$~interaction
region, 
a~large-area silicon-strip detector located
upstream of a~dipole magnet with a~bending power of about
$4{\mathrm{\,Tm}}$, and three stations of silicon-strip detectors and straw
drift tubes placed downstream of the magnet.
The~tracking system provides a measurement of the~momentum, 
\ptot, of charged particles with
a~relative uncertainty that varies from 0.5\% at low momentum to 1.0\% at 200\gevc.
The~minimum distance of a~track to a~primary vertex, the impact parameter, 
is measured with a~resolution of~\mbox{$(15+29/\pt)\mum$},
where \pt is the component of the momentum transverse to the~beam, in\,\gevc.
Different types of charged hadrons are distinguished using information
from two ring-imaging Cherenkov detectors. 
Photons, electrons and hadrons are identified by a calorimeter system consisting of
scintillating-pad and preshower detectors, an electromagnetic
calorimeter and a hadronic calorimeter. 
Muons~are identified by a
system composed of alternating layers of iron and multiwire
proportional chambers.
The online event selection is performed by a trigger~\cite{LHCb-DP-2012-004}, 
which consists of a hardware stage, based on information from the calorimeter and muon
systems, followed by a software stage, which applies a full event
reconstruction.
At~the~hardware stage, events for this analysis are selected requiring
dimuon candidates with a~product of their transverse momenta~\pt
larger than
\mbox{$1.7\,(2.6)\gev^2/c^2$} for data 
collected at~\mbox{$\sqs=7\,(8)\tev$}. 
In the~subsequent software trigger, two well
reconstructed tracks are required to have hits 
in the~muon system,
to~have \mbox{$\pt>500\mevc$} and \mbox{$\ptot>6\gevc$} 
and 
to~form a~common vertex.  
Only events with a~dimuon candidate with 
a~mass~$m_{\mumu}$ larger than~$4.7\gevcc$ 
are retained for further analysis.

The simulation is performed using the \lhcb configuration~\cite{LHCb-PROC-2010-056} 
of the {\sc Pythia}\,6 event generator~\cite{Sjostrand:2006za}.
Decays of hadronic particles are described by \evtgen~\cite{Lange:2001uf}
in which final\nobreakdash-state photons are generated using \photos~\cite{Golonka:2005pn}.
The~interaction of the~generated particles with the~detector, 
and its response,
 are implemented using 
the~\geant toolkit~\cite{Allison:2006ve, Agostinelli:2002hh}
as described in Ref.~\cite{LHCb-PROC-2011-006}.

\section{Event selection}
\label{sec:evsec}

The~event selection strategy is based on the~independent 
selection of \YoneS, \YtwoS and \YthreeS mesons\,(jointly 
referred to by the symbol~\ups throughout  the~paper) and 
charmed hadrons, namely  \Dz, \Dp and \Ds~mesons and 
\Lc~baryons\,(jointly referred to  by the symbol~\Charm herafter) originating 
from the~same $\proton\proton$~collision vertex.
The~\ups~candidates are reconstructed via their dimuon decays, and 
the~\mbox{$\Dz\to\Km\pip$},
\mbox{$\Dp\to\Km\pip\pip$},
\mbox{$\Ds\to\Kp\Km\pip$} and 
\mbox{$\Lc\to\proton\Km\pip$}~decay modes are used 
for the~reconstruction of charm 
hadrons. Charge conjugate processes are implied throughout the~paper.
The~fiducial region for this analysis is defined in terms  
of the~\pt and~the~rapidity~$y$ of \ups~and \Charm~hadrons
to be 
\mbox{$\pty<15\gevc$}, 
\mbox{$2.0<\yy<4.5$},
\mbox{$1<\ptC<20\gevc$} and 
\mbox{$2.0<\yC<4.5$}.

The~event selection for $\ups\to\mumu$ candidates follows previous LHCb 
studies~\cite{LHCb-PAPER-2015-045}, and the~selection of \Charm~hadrons 
follows Refs.~\cite{LHCb-PAPER-2012-003,LHCb-PAPER-2013-062}.
Only good quality tracks~\cite{LHCb-DP-2013-002}, 
identified as muons~\cite{LHCb-DP-2013-001}, 
kaons, pions or protons~\cite{LHCb-DP-2012-003} are used in the~analysis. 
A~good quality vertex is required for  
\mbox{$\ups\to\mumu$},
\mbox{$\Dz\to\Km\pip$},
\mbox{$\Dp\to\Km\pip\pip$},
\mbox{$\Ds\to\Kp\Km\pip$} and 
\mbox{$\Lc\to\proton\Km\pip$}~candidates. 
For~$\Ds\to\Kp\Km\pip$ candidates,
the~mass of the~$\Kp\Km$~pair
is required to be in the~region
\mbox{$\mathrm{m}_{\Kp\Km}<1.04~\mathrm{GeV}/c^2$},
which is~dominated by the~\mbox{$\Ds\to\Pphi\pip$}~decay.
To~suppress combinatorial background the~decay time
of \Charm~hadrons 
is required to exceed $100\mum/c$.
Full decay chain fits are applied separately 
for selected \ups~and \Charm~candidates~\cite{Hulsbergen:2005pu}.
For~\ups~mesons it is required
that the~vertex is compatible with one of the~reconstructed 
$\proton\proton$~collision vertices. 
In~the~case of long-lived charm hadrons,
the~momentum direction is required to be consistent 
with the~flight direction calculated from the~locations  
of the~primary and secondary vertices.  
The~reduced $\chi^2$ of these fits, 
both $\chi^2_{\mathrm{fit}}\left(\ups\right)/\mathrm{ndf}$ and 
$\chi^2_{\mathrm{fit}}\left(\Charm\right)/\mathrm{ndf}$,  
are required to be
less than~5, where $\mathrm{ndf}$~is the~number of 
degrees of freedom in the~fit.
The requirements favour the~selection of 
charm hadrons produced promptly 
at the~$\proton\proton$~collision vertex 
and  significantly suppress
the~feed down from charm hadrons produced 
in decays of beauty hadrons.
The~contamination
  of such \Charm~hadrons in the~selected sample
  varies between $(0.4\pm0.2)\%$ for \Dp~mesons
  to $(1.5\pm0.5)\%$ for \Lc~baryons.

\begin{figure}[t]
  \setlength{\unitlength}{1mm}
  \centering
  \begin{picture}(155,120)
    %
    \put(  0, 60){ 
      \includegraphics*[width=75mm,height=60mm,%
      ]{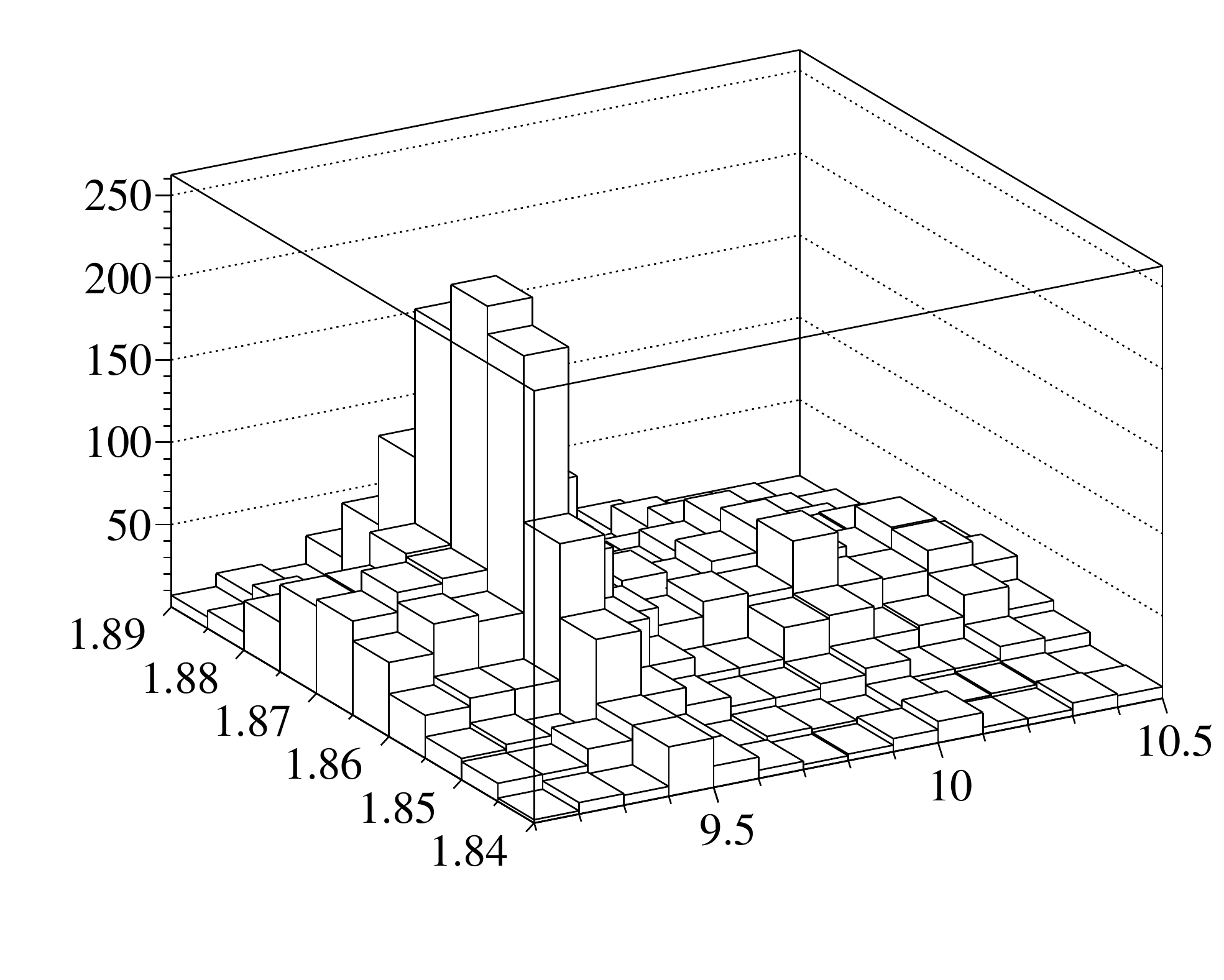}
    }
    \put( 80, 60){ 
      \includegraphics*[width=75mm,height=60mm,%
      ]{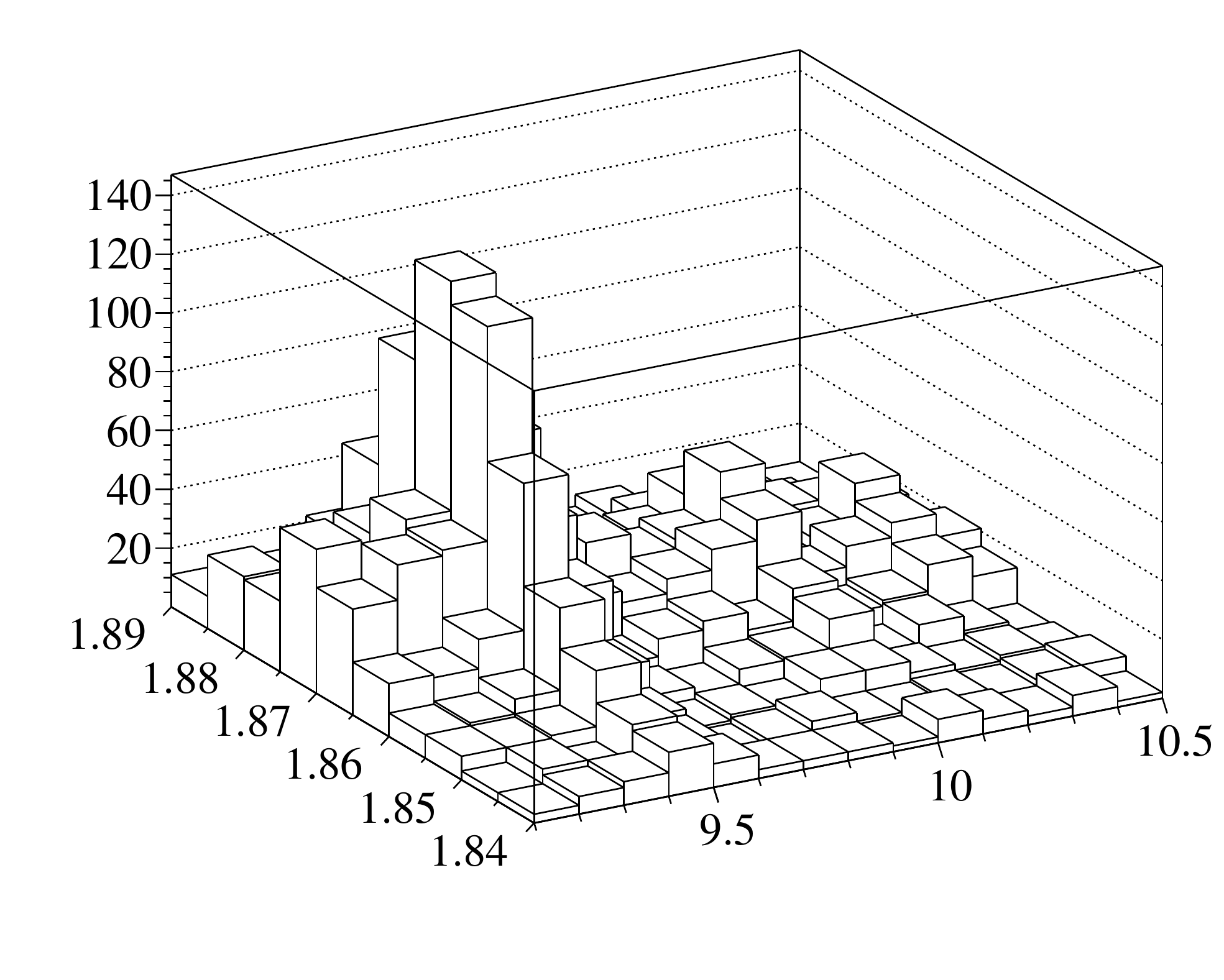}
    }
    \put(  0,  0){ 
      \includegraphics*[width=75mm,height=60mm,%
      ]{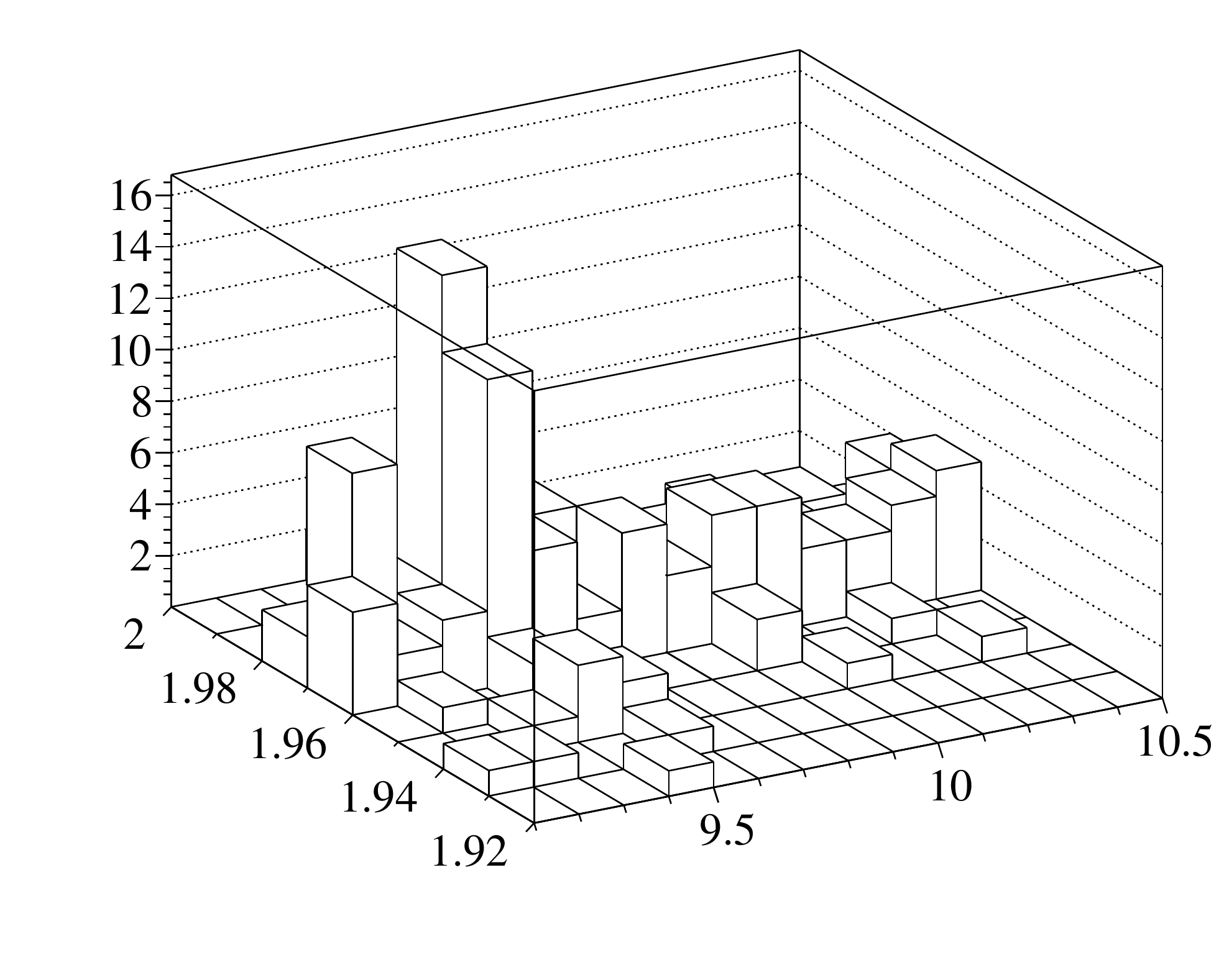}
    }
    \put( 80,  0){ 
      \includegraphics*[width=75mm,height=60mm,%
      ]{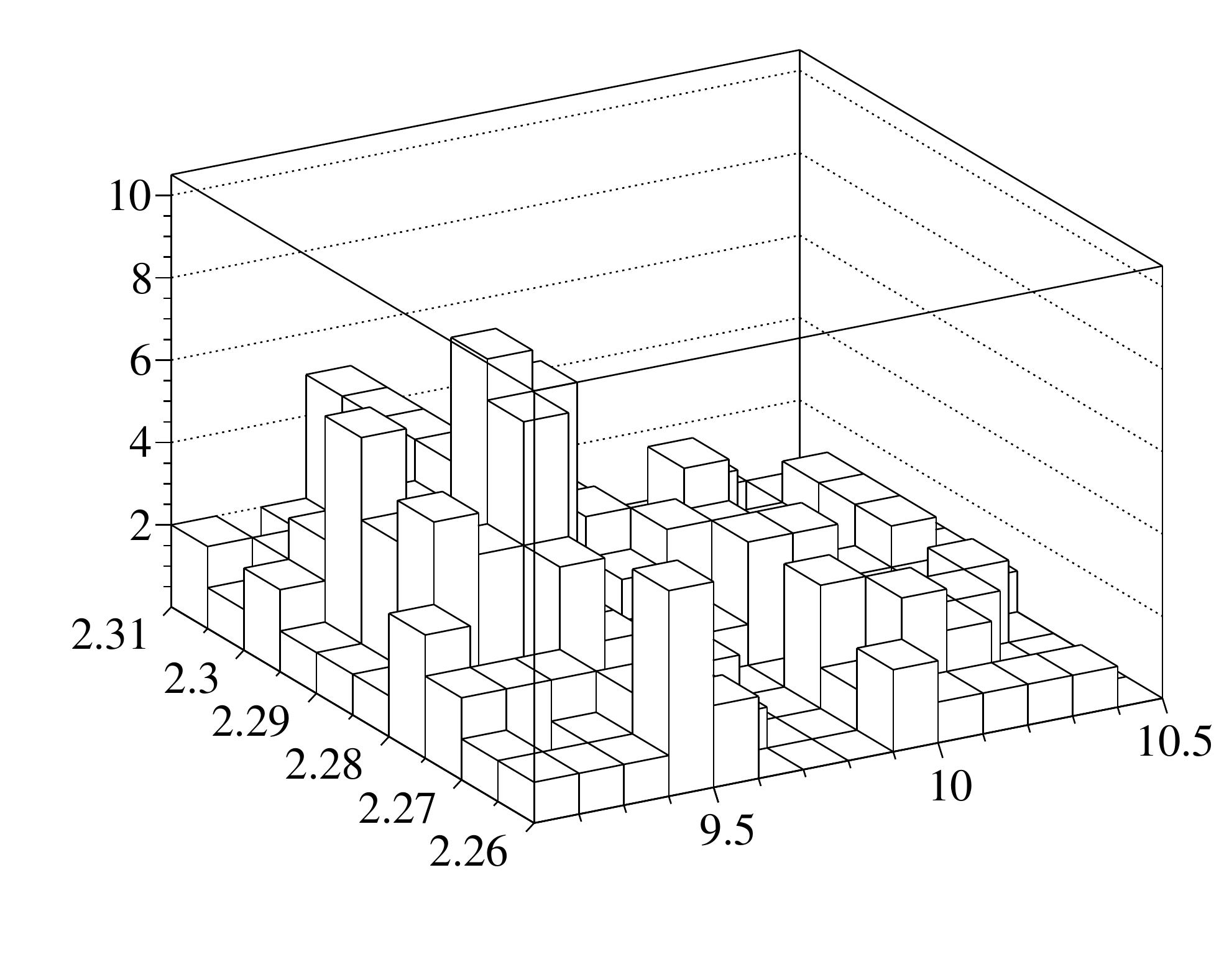}
    }
    \put(  5,115) { a)~$\begin{array}{l}\lhcb\\ \ups\Dz \end{array}$}
    \put( 85,115) { b)~$\begin{array}{l}\lhcb\\ \ups\Dp \end{array}$}
    \put(  5, 55) { c)~$\begin{array}{l}\lhcb\\ \ups\Ds \end{array}$}
    \put( 85, 55) { d)~$\begin{array}{l}\lhcb\\ \ups\Lc \end{array}$}
    \put( 3.5,76)  { 
      \begin{rotate}{-32} 
        $m_{\Km\pip}~\left[\!\gevcc\right]$
      \end{rotate}
    }
    \put(83.5,76)  { 
      \begin{rotate}{-32} 
        $m_{\Km\pip\pip}~\left[\!\gevcc\right]$
      \end{rotate}
    }
    \put( 3.5,16)  { 
      \begin{rotate}{-32} 
        $m_{\Km\Kp\pip}~\left[\!\gevcc\right]$
      \end{rotate}
    }
    \put(83.5,16)  { 
      \begin{rotate}{-32} 
        $m_{\mathrm{p}\Km\pip}~\left[\!\gevcc\right]$
      \end{rotate}
    }
    \put( 45,64.0)  { 
      \begin{rotate}{10} 
        $m_{\mumu}~\left[\!\gevcc\right]$
      \end{rotate}
    }
    \put(125,64.0)  { 
      \begin{rotate}{10} 
        $m_{\mumu}~\left[\!\gevcc\right]$
      \end{rotate}
    }
    \put( 45, 4.0)  { 
      \begin{rotate}{10} 
        $m_{\mumu}~\left[\!\gevcc\right]$
      \end{rotate}
    }
    \put(125, 4.0)  { 
      \begin{rotate}{10} 
        $m_{\mumu}~\left[\!\gevcc\right]$
      \end{rotate}
    }
    \put ( -2,77) { \small \begin{sideways}\mbox{$N/(100\times5\,\mathrm{MeV}^2/c^4)$}\end{sideways}} 
    \put ( -2,19) { \small \begin{sideways}\mbox{$N/(100\times10\,\mathrm{MeV}^2/c^4)$}\end{sideways}} 
    \put ( 78,79) { \small \begin{sideways}\mbox{$N/(100\times5\,\mathrm{MeV}^2/c^4)$}\end{sideways}} 
    \put ( 79,19) { \small \begin{sideways}\mbox{$N/(100\times10\,\mathrm{MeV}^2/c^4)$}\end{sideways}} 
  \end{picture}
  \caption { \small
    Invariant mass distributions for selected 
    combination of 
    \ups~mesons and \Charm~hadrons:
    a)\,$\ups\Dz$,
    b)\,$\ups\Dp$,
    c)\,$\ups\Ds$ and 
    d)\,$\ups\Lc$.
  }
  \label{fig:signal_lego}
\end{figure}

The~selected \ups~and \Charm~candidates are paired to form $\ups\Charm$~candidates. 
A~global fit to the~$\ups\Charm$~candidates  is performed~\cite{Hulsbergen:2005pu},
similar to that described above, which requires both hadrons to be consistent 
with originating from a~common vertex. The~reduced $\chi^2$~of this fit, 
$\chi^2_{\mathrm{fit}}\left(\ups\Charm\right)/\mathrm{ndf}$, 
is required to be less than 5.
This reduces the~background from  the~pile\nobreakdash-up of 
two independent 
$\proton\proton$~interactions producing 
separately a~\ups~meson and~\Charm~hadron to a~negligible level, 
keeping 100\% of the~signal \ups~mesons 
and \Charm~hadrons from the~same primary vertex.
The~two\nobreakdash-dimensional mass distributions for $\ups\Charm$~pairs 
after the~selection are displayed 
in~Fig.~\ref{fig:signal_lego}.

\section{Signal extraction and cross-section determination}
\label{sec:signal}

The event yields are determined using unbinned extended maximum likelihood
fits to the two-dimensional $\ups\Charm$~mass distributions of the~selected candidates.
The~fit model is a~sum of several components, each of which 
is the~product of a~dimuon mass distribution, corresponding to an~individual
\ups~state or combinatorial background,
and a~\Charm~candidate mass 
distribution,
corresponding to a~\Charm~signal 
or combinatorial
background component.
The~\mbox{$\YoneS\to\mumu$},
    \mbox{$\YtwoS\to\mumu$} and 
    \mbox{$\YthreeS\to\mumu$}~signals are each modelled by a~double-sided Crystal~Ball
function~\cite{LHCb-PAPER-2011-013,Skwarnicki:1986xj,LHCb-PAPER-2013-066}
and referred to as
$S_{\ups}$ in this section.
A~modified Novosibirsk
function~\cite{Lees:2011gw}\,(referred to as $S_{\Charm}$) is used to describe 
the~\mbox{$\Dz\to\Km\pip$}, 
\mbox{$\Dp\to\Km\pip\pip$},
\mbox{$\Ds\to\Kp\Km\pip$}~and 
\mbox{$\Lc\to\proton\Km\pip$}~signals.
All shape parameters and signal peak positions 
are fixed from fits to large   
inclusive $\ups\to\mumu$ and \Charm~hadron data samples. 
Combinatorial background components 
$B_{\mumu}$ and 
$B_{\Charm}$ 
are modelled with a~product of exponential and polynomial functions
\begin{equation}
B ( m )\propto \mathrm{e}^{-\upbeta m} \times {\cal{P}}_{n}(m),
\label{eq:b}
\end{equation}
with a~slope parameter $\upbeta$~and
a~polynomial function ${\mathcal{P}}_{n}$, which is represented as 
a~B\'ezier sum of basic Bernstein polynomials of order~$n$
with non\nobreakdash-negative coefficients~\cite{Farouki2012379}.
For~the~large yield $\ups\Dz$~and $\ups\Dp$~samples, 
the~second\nobreakdash-order 
polynomials\,\mbox{($n=2$)} are used in the~fit, 
while \mbox{$n=1$}~is used for the~$\ups\Ds$~and $\ups\Lc$~cases.

These basic functions
are used to build the components of  the two dimensional 
mass fit following Ref.~\cite{LHCb-PAPER-2012-003}. 
For each \Charm~hadron the~reconstructed signal sample consists
of the~following components:
\begin{itemize}
\item[-] Three $\ups\Charm$~signal components:
  each is modelled by a product
  of the~individual signal \ups~components, $S_{\YoneS}(m_{\mumu})$, 
  $S_{\YtwoS}(m_{\mumu})$ or $S_{\YthreeS}(m_{\mumu})$, and 
  signal \Charm~hadron component, $S_{\Charm}(m_{\Charm})$.
\item[-] Three components describing the~production of single \ups~mesons 
  together with combinatorial
  background for the~\Charm~signal:
  each component is modelled by a~product of
  the~signal \ups~component, $S_{\ups}(m_{\mumu})$
  and the~background component $B_{\Charm}(m_{\Charm})$.
\item[-] Single production of \Charm~hadrons together with 
  combinatorial background  for the~\ups~component: 
  this is modelled by a~product of
  the~signal \Charm~component, $S_{\Charm}(m_{\Charm})$, 
  and the~background component 
  $B_{\mumu}(m_{\mumu})$.
\item[-] Combinatorial background: 
  this is modelled by a~product of the~individual background components
  $B_{\mumu}(m_{\mumu})$ and $B_{\Charm}(m_{\Charm})$.
\end{itemize}
For each \Charm~hadron the~complete
fit function
$\mathcal{F}( m_{\mumu}, m_{\Charm})$ is 
\begin{equation} \label{eq:pdf}
\begin{split}
  {\mathcal{F}}( m_{\mumu}, m_{\Charm}) & = 
  N^{\YoneS\Charm}         \times   S_{\YoneS}(m_{\mumu}) \times  S_{\Charm}(m_{\Charm}) \\ 
  & +  
  N^{\YtwoS\Charm}         \times   S_{\YtwoS}(m_{\mumu}) \times  S_{\Charm}(m_{\Charm}) \\ 
  & +  
  N^{\YthreeS\Charm}         \times   S_{\YthreeS}(m_{\mumu}) \times  S_{\Charm}(m_{\Charm}) \\ 
  & + 
  N^{\YoneS B}   \times   S_{\YoneS}(m_{\mumu}) \times  B_{\Charm}(m_{\Charm})  \\ 
  & +  
  N^{\YtwoS B}   \times   S_{\YtwoS}(m_{\mumu}) \times  B_{\Charm}(m_{\Charm})  \\ 
  & +  
  N^{\YthreeS B }   \times   S_{\YthreeS}(m_{\mumu}) \times  B_{\Charm}(m_{\Charm}) \\ 
  & +  
  N^{ B \Charm} \times   B_{\mumu}(m_{\mumu}) \times  S_{\Charm}(m_{\Charm}) \\
  & +  
  N^{BB}      \times  B_{\mumu}(m_{\mumu}) \times  B_{\Charm}(m_{\Charm}), \\
\end{split}
\end{equation}
where the~different coefficients 
$N^{\ups\Charm}$,
$N^{\ups B}$,
$N^{ B\Charm}$ and 
$N^{ BB}$
are the~yields of the~eight components described above.

\begin{figure}[t]
  \setlength{\unitlength}{1mm}
  \centering
  \begin{picture}(150,120)
    %
    \put(  0, 60){ 
      \includegraphics*[width=75mm,height=60mm,%
      ]{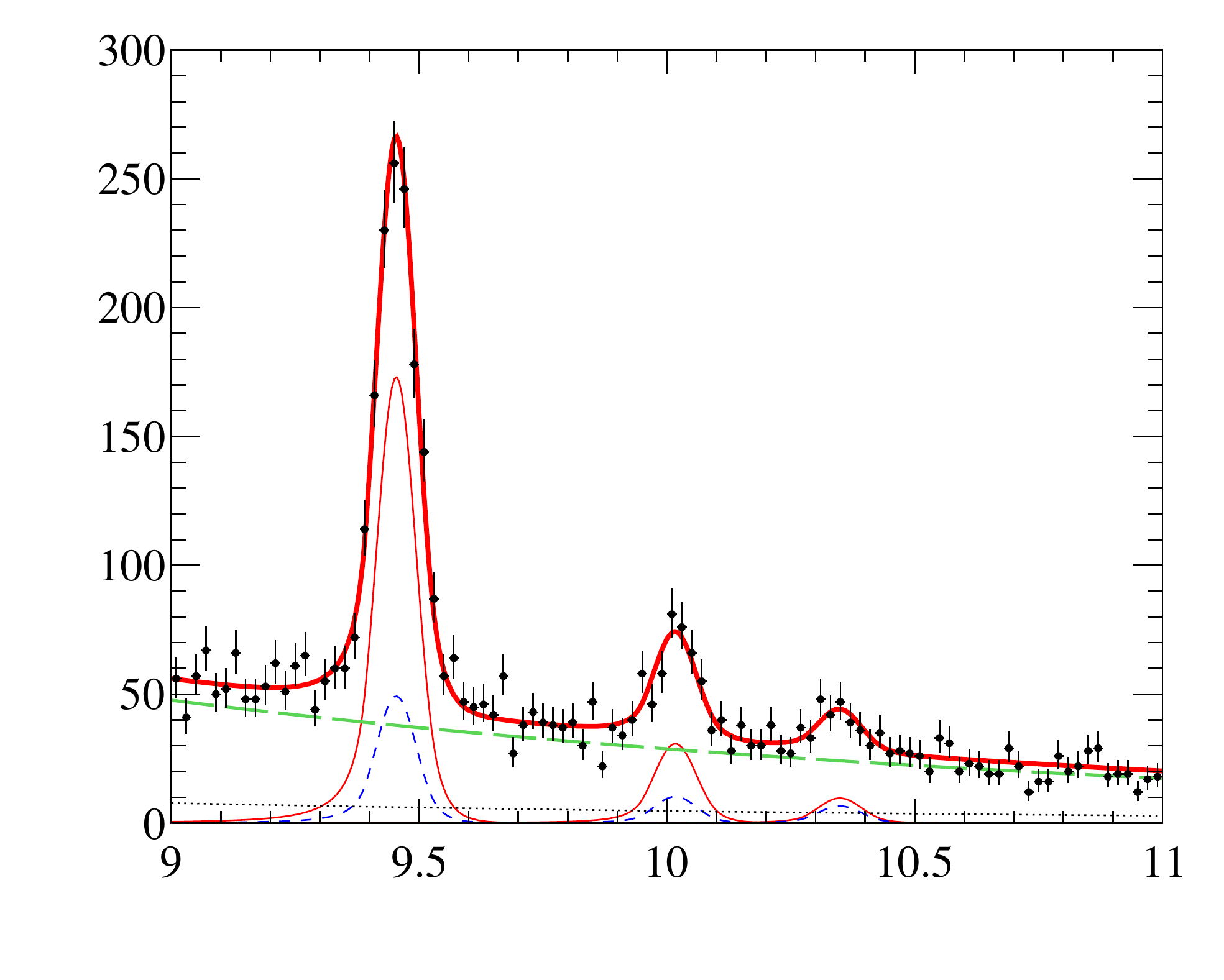}
    }
    \put( 75, 60){ 
      \includegraphics*[width=75mm,height=60mm,%
      ]{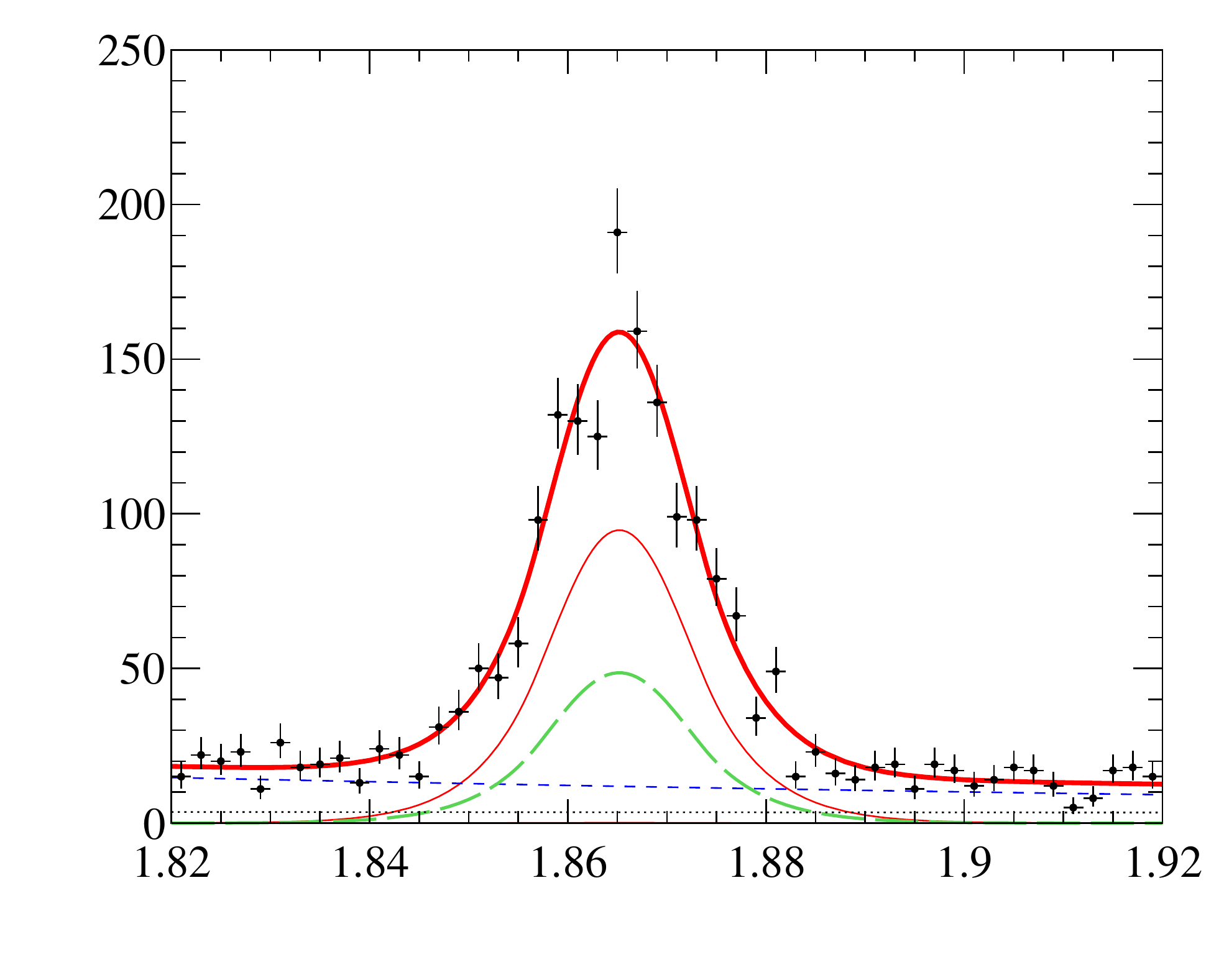}
    }
    \put(  0,  0){ 
      \includegraphics*[width=75mm,height=60mm,%
      ]{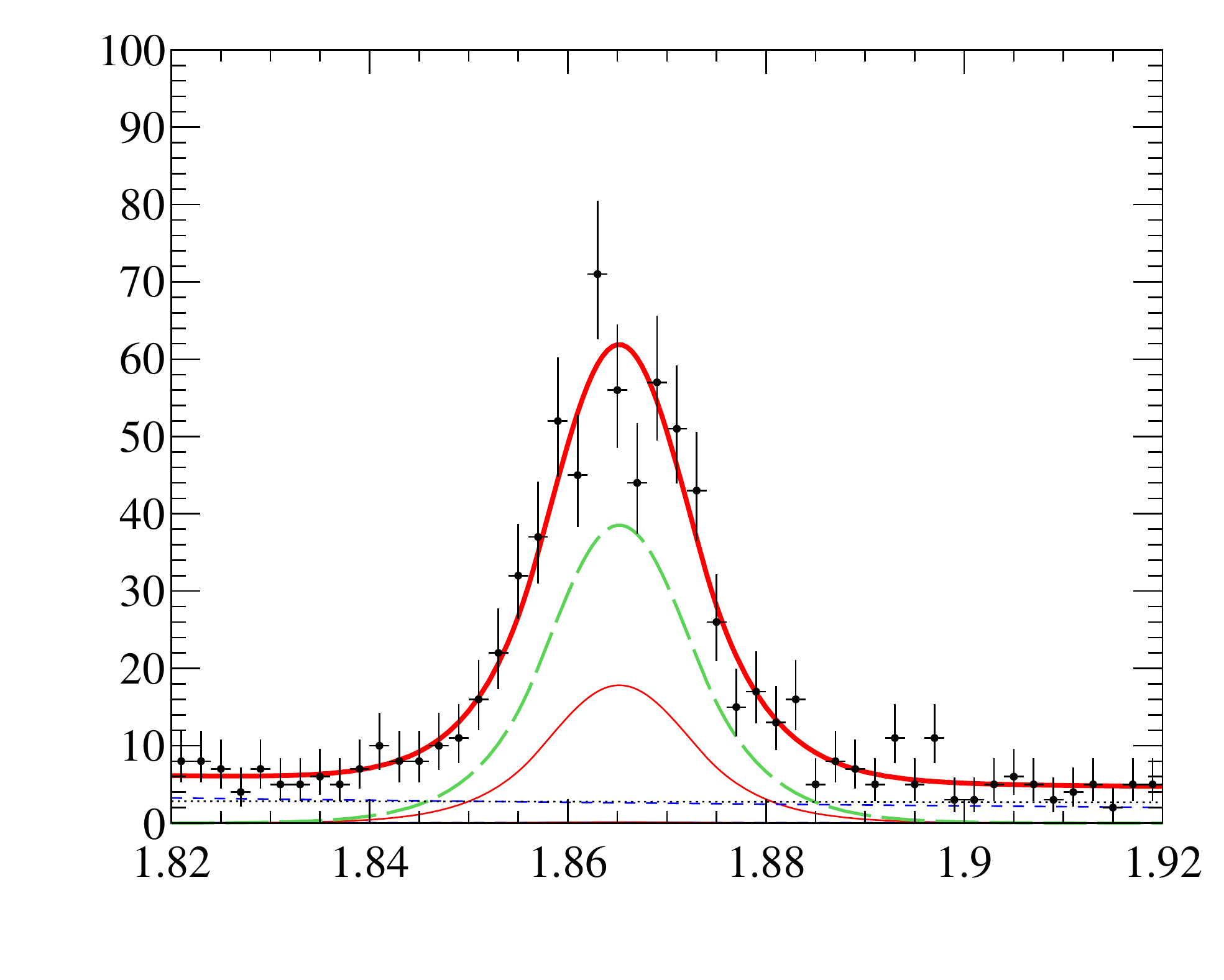}
    }
    \put( 75,  0){ 
      \includegraphics*[width=75mm,height=60mm,%
      ]{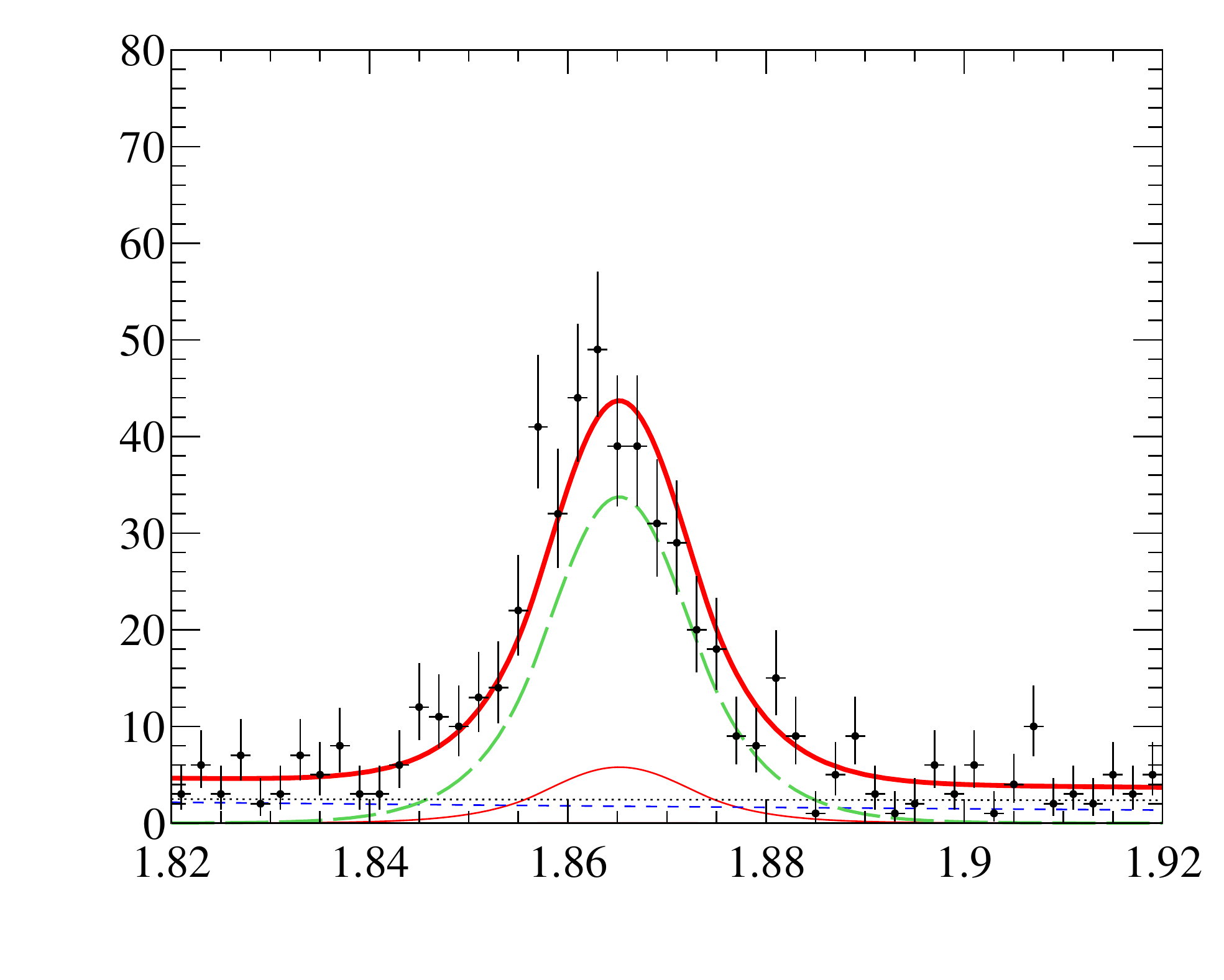}
    }
    \put(  1, 75) { \begin{sideways}Candidates/(20\mevcc)\end{sideways}} 
    \put( 76, 75) { \begin{sideways}Candidates/(2\mevcc) \end{sideways}} 
    \put(  1, 15) { \begin{sideways}Candidates/(2\mevcc) \end{sideways}} 
    \put( 76, 15) { \begin{sideways}Candidates/(2\mevcc) \end{sideways}} 
    \put( 35, 61) { $m_{\mumu}$  } \put(  57, 61) { $\left[\!\gevcc\right]$ }  
    \put(110, 61) { $m_{\Km\pip}$ } \put(132, 61) { $\left[\!\gevcc\right]$ }  
    \put( 35,  1) { $m_{\Km\pip}$ } \put( 57,  1) { $\left[\!\gevcc\right]$ }  
    \put(110,  1) { $m_{\Km\pip}$ } \put(132,  1) { $\left[\!\gevcc\right]$ }  
    \put( 47,108) { a)~$\begin{array}{l} \lhcb \\ \ups\Dz\end{array}$ }
    \put(122,108) { b)~$\begin{array}{l} \lhcb \\ \YoneS\Dz\end{array}$ }
    \put( 47, 48) { c)~$\begin{array}{l} \lhcb \\ \YtwoS\Dz\end{array}$ }
    \put(122, 48) { d)~$\begin{array}{l} \lhcb \\ \YthreeS\Dz\end{array}$ }
  \end{picture}
  \caption { \small
    Projections from two-dimensional extended 
    unbinned maximum likelihood fits in bands 
    a)~\mbox{$1.844<m_{\Km\pip}<1.887 \mevcc$},    
    b)~\mbox{$9.332<m_{\mumu}  <9.575 \gevcc$},    
    c)~\mbox{$9.889<m_{\mumu}  <10.145\gevcc$} and   
    d)~\mbox{$10.216<m_{\mumu} <10.481\gevcc$}.  
    The total fit function is shown by a~solid thick\,(red) curve;
    three individual $\ups\Dz$~signal components 
    are shown by solid thin\,(red) curves;
    three components describing \ups~signals and combinatorial 
    background in $\Km\pip$~mass 
    are shown with short-dashed\,(blue) curves; 
    the component modelling the true \Dz~signal and combinatorial
    background in \mumu~mass is shown with a~long-dashed\,(green) curve 
    and the component describing combinatorial background
    is shown with a~thin dotted\,(black) line.
  }
  \label{fig:signal_d0_proj_bands}
\end{figure}

\begin{figure}[t]
  \setlength{\unitlength}{1mm}
  \centering
  \begin{picture}(150,120)
    %
    \put(  0, 60){ 
      \includegraphics*[width=75mm,height=60mm,%
      ]{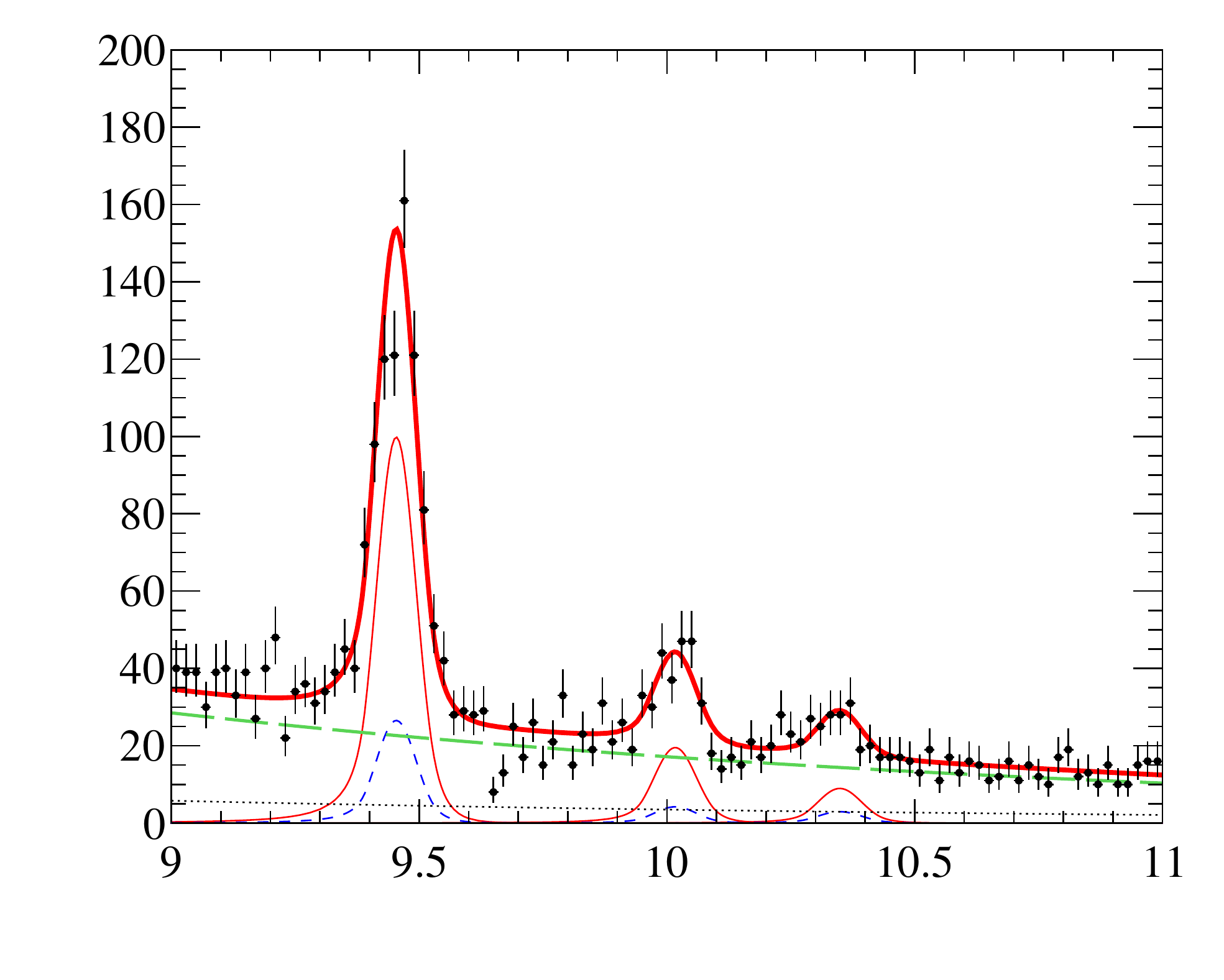}
    }
    \put( 75, 60){ 
      \includegraphics*[width=75mm,height=60mm,%
      ]{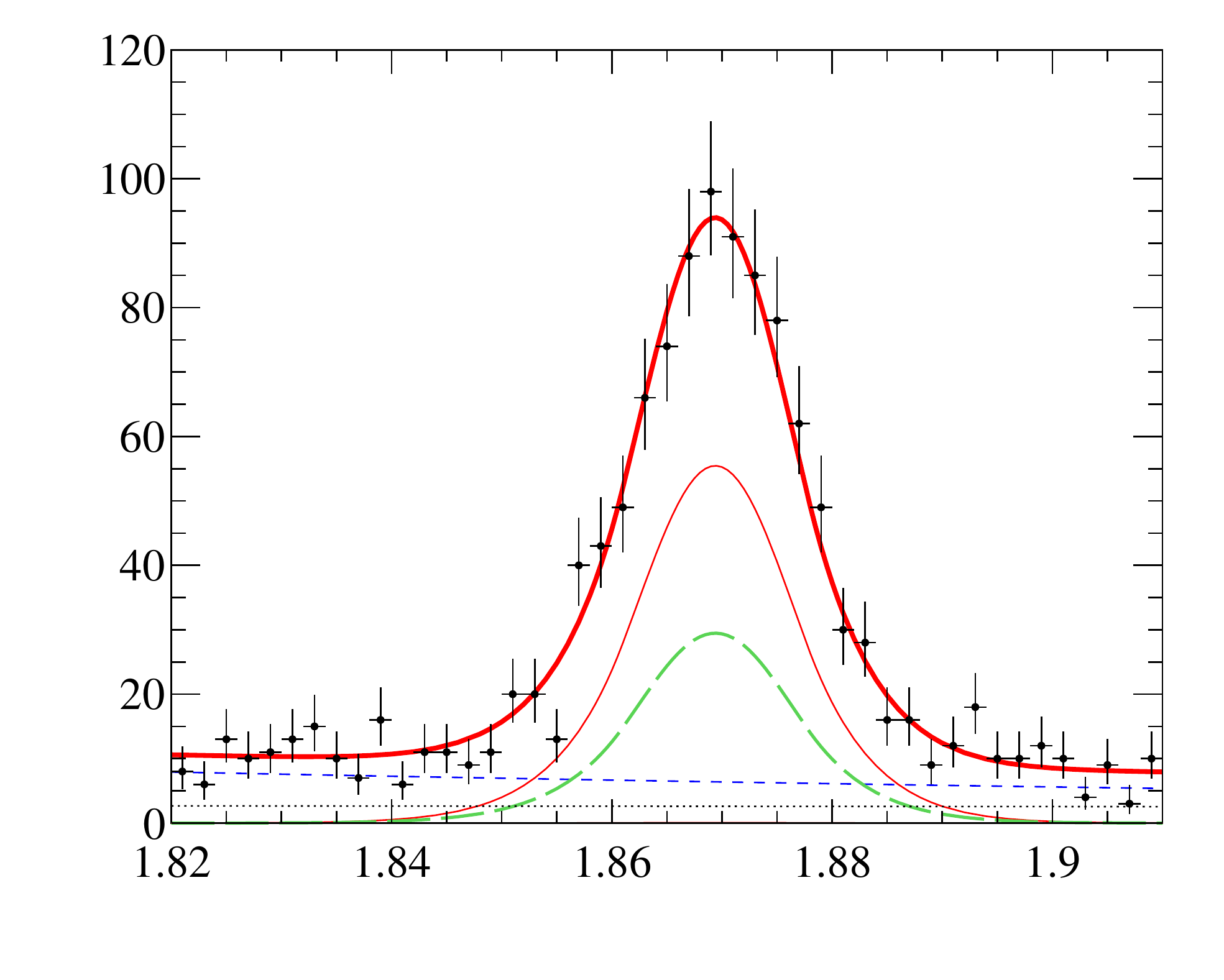}
    }
    \put(  0,  0){ 
      \includegraphics*[width=75mm,height=60mm,%
      ]{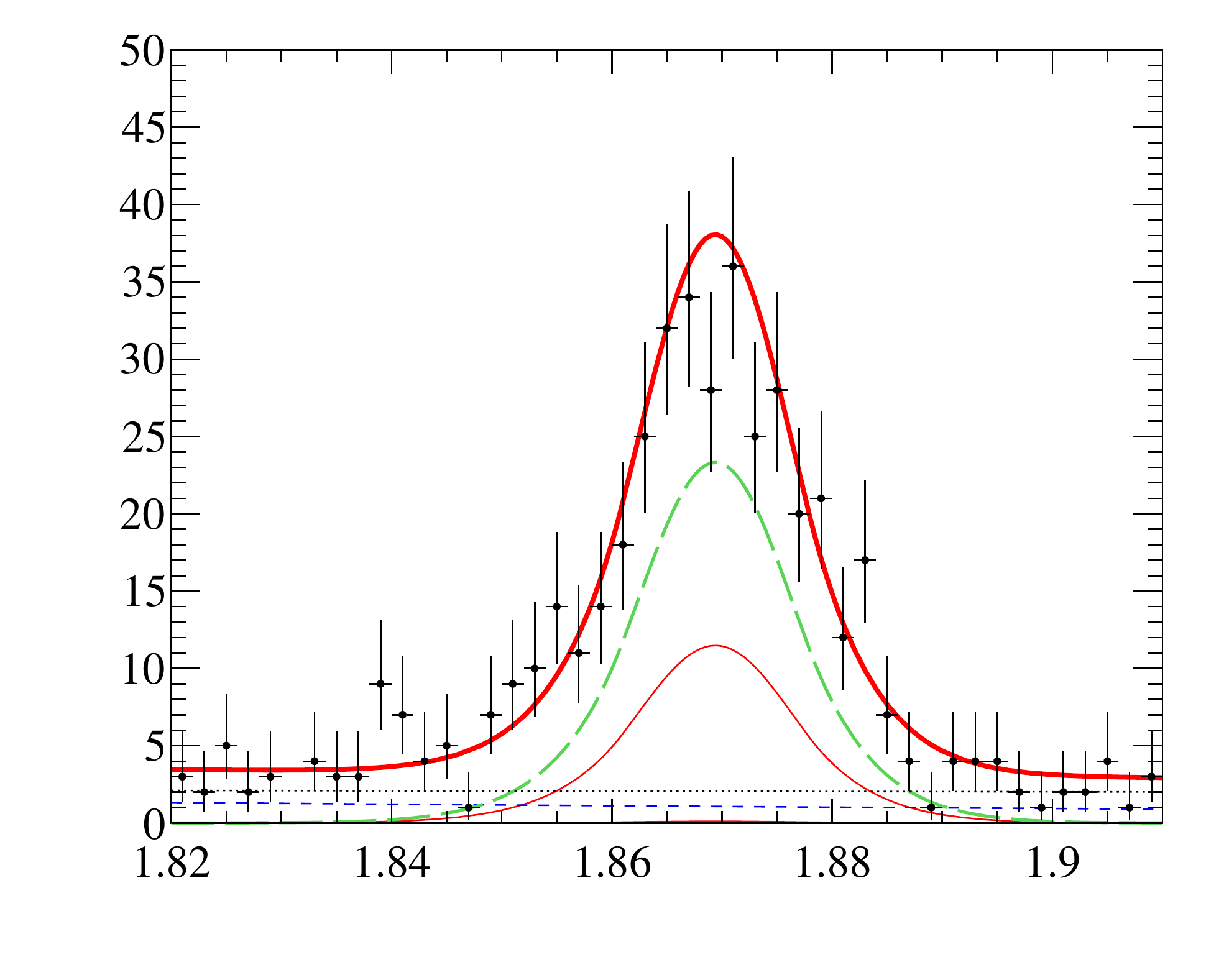}
    }
    \put( 75,  0){ 
      \includegraphics*[width=75mm,height=60mm,%
      ]{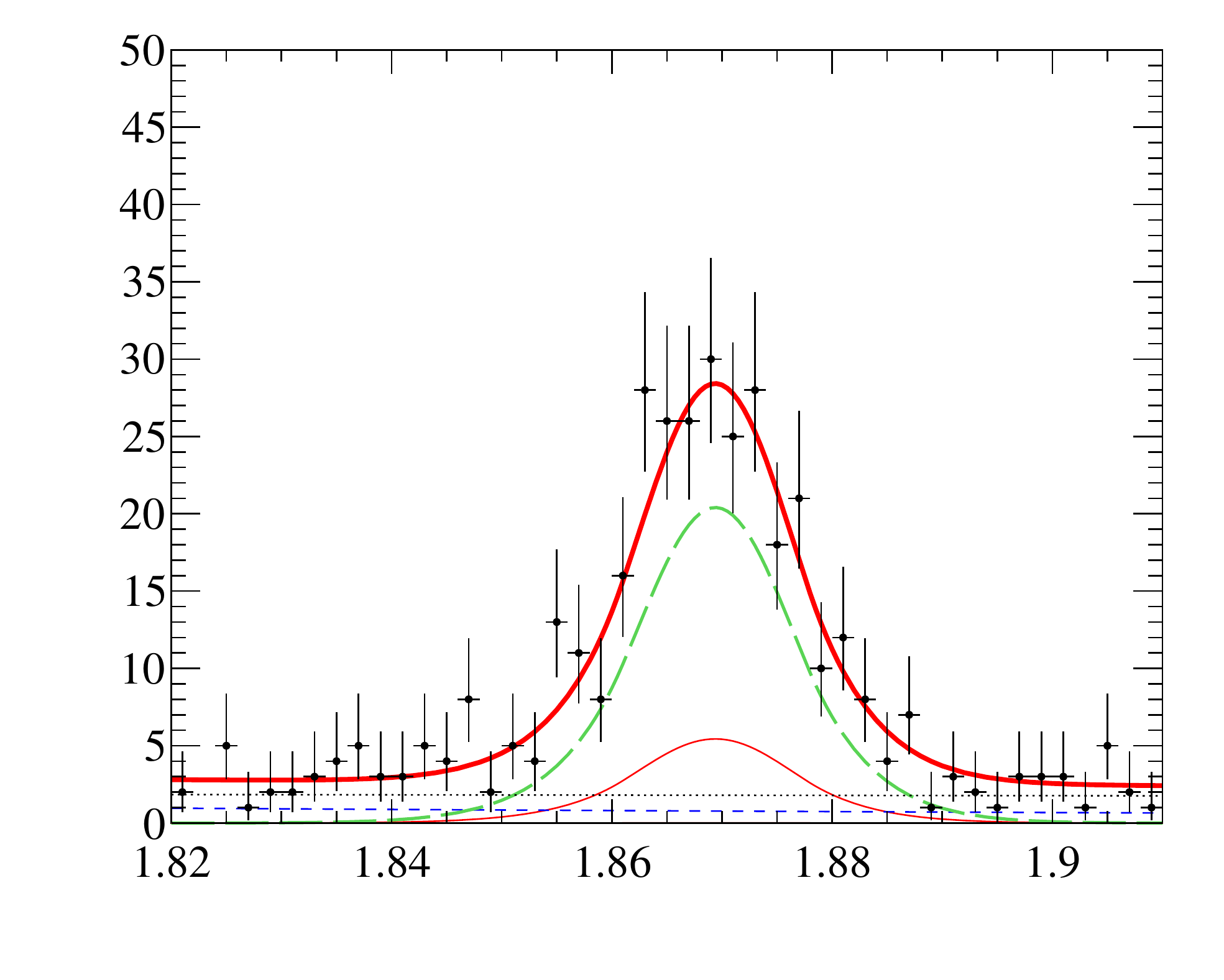}
    }
    \put(  1, 75) { \begin{sideways}Candidates/(20\mevcc)\end{sideways}} 
    \put( 76, 75) { \begin{sideways}Candidates/(2\mevcc) \end{sideways}} 
    \put(  1, 15) { \begin{sideways}Candidates/(2\mevcc) \end{sideways}} 
    \put( 76, 15) { \begin{sideways}Candidates/(2\mevcc) \end{sideways}} 
    \put( 35, 61) { $m_{\mumu}$      } \put( 57, 61) { $\left[\!\gevcc\right]$ }  
    \put(110, 61) { $m_{\Km\pip\pip}$ } \put(132, 61) { $\left[\!\gevcc\right]$ }  
    \put( 35,  1) { $m_{\Km\pip\pip}$ } \put( 57,  1) { $\left[\!\gevcc\right]$ }  
    \put(110,  1) { $m_{\Km\pip\pip}$ } \put(132,  1) { $\left[\!\gevcc\right]$ }  
    %
    %
    \put( 47,108) { a)~$\begin{array}{l} \lhcb \\ \ups\Dp\end{array}$ }
    \put(122,108) { b)~$\begin{array}{l} \lhcb \\ \YoneS\Dp\end{array}$ }
    \put( 47, 48) { c)~$\begin{array}{l} \lhcb \\ \YtwoS\Dp\end{array}$ }
    \put(122, 48) { d)~$\begin{array}{l} \lhcb \\ \YthreeS\Dp\end{array}$ }
  \end{picture}
  \caption { \small
    Projections from two-dimensional extended 
    unbinned maximum likelihood fits in bands
    a)~\mbox{$1.848<m_{\Km\pip\pip}<1.891 \mevcc$},   
    b)~\mbox{$9.332<m_{\mumu}  <9.575 \gevcc$ },   
    c)~\mbox{$9.889<m_{\mumu}  <10.145\gevcc$ } and    
    d)~\mbox{$10.216<m_{\mumu} <10.481\gevcc$}.   
    The total fit function is shown by a~solid thick\,(red) curve;
    three individual $\ups\Dp$~signal components 
    are shown by solid thin\,(red) curves;
    three components describing \ups~signals and combinatorial 
    background in $\Km\pip\pip$~mass 
    are shown with short-dashed\,(blue) curves; 
    the component modelling the true \Dp~signal and combinatorial
    background in \mumu~mass is shown with a~long-dashed\,(green) curve 
    and the component describing combinatorial background
    is shown with a~thin dotted\,(black) line.
  }
  \label{fig:signal_dp_proj_bands}
\end{figure}

\begin{figure}[t]
  \setlength{\unitlength}{1mm}
  \centering
  \begin{picture}(150,120)
    %
    \put(  0, 60){ 
      \includegraphics*[width=75mm,height=60mm,%
      ]{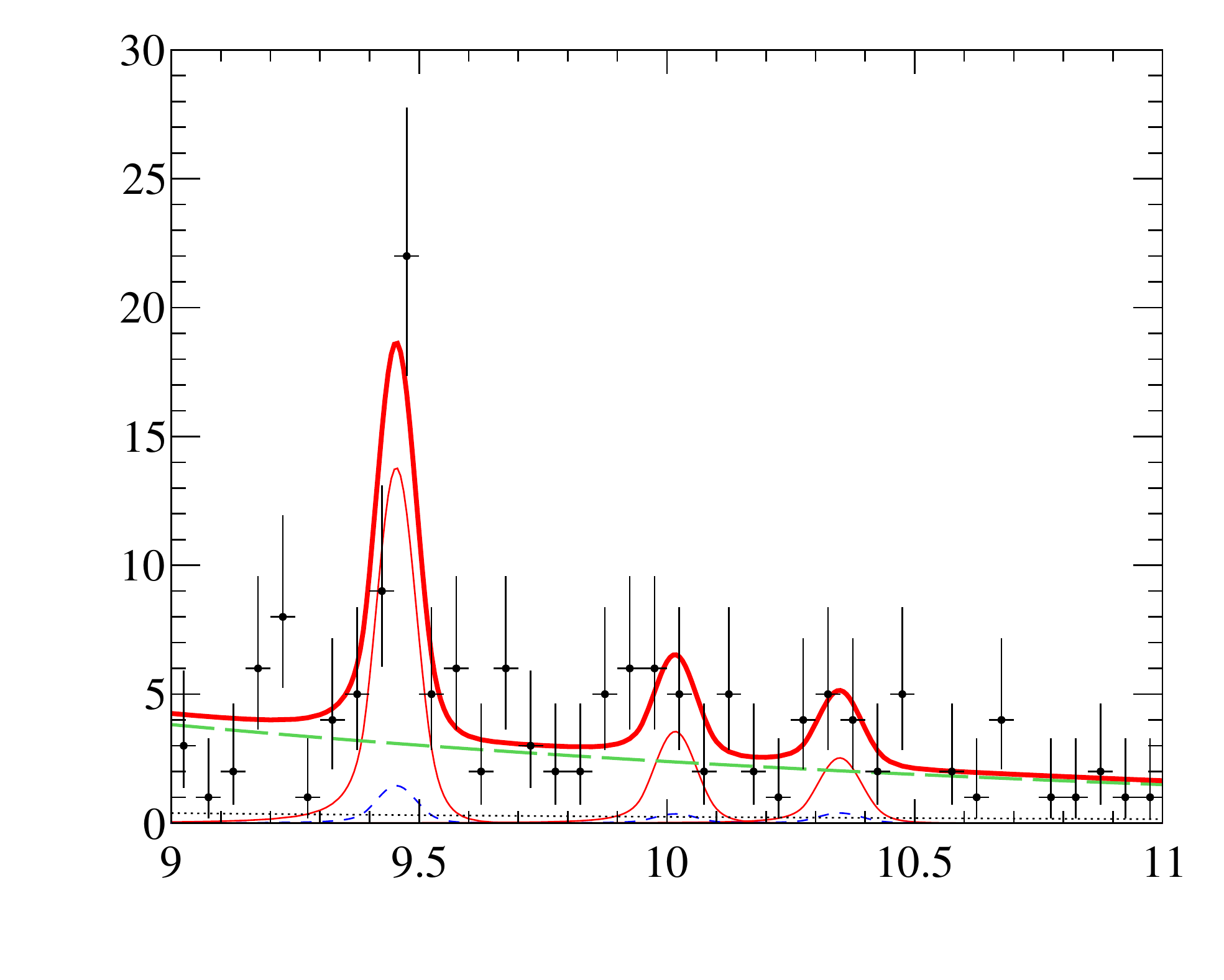}
    }
    \put( 75, 60){ 
      \includegraphics*[width=75mm,height=60mm,%
      ]{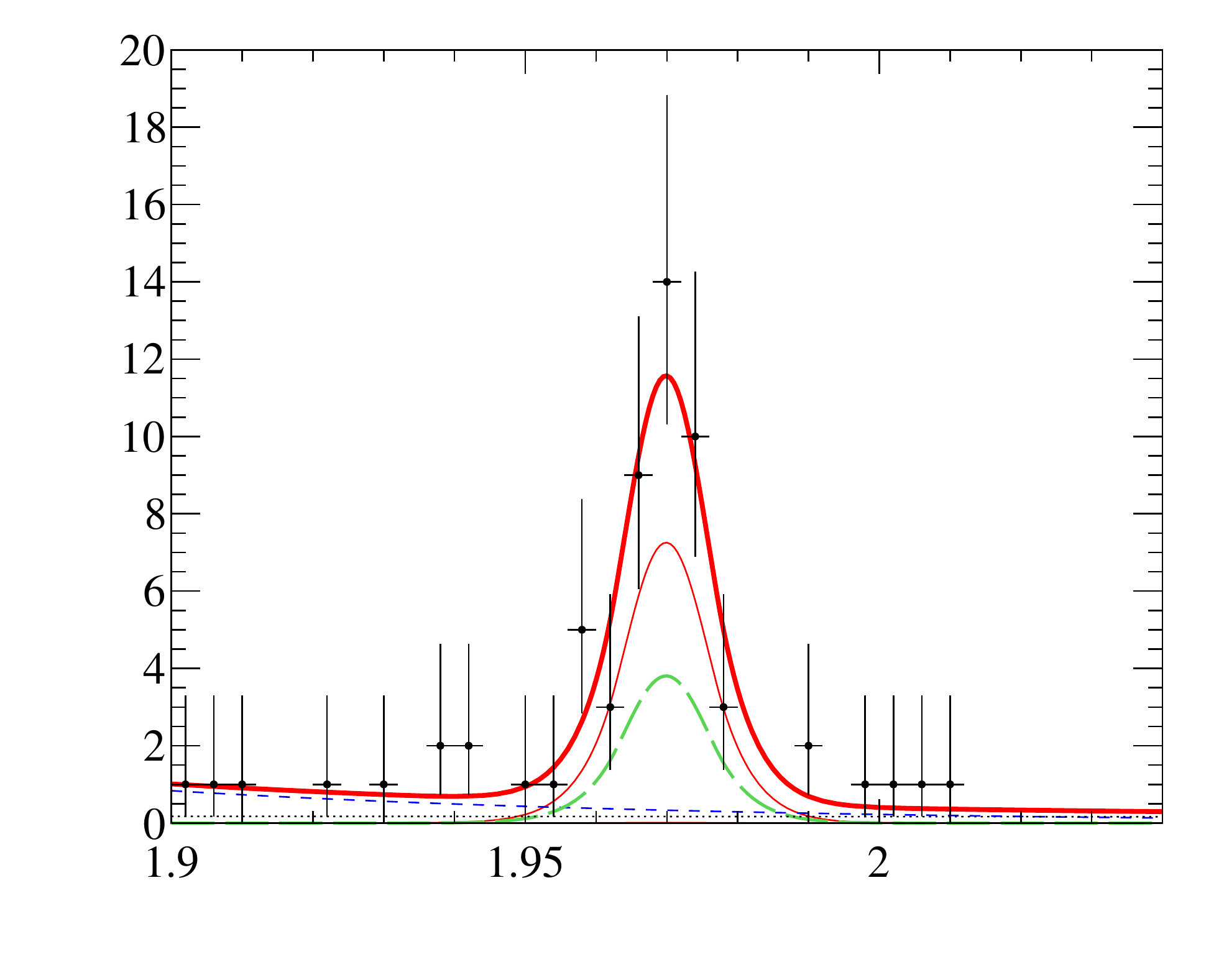}
    }
    \put(  0,  0){ 
      \includegraphics*[width=75mm,height=60mm,%
      ]{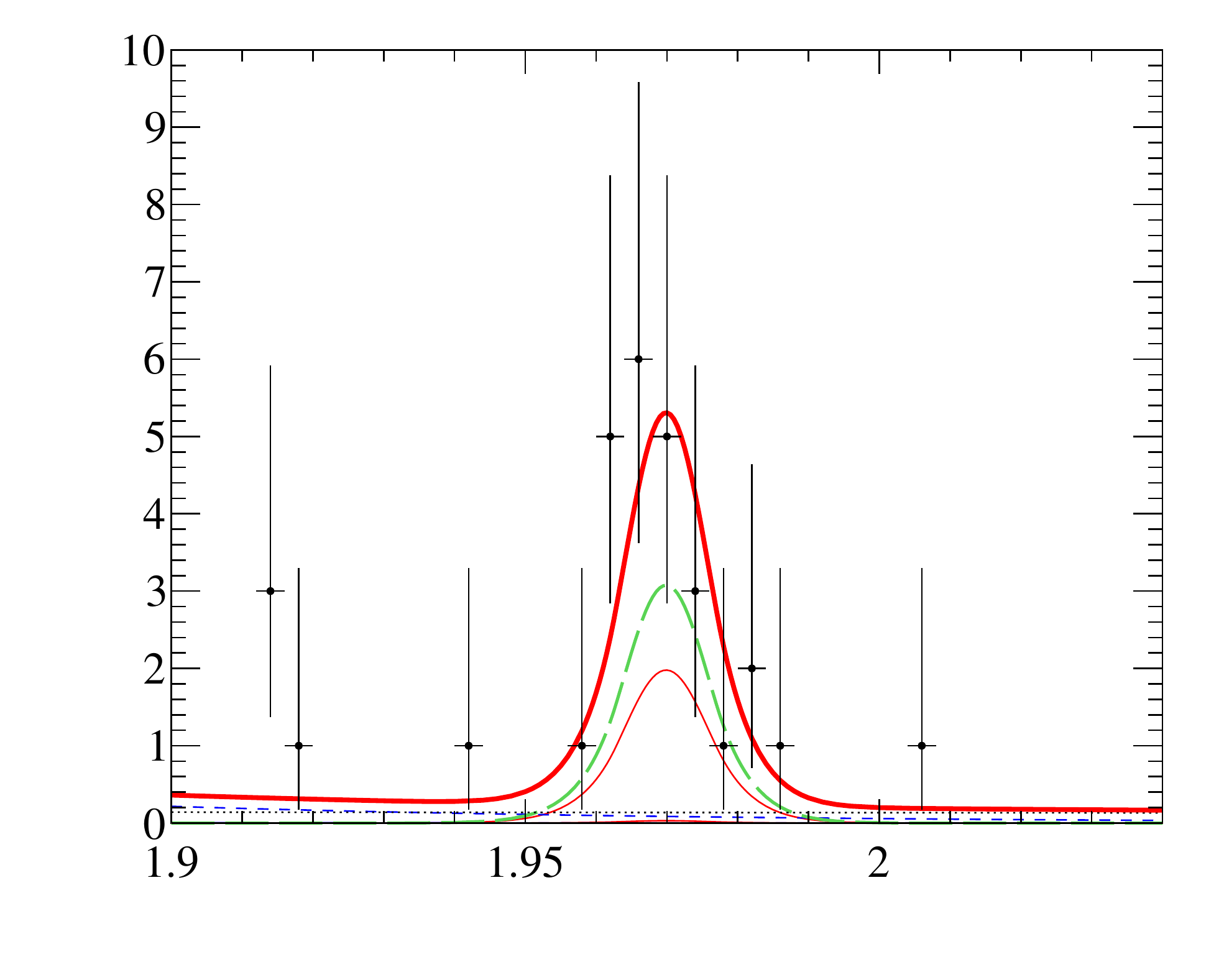}
    }
    \put( 75,  0){ 
      \includegraphics*[width=75mm,height=60mm,%
      ]{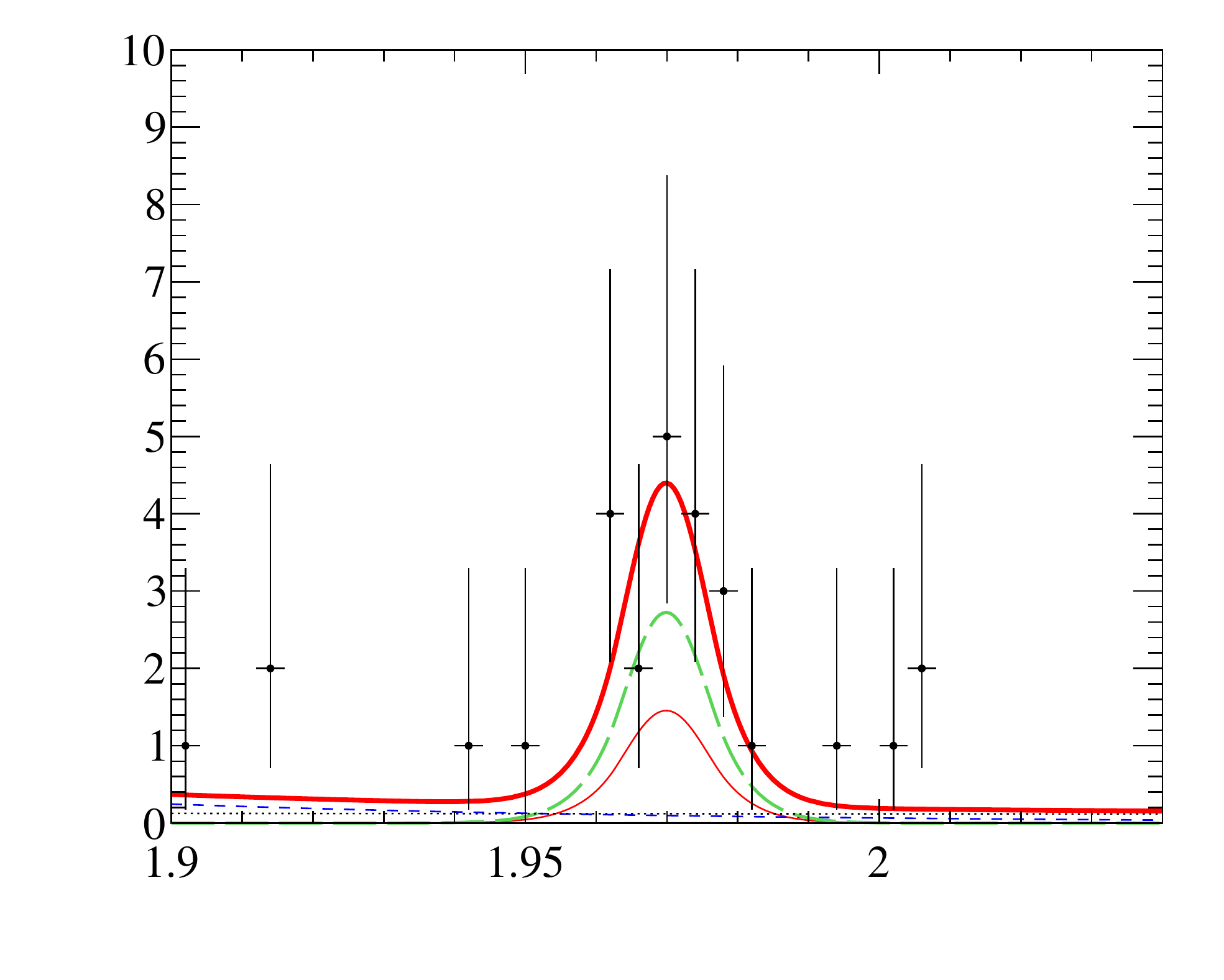}
    }
    \put(  2, 75) { \begin{sideways}Candidates/(50\mevcc)\end{sideways}} 
    \put( 77, 75) { \begin{sideways}Candidates/(4\mevcc) \end{sideways}} 
    \put(  2, 15) { \begin{sideways}Candidates/(4\mevcc) \end{sideways}} 
    \put( 77, 15) { \begin{sideways}Candidates/(4\mevcc) \end{sideways}} 
    \put( 35, 61) { $m_{\mumu}$   } \put( 57, 61) { $\left[\!\gevcc\right]$ }  
    \put(110, 61) { $m_{\left(\Km\Kp\right)_{\Pphi}\pip}$ } \put(132, 61) { $\left[\!\gevcc\right]$ }  
    \put( 35,  1) { $m_{\left(\Km\Kp\right)_{\Pphi}\pip}$ } \put( 57,  1) { $\left[\!\gevcc\right]$ }  
    \put(110,  1) { $m_{\left(\Km\Kp\right)_{\Pphi}\pip}$ } \put(132,  1) { $\left[\!\gevcc\right]$ }  
    %
    %
    \put( 46,108) { a)~$\begin{array}{l} \lhcb \\ \ups\Ds\end{array}$ }
    \put(121,108) { b)~$\begin{array}{l} \lhcb \\ \YoneS\Ds\end{array}$ }
    \put( 46, 48) { c)~$\begin{array}{l} \lhcb \\ \YtwoS\Ds\end{array}$ }
    \put(121, 48) { d)~$\begin{array}{l} \lhcb \\ \YthreeS\Ds\end{array}$ }
  \end{picture}
  \caption { \small
    Projections from two-dimensional extended 
    unbinned maximum likelihood fits in bands
    a)~\mbox{$1.952<m_{\left(\Km\Kp\right)_{\Pphi}\pip}<1.988 \mevcc$},   
    b)~\mbox{$9.332<m_{\mumu}  <9.575 \gevcc$ },
    c)~\mbox{$9.889<m_{\mumu}  <10.145\gevcc$ } and  
    d)~\mbox{$10.216<m_{\mumu} <10.481\gevcc$}.   
    The~total fit function is shown by a~solid thick\,(red) curve;
    three individual $\ups\Ds$~signal components 
    are shown by solid thin\,(red) curves;
    three components describing \ups~signals and combinatorial 
    background in $\left(\Km\Kp\right)_{\Pphi}\pip$~mass 
    are shown with short-dashed\,(blue) curves; 
    the component modelling the true \Ds~signal and combinatorial
    background in \mumu~mass is shown with a~long-dashed\,(green) curve 
    and the component describing combinatorial background
    is shown with a~thin dotted\,(black) line.
  }
  \label{fig:signal_ds_proj_bands}
\end{figure}

\begin{figure}[t]
  \setlength{\unitlength}{1mm}
  \centering
  \begin{picture}(150,120)
    %
    \put(  0, 60){ 
      \includegraphics*[width=75mm,height=60mm,%
      ]{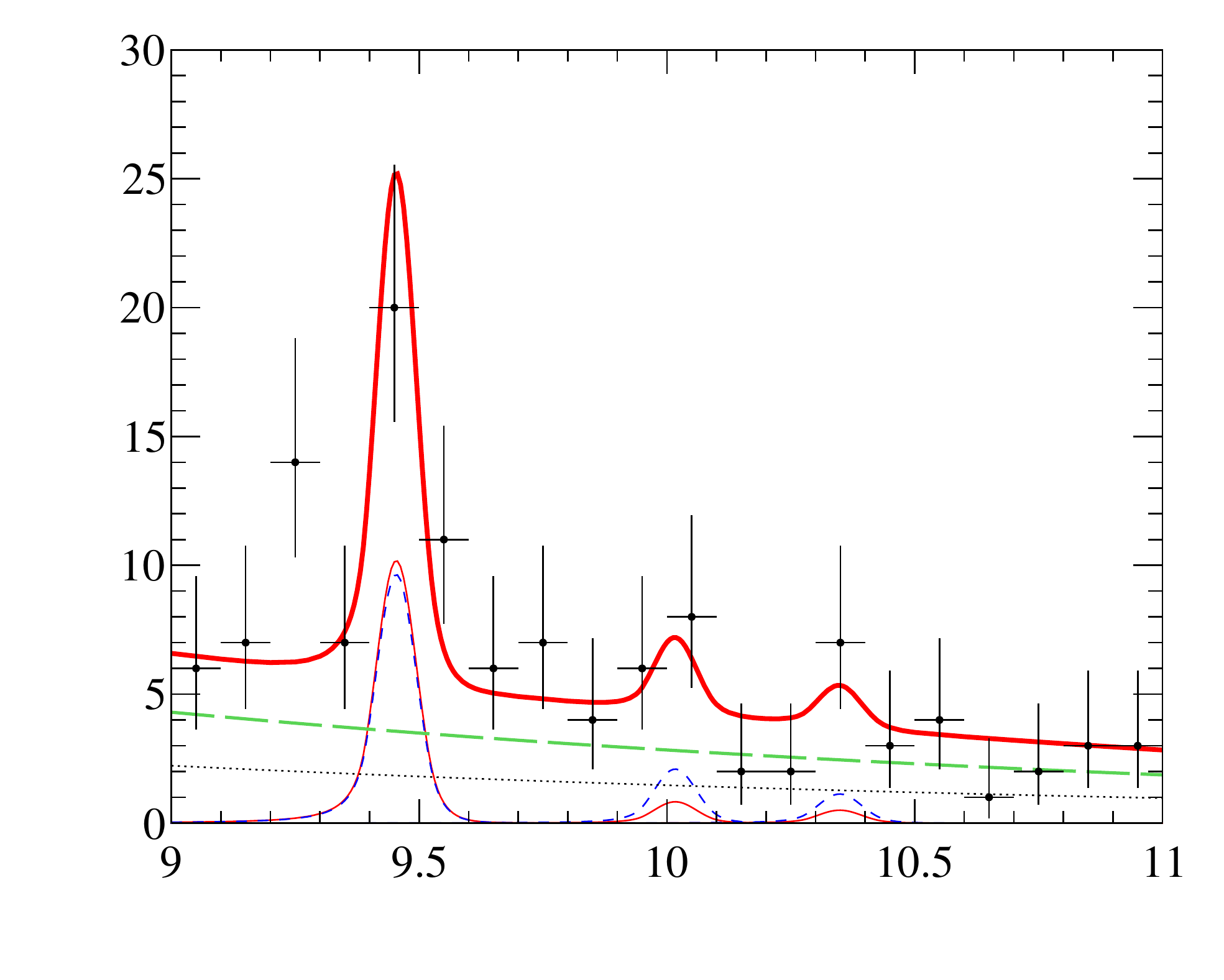}
    }
    \put( 75, 60){ 
      \includegraphics*[width=75mm,height=60mm,%
      ]{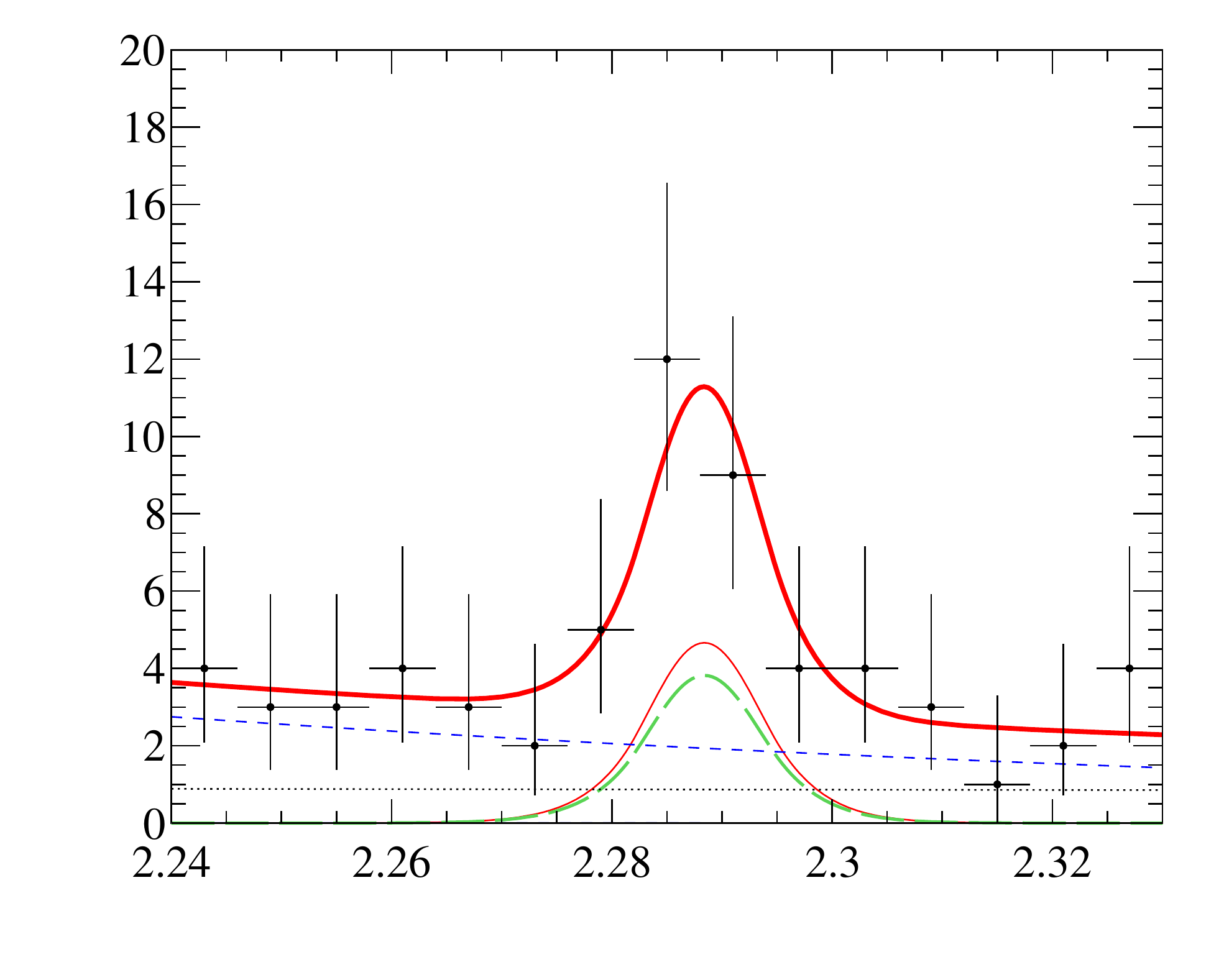}
    }
    \put(  0,  0){ 
      \includegraphics*[width=75mm,height=60mm,%
      ]{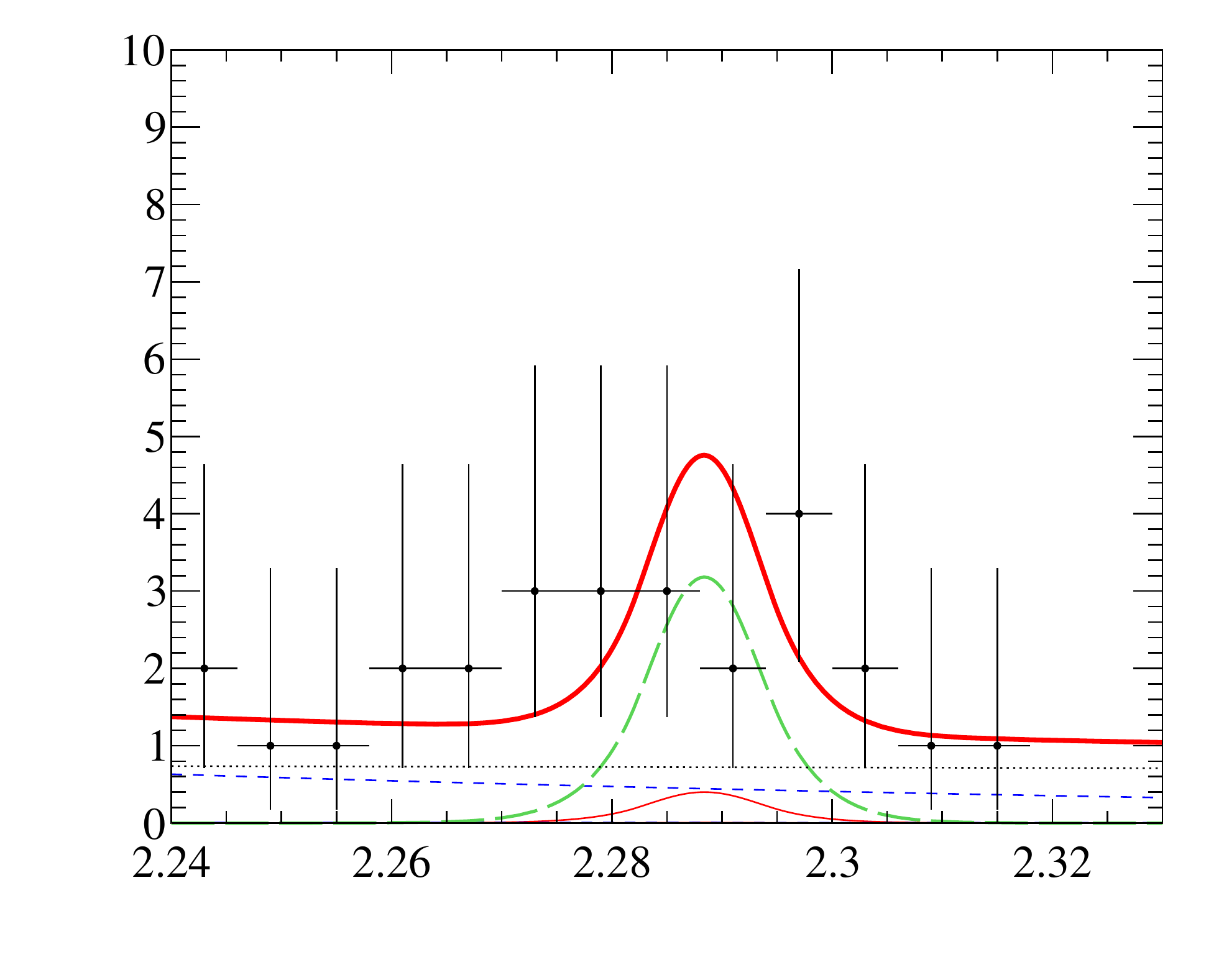}
    }
    \put( 75,  0){ 
      \includegraphics*[width=75mm,height=60mm,%
      ]{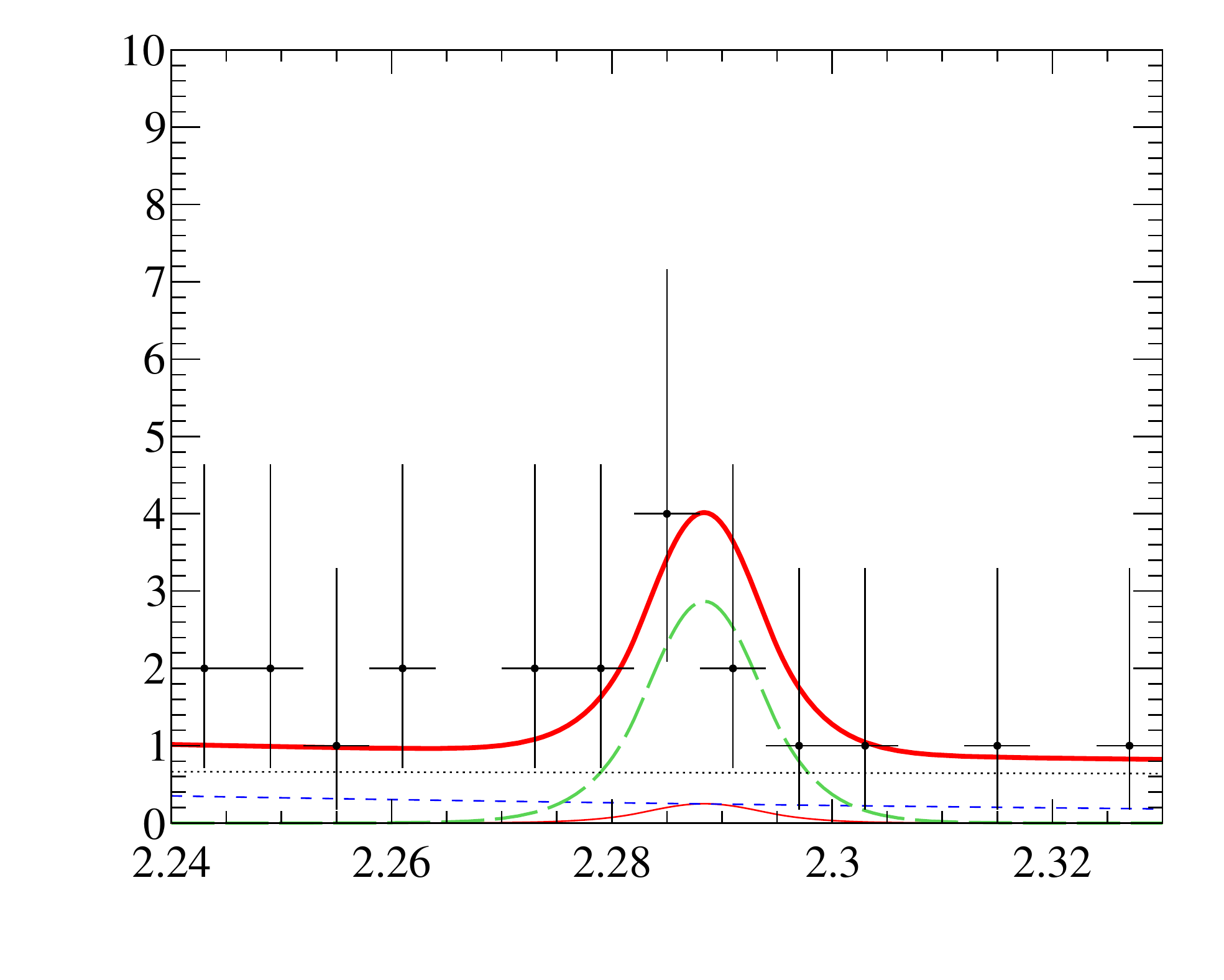}
    }
    \put(  2, 72) { \begin{sideways}Candidates/(100\mevcc)\end{sideways}} 
    \put( 77, 75) { \begin{sideways}Candidates/(6\mevcc) \end{sideways}} 
    \put(  2, 15) { \begin{sideways}Candidates/(6\mevcc) \end{sideways}} 
    \put( 77, 15) { \begin{sideways}Candidates/(6\mevcc) \end{sideways}} 
    \put( 35, 61) { $m_{\mumu}$   }       \put( 57, 61) { $\left[\!\gevcc\right]$ }  
    \put(110, 61) { $m_{\proton\Km\pip}$ } \put(132, 61) { $\left[\!\gevcc\right]$ }  
    \put( 35,  1) { $m_{\proton\Km\pip}$ } \put( 57,  1) { $\left[\!\gevcc\right]$ }  
    \put(110,  1) { $m_{\proton\Km\pip}$ } \put(132,  1) { $\left[\!\gevcc\right]$ }  
    %
    \put( 46,108) { a)~$\begin{array}{l} \lhcb \\ \ups\Lc\end{array}$ }
    \put(121,108) { b)~$\begin{array}{l} \lhcb \\ \YoneS\Lc\end{array}$ }
    \put( 46, 48) { c)~$\begin{array}{l} \lhcb \\ \YtwoS\Lc\end{array}$ }
    \put(121, 48) { d)~$\begin{array}{l} \lhcb \\ \YthreeS\Lc\end{array}$ }
  \end{picture}
  \caption { \small
    Projections from two-dimensional extended 
    unbinned maximum likelihood fits in bands
    a)~\mbox{$2.273<m_{\proton\Km\pip}<2.304 \mevcc$ },   
    b)~\mbox{$9.332<m_{\mumu}  <9.575 \gevcc$ },   
    c)~\mbox{$9.889<m_{\mumu}  <10.145\gevcc$ } and    
    d)~\mbox{$10.216<m_{\mumu} <10.481\gevcc$ }.   
    The~total fit function is shown by a~solid thick\,(red) curve;
    three individual $\ups\Lc$~signal components 
    are shown by solid thin\,(red) curves;
    three components describing \ups~signals and combinatorial 
    background in $\proton\Km\pip$~mass 
    are shown with short-dashed\,(blue) curves; 
    the component modelling the true \Lc~signal and combinatorial
    background in \mumu~mass is shown with a~long-dashed green curve 
    and the~component describing combinatorial background
    is shown with a~thin dotted\,(black) line.
  }
  \label{fig:signal_lc_proj_bands}
\end{figure}

The~fit results are summarized in Table~\ref{tab:fits}, and 
the fit projections are presented in Figs.~\ref{fig:signal_d0_proj_bands},
\ref{fig:signal_dp_proj_bands}, 
\ref{fig:signal_ds_proj_bands} and~\ref{fig:signal_lc_proj_bands}.
The statistical significances of the signal components
are determined using 
a~Monte\nobreakdash-Carlo~technique with a~large number of pseudoexperiments.
They are presented in 
Table~\ref{tab:signiftoy}.
For~the~five modes, 
$\YoneS\Dz$,
$\YtwoS\Dz$,
$\YoneS\Dp$,
$\YtwoS\Dp$ and 
$\YoneS\Ds$, the~significances exceed five standard deviations.
No significant signals are found for the~associated production of
\ups~mesons and \Lc~baryons.

\begin{table}[t]
  \centering
  \caption{
    Signal yields $N^{\ups\Charm}$~for $\ups\Charm$~production,
    determined with two-dimensional extended unbinned maximum likelihood fits
    to the~candidate $\ups\Charm$~samples.
  } \label{tab:fits}
  \vspace*{3mm}
  \begin{tabular*}{0.9\textwidth}{@{\hspace{10mm}}l@{\extracolsep{\fill}}ccc@{\hspace{10mm}}}
    & $\YoneS$
    & $\YtwoS$
    & $\YthreeS$
    \\[1mm]
    \hline 
    \\[-2mm]
    \Dz     &  $980 \pm  50$   &  $ 184   \pm 27$  &  $ 60  \pm 22$ 
    \\ 
    \Dp     &  $556 \pm  35$   &  $ 116   \pm 20$  &  $ 55  \pm 17$ 
    \\ 
    \Ds     
    &  $\phantom{0}31\pm7\phantom{0}$ 
    &  $\phantom{00}9\pm5\phantom{0}$   
    &  $\phantom{0}6\pm4\phantom{0}$
    \\ 
    \Lc     
    &  $\phantom{0}11\pm6\phantom{0}$ 
    &  $\phantom{00}1\pm4\phantom{0}$   
    &  $\phantom{0} 1\pm3\phantom{0}$  
  \end{tabular*}   
\end{table}

\begin{table}[t]
  \centering
  \caption{
    Statistical significances of the observed $\ups\Charm$~signals 
    in units of standard deviations 
    determined using pseudoexperiments.
    The~values in parentheses indicate the~statistical 
    significance calculated 
    using Wilks' theorem~\cite{Wilks:1938dza}.
  } \label{tab:signiftoy}
  \vspace*{3mm}
  \begin{tabular*}{0.9\textwidth}{@{\hspace{10mm}}l@{\extracolsep{\fill}}ccc@{\hspace{10mm}}}
    & $\YoneS$
    & $\YtwoS$
    & $\YthreeS$
    \\[1mm]
    \hline 
    \\[-2mm]
    \Dz     &  $>5\,(26)$                        &  $>5\,(7.7)$                         &  3.1   \\ 
    \Dp     &  $>5\,(19)$                        &  $>5\,(6.4)$                         &  4.0   \\ 
    \Ds     &  $>5\,(6)\phantom{0}$              &  $\phantom{>}2.5\phantom{\,(0.0)}$   &  1.9   \\
    \Lc     &  $\phantom{>}2.5\phantom{\,(00)}$  &  $\phantom{>}0.9\phantom{\,(0.0)}$   &  0.9             
  \end{tabular*}   
\end{table}

\begin{figure}[t]
  \setlength{\unitlength}{1mm}
  \centering
  \begin{picture}(150,120)
    %
    \put(  0, 60){ 
      \includegraphics*[width=75mm,height=60mm,%
      ]{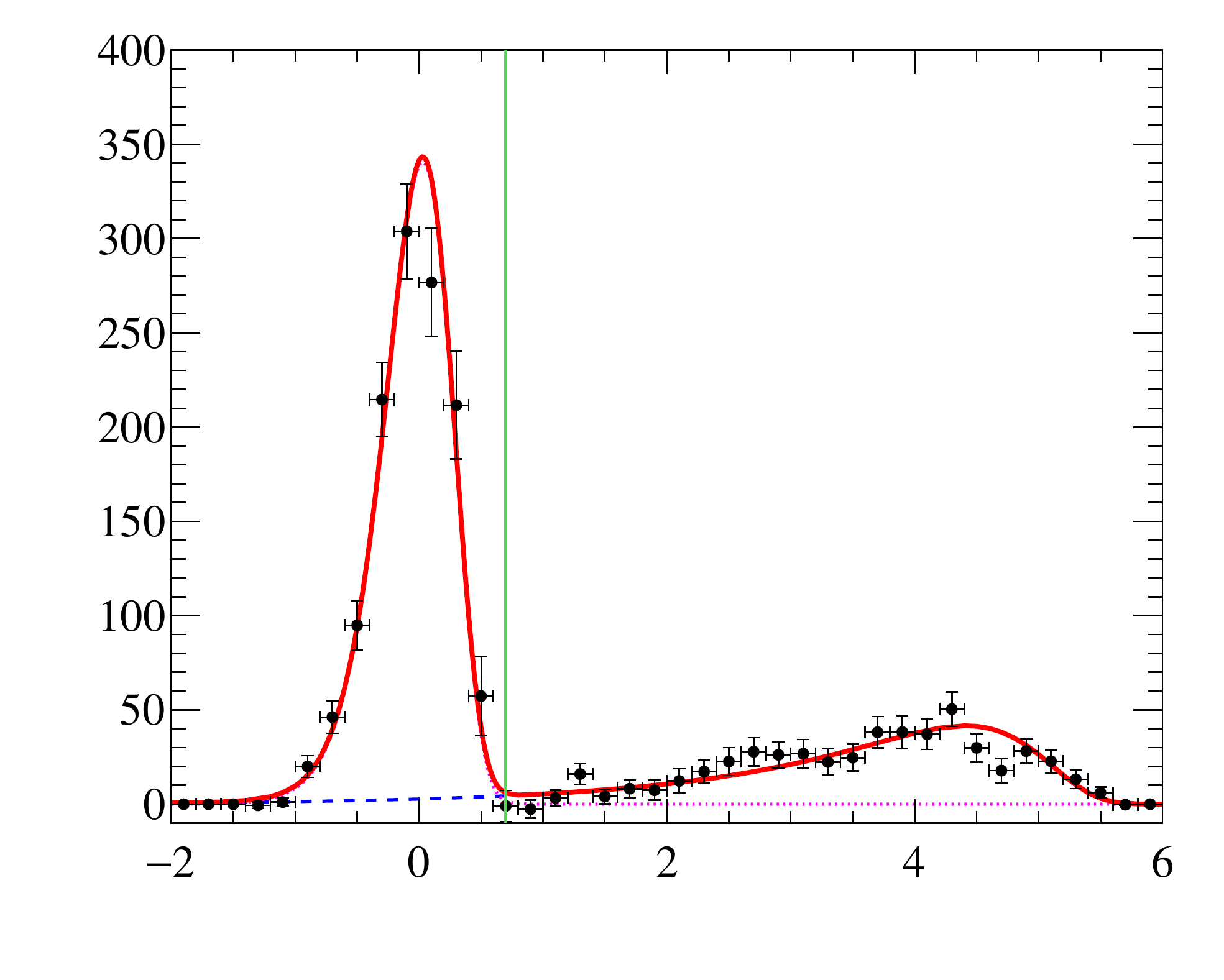}
    }
    \put( 75, 60){ 
      \includegraphics*[width=75mm,height=60mm,%
      ]{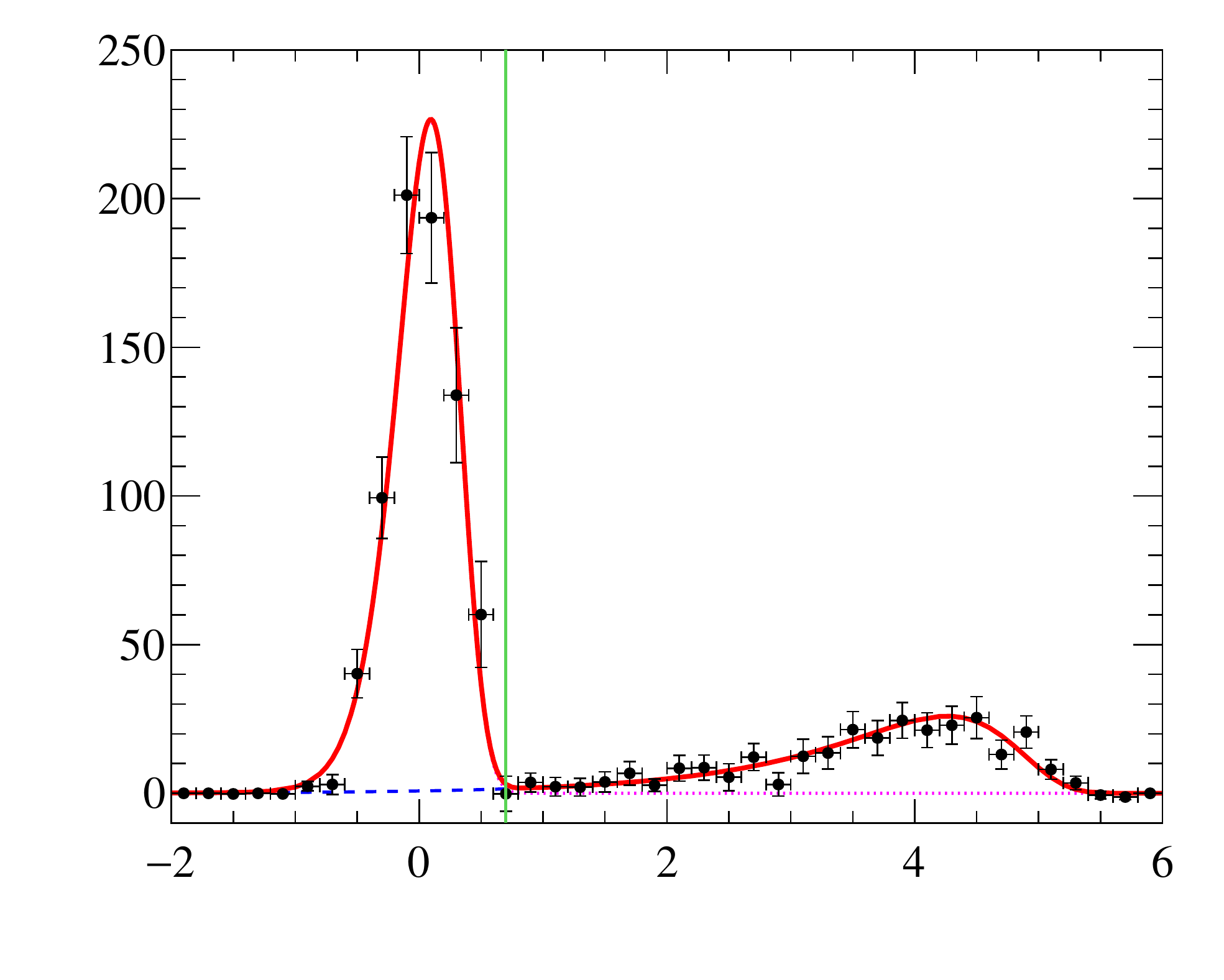}
    }
    \put( 25, 61) { $\log_{10}\left( \chisq_{\mathrm{fit}}\left(\ups\Dz\right)/\mathrm{ndf}\right)$ }   
    \put(100, 61) { $\log_{10}\left( \chisq_{\mathrm{fit}}\left(\ups\Dp\right)/\mathrm{ndf}\right)$ }
    \put(  0,  0){ 
      \includegraphics*[width=75mm,height=60mm,%
      ]{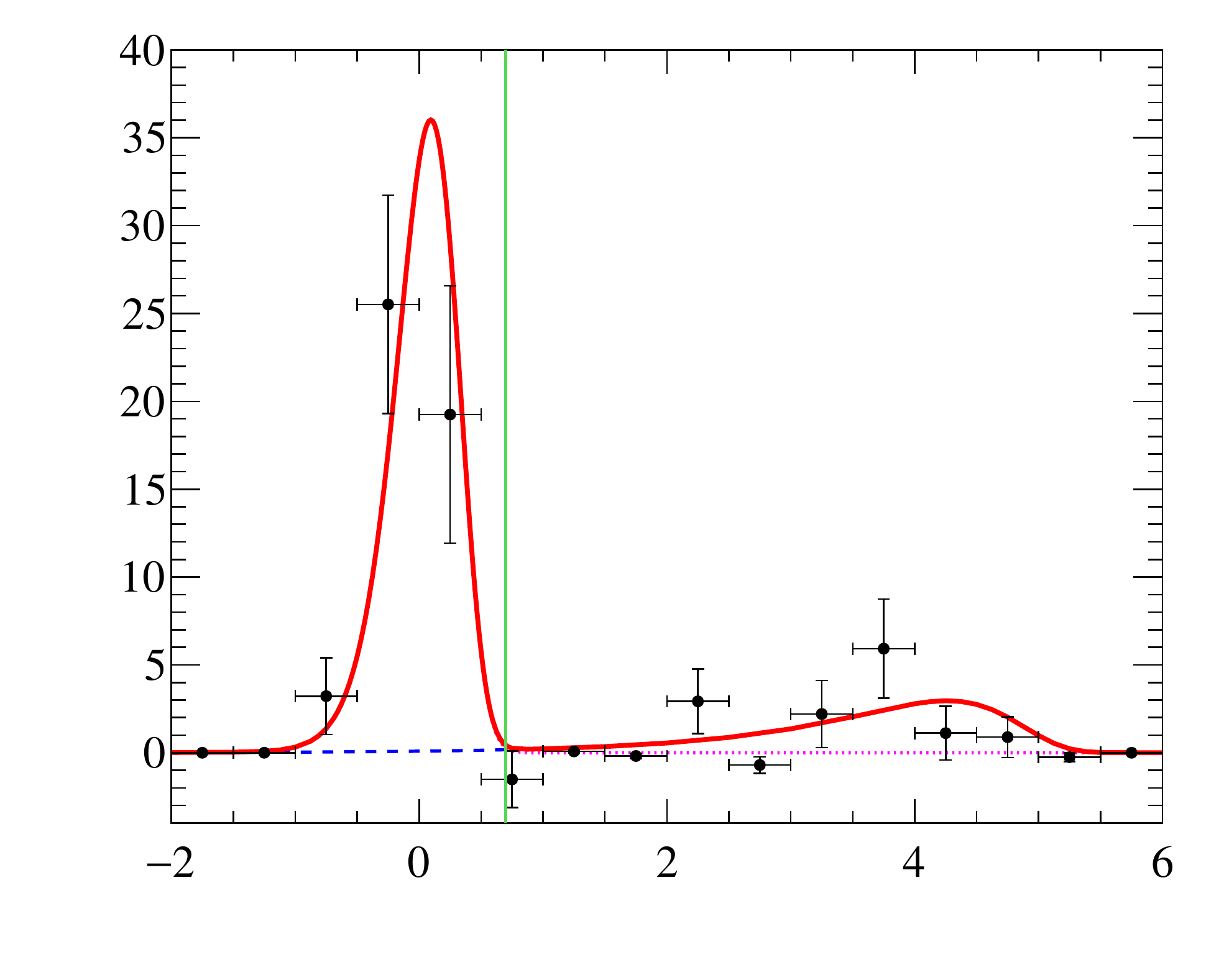}
    }
    \put( 75,  0){ 
      \includegraphics*[width=75mm,height=60mm,%
      ]{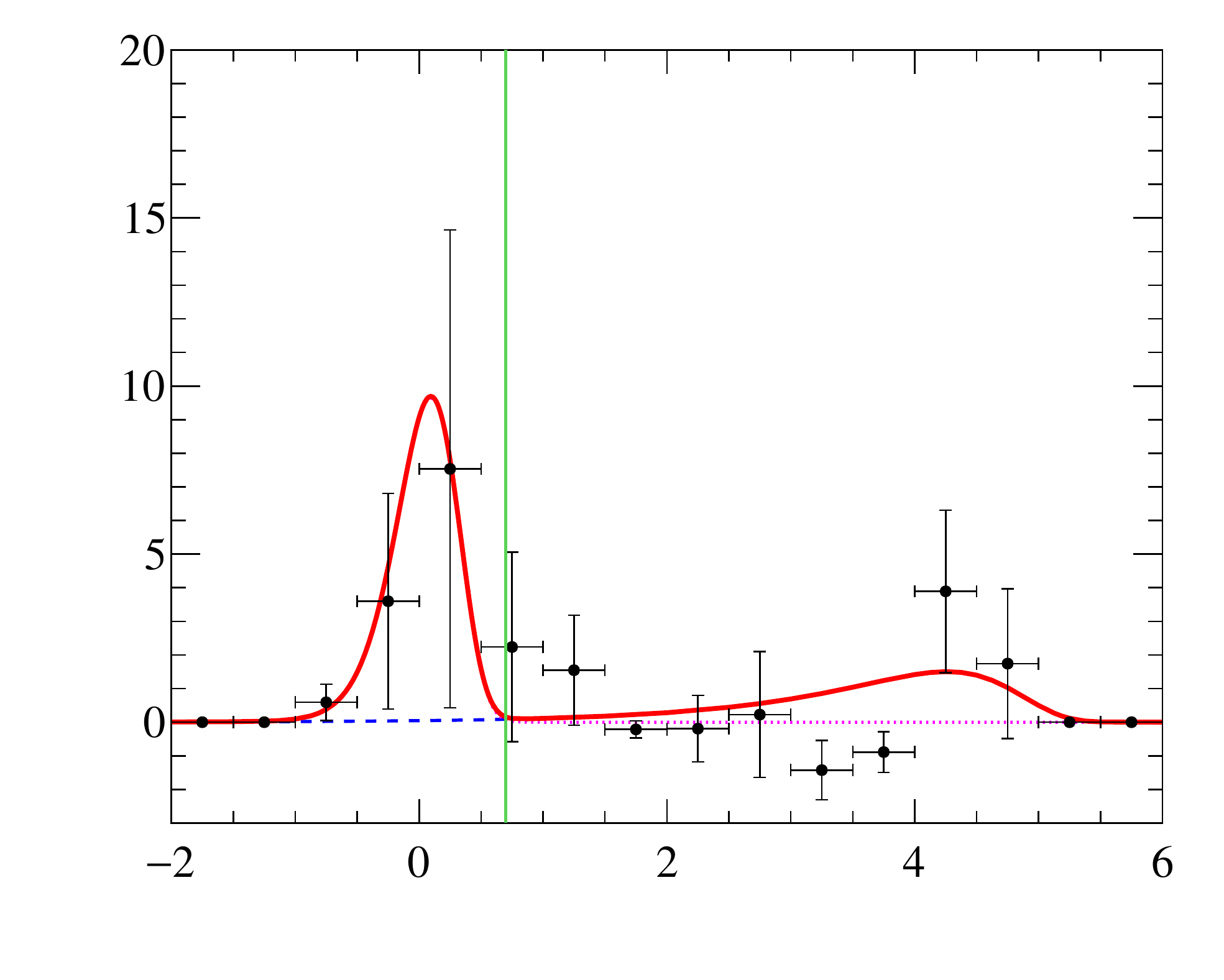}
    }
    \put ( -1, 90) { \begin{sideways}$\chisq_{\mathrm{fit}}\tfrac{\deriv N }{\deriv\chisq_{\mathrm{fit}}}~~~\left[\tfrac{\ln10}{0.2}\right]$\end{sideways}} 
    \put ( 74, 90) { \begin{sideways}$\chisq_{\mathrm{fit}}\tfrac{\deriv N }{\deriv\chisq_{\mathrm{fit}}}~~~\left[\tfrac{\ln10}{0.2}\right]$\end{sideways}} 
    \put (  0, 30) { \begin{sideways}$\chisq_{\mathrm{fit}}\tfrac{\deriv N }{\deriv\chisq_{\mathrm{fit}}}~~~\left[\tfrac{\ln10}{0.5}\right]$\end{sideways}} 
    \put ( 75, 30) { \begin{sideways}$\chisq_{\mathrm{fit}}\tfrac{\deriv N }{\deriv\chisq_{\mathrm{fit}}}~~~\left[\tfrac{\ln10}{0.5}\right]$\end{sideways}} 
    \put( 25,  1) { $\log_{10}\left( \chisq_{\mathrm{fit}}\left(\ups\Ds\right)/\mathrm{ndf}\right)$ }   
    \put(100,  1) { $\log_{10}\left( \chisq_{\mathrm{fit}}\left(\ups\Lc\right)/\mathrm{ndf}\right)$ }
    \put( 45,108) { a)~$\begin{array}{l} \lhcb \\ \ups\Dz \end{array}$ }
    \put(120,108) { b)~$\begin{array}{l} \lhcb \\ \ups\Dp \end{array}$ } 
    \put( 45, 48) { c)~$\begin{array}{l} \lhcb \\ \ups\Ds\end{array}$ }
    \put(120, 48) { d)~$\begin{array}{l} \lhcb \\ \ups\Lc\end{array}$ }
  \end{picture}
  
  \caption { \small
    Background-subtracted distributions 
    of $\chisq_{\mathrm{fit}}\left(\Charm\right)/\mathrm{ndf}$
    for 
    a)\,$\ups\Dz$,
    b)\,$\ups\Dp$,
    c)\,$\ups\Ds$ and 
    d)\,$\ups\Lc$~cases. 
    A~thin vertical\,(green) line indicates the requirement 
    \mbox{$\chisq_{\mathrm{fit}}\left(\ups\Charm\right)/\mathrm{ndf}<5$} 
    used in the analysis.
    The~solid\,(red) curves indicate a~fit
    to a~sum of two components, 
    each described by $\Gamma$-distribution~shape. 
    The~pileup component is shown with a~dashed \,(blue) line.
  }
  \label{fig:pileup}
\end{figure}

The~possible contribution from pile\nobreakdash-up events is estimated from data 
following the~method from Refs.~\cite{LHCb-PAPER-2012-003,LHCb-PAPER-2013-062} 
by relaxing the~requirement on $\chisq_{\mathrm{fit}}\left(\ups\Charm\right)/\mathrm{ndf}$.
Due to the~requirements 
$\chisq_{\mathrm{fit}}\left(\ups\right)/\mathrm{ndf}<5$ and 
$\chisq_{\mathrm{fit}}\left(\Charm\right)/\mathrm{ndf}<5$, 
the~value of  
\mbox{$\chisq_{\mathrm{fit}}\left(\ups\Charm\right)/\mathrm{ndf}$} does not exceed 5~units
for signal events with \ups~and \Charm~hadron from the~same 
$\proton\proton$~collision vertex.
The~background is subtracted using the~\sPlot technique~\cite{Pivk:2004ty}. 
The~$\chisq_{\mathrm{fit}}\left(\ups\Charm\right)/\mathrm{ndf}$
distributions are shown in~Fig.~\ref{fig:pileup}.
The~distributions exhibit two components:
the~peak at low \chisq~is attributed 
to associated $\ups\Charm$~production,
and the~broad structure at large values 
of \chisq~corresponds to the~contribution from pile\nobreakdash-up~events.
The~distributions are fitted with 
a~function that has two components,
each described 
by a~$\Gamma$-distribution. 
The~shape is motivated by the~observation that 
$\chisq_{\mathrm{fit}}/\mathrm{ndf}$ should follow 
a~scaled\nobreakdash-\chisq~distribution. 
The~possible contribution from pile\nobreakdash-up events is 
estimated by integrating the~pile\nobreakdash-up component in the~region 
\mbox{$\chisq_{\mathrm{fit}}\left(\ups\Charm\right)/\mathrm{ndf}<5$}.
It does not exceed 1.5\% for all four cases and is neglected. 

The production cross\nobreakdash-section is determined for the~four modes
with the~largest yield:  
$\YoneS\Dz$, $\YtwoS\Dz$, $\YoneS\Dp$ and~$\YtwoS\Dp$.
The~cross\nobreakdash-section is calculated using a~subsample of events 
where the~reconstructed \ups~candidate is explicitly matched to 
the~dimuon~candidate that triggers the~event.  
This~requirement reduces signal yields by approximately 
20\%, 
but allows a~robust determination 
of trigger efficiencies.   
The~cross\nobreakdash-section for 
the~associated production of \ups~mesons
with \Charm~hadrons
in the~kinematic range of~LHCb
is calculated as
\begin{equation}
  \mathscr{B}_{\mumu}
  \times\upsigma^{\ups\Charm} = 
  \dfrac{1}{\mathscr{L}\times\mathscr{B}_{\Charm}} N_{\mathrm{corr}}^{\ups\Charm},
\end{equation}
where $\mathscr{L}$~is the~integrated luminosity~\cite{LHCb-PAPER-2014-047}, 
$\mathscr{B}_{\mumu}$~and $\mathscr{B}_{\Charm}$~
are the~world average branching  fractions of 
$\ups\to\mumu$ and the~charm decay modes~\cite{PDG2014},  
and $N_{\mathrm{corr}}^{\ups\Charm}$~is 
the~efficiency-corrected  yield 
of the~signal $\ups\Charm$~events 
in the~kinematic range of this analysis. 
Production cross\nobreakdash-sections are determined separately for 
data sets accumulated at $\sqs=7$~and~$8\tev$.

The efficiency-corrected signal yields $N_{\mathrm{corr}}^{\ups\Charm}$ are determined 
using an~extended unbinned maximum likelihood 
fit to the~weighted two-dimensional invariant 
mass distributions of the~selected $\ups\Charm$~candidates. 
The~weight $\upomega$ for each event is calculated as 
$\upomega = 1/\varepsilon^{\mathrm{tot}}$,
where  $\varepsilon^{\mathrm{tot}}$ is the~total efficiency 
for the~given event.

The effective DPS~cross\nobreakdash-section and 
the~ratios $R^{\ups\Charm}$ are calculated as
\begin{subequations}
\begin{eqnarray}
  \upsigma_{\mathrm{eff}}            
  & = & 
  \dfrac{ \upsigma^{\ups} \times \upsigma^{\Charm} }{ \upsigma^{\ups\Charm} }, \label{eq:sdps} \\ 
  R^{\ups\Charm}           
  & = &
  \dfrac{ \upsigma^{\ups\Charm} }{\upsigma^{\ups}}, 
\end{eqnarray}\end{subequations}
where  $\upsigma^{\ups}$ is the~production cross\nobreakdash-section of \ups~mesons 
taken from Ref.~\cite{LHCb-PAPER-2015-045}.
The~double\nobreakdash-differential production 
cross\nobreakdash-sections of charm mesons
has been measured at $\sqs=7\tev$ in the~region 
~\mbox{$2.0<\yC<4.5$}, 
\mbox{$\ptC<8\gevc$}~\cite{LHCb-PAPER-2012-041}.
According to FONLL~calculations~\cite{Cacciari:1998it,Cacciari:2001td,Cacciari:2012ny}, 
the~contribution from 
the~region $8<\ptC<20\gevc$ is significantly 
smaller than the~uncertainty for the~measured  
cross\nobreakdash-section in the~region $1<\ptC<8\gevc$. 
It~allows to estimate  
the~production cross\nobreakdash-section of charm mesons 
in the~region~\mbox{$2.0<\yC<4.5$}, \mbox{$1<\ptC<20\gevc$}, 
used in Eq.~\eqref{eq:sdps}.
For~the~production cross\nobreakdash-section of charm mesons 
at $\sqs=8\tev$, the~measured cross\nobreakdash-section at $\sqs=7\tev$
    is rescaled by the~ratio 
    $R^{\mathrm{FONLL}}_{8/7}(\pt,y)$ of the~double\nobreakdash-differential
    cross\nobreakdash-sections, as calculated 
    with FONLL~\cite{Cacciari:1998it,Cacciari:2001td,Cacciari:2012ny}
    at~$\sqs=8$~and~$7\tev$.

The ratios $R^{\Dz/\Dp}$ and 
$R_{\Charm}^{\YtwoS/\YoneS}$,
defined in Eq.~\eqref{eq:dpsth2}, are calculated as
\begin{subequations}
\begin{eqnarray}
  R^{\Dz/\Dp}
  & = & 
  \dfrac{\upsigma^{\ups\Dz}}{\upsigma^{\ups\Dp}}
  = 
  \dfrac { N_{\mathrm{corr}}^{\ups\Dz}}
  { N_{\mathrm{corr}}^{\ups\Dp}},    \\ 
  R_{\Charm}^{\YtwoS/\YoneS}
  & = & 
  \mathscr{B}_{2/1}
  \dfrac{\upsigma^{\YtwoS\Charm}}{\upsigma^{\YoneS\Charm}}
  = 
  \dfrac {   N^{ \YtwoS \Charm} }
  {   N^{ \YoneS \Charm} }  \times 
  \dfrac
  {  \left\langle \varepsilon_{ \YoneS \Charm} \right\rangle}
  {  \left\langle \varepsilon_{ \YtwoS \Charm} \right\rangle}, 
\end{eqnarray}\end{subequations}
where $\langle \varepsilon_{\ups\Charm}\rangle$
denotes the~average efficiency.
Within the~DPS mechanism, the~transverse momenta and rapidity spectra of
\Charm~mesons for the~signal $\YoneS\Charm$ and $\YtwoS\Charm$~events
are expected to be the~same.
This~allows to express the ratio
of the~average $\langle \varepsilon_{\ups\Charm}\rangle$~efficiencies
in terms of ratio of average efficiencies for inclusive \ups~mesons
\begin{equation}
  \dfrac
  {  \left\langle \varepsilon_{ \YoneS \Charm} \right\rangle}
  {  \left\langle \varepsilon_{ \YtwoS \Charm} \right\rangle} = 
  \dfrac
  {  \left\langle \varepsilon_{ \YoneS } \right\rangle}
  {  \left\langle \varepsilon_{ \YtwoS } \right\rangle}, 
\end{equation}
and the~latter is taken from Ref.~\cite{LHCb-PAPER-2015-045}. 


The~total efficiency $\varepsilon^{\mathrm{tot}}$, for each $\ups\Charm$~candidate
is calculated following Ref.~\cite{LHCb-PAPER-2012-003} as 
\begin{equation}
  \varepsilon^{\mathrm{tot}}_{\ups\Charm} = 
  \varepsilon^{\mathrm{tot}}_{\ups} \times
  \varepsilon^{\mathrm{tot}}_{\Charm},       \label{eq:effic}
\end{equation}
and applied individually an~on event\nobreakdash-by\nobreakdash-event basis,
where 
$\varepsilon^{\mathrm{tot}}_{\ups}$ and 
$\varepsilon^{\mathrm{tot}}_{\Charm}$ are
the~total efficiencies for \ups~and charm hadrons respectively.
These efficiencies are calculated as 
\begin{subequations}
  \begin{eqnarray}
    \varepsilon^{\mathrm{tot}}_{\ups}  & = &   
    \varepsilon^{\mathrm{rec}}_{\ups} \times 
    \varepsilon^{\mathrm{trg}}_{\ups} \times 
    \varepsilon^{\Pmu\mathrm{ID}}_{\ups}, \label{eq:effups}\\ 
    \varepsilon^{\mathrm{tot}}_{\Charm} & =  & 
    \varepsilon^{\mathrm{rec}}_{\Charm} \times 
    \varepsilon^{\mathrm{h}\mathrm{ID}}_{\Charm} , \label{eq:effcharm}
  \end{eqnarray}
\end{subequations}
where 
$\varepsilon^{\mathrm{rec}}$~is
the~detector acceptance, 
reconstruction and event selection efficiency and  
$\varepsilon^{\mathrm{trg}}$~is the~trigger efficiency for selected events.
The~particle identification efficiencies for $\ups$~and $\Charm$~candidates
$\varepsilon^{\Pmu\mathrm{ID}}_{\ups}$  and 
$\varepsilon^{\mathrm{h}\mathrm{ID}}_{\Charm}$
are calculated as 
\begin{subequations}
\begin{eqnarray}
  \varepsilon^{\Pmu\mathrm{ID}}_{\ups} & = & 
  \varepsilon^{\mathrm{ID}}_{\mup} \times  
  \varepsilon^{\mathrm{ID}}_{\mun} , \\ 
  \varepsilon^{\mathrm{h}\mathrm{ID}}_{\Charm} 
  & = & 
  \prod_{\kaon}    \varepsilon^{\mathrm{ID}}_{\kaon}   \times 
  \prod_{\pion}    \varepsilon^{\mathrm{ID}}_{\pion}  
\end{eqnarray} \end{subequations}
where
$\varepsilon^{\mathrm{ID}}_{\Pmu^{\pm}}$,
$\varepsilon^{\mathrm{ID}}_{\kaon}$ and 
$\varepsilon^{\mathrm{ID}}_{\pion}$
are the~efficiencies for the single~muon, kaon and pion 
identification, respectively.

The~efficiencies $\varepsilon^{\mathrm{rec}}$ and $\varepsilon^{\mathrm{trg}}$
are determined using simulated samples of \ups, \Dz~and \Dp~events as
a~function of $\pt$~and~$y$ of the~\ups~and the~\Charm~hadron.
The~differential treatment results in a~robust determination
of the~efficiency\nobreakdash-corrected signal yields, with no~dependence
on the~particle spectra in the~simulated samples.
The~derived values of the~efficiencies are corrected to account
for small discrepancies in the~detector response between data and simulation.
These~corrections are obtained using data\nobreakdash-driven
techniques~\cite{LHCb-DP-2013-002,LHCb-DP-2013-001}.

The efficiencies for muon, kaon and pion identification 
are determined directly from data using 
large samples of  low-background  
$\jpsi\to\mumu$ and 
$\Dstarp\to\left( \Dz\to\Km\pip\right)\pip$~decays.
The identification~efficiencies are evaluated 
as a~function of the~kinematic parameters of the~final\nobreakdash-state particles, 
and the~track multiplicity in the~event~\cite{LHCb-DP-2012-003}.

The efficiency is dependent on the~polarisation of
the~\ups~mesons~\cite{LHCb-PAPER-2011-036,LHCb-PAPER-2013-016,LHCb-PAPER-2013-066,LHCb-PAPER-2015-045}
The~polarisation of the~\ups~mesons  produced in $\proton\proton$~collisions
at \mbox{$\sqs=7\tev$} at high~\pty and central rapidity 
has been studied by the~CMS collaboration~\cite{Chatrchyan:2012woa}
in the~centre-of-mass helicity, 
Collins\nobreakdash-Soper~\cite{Collins:1977iv}
and the~perpendicular helicity frames.
No evidence of significant  transverse or longitudinal polarisation
has been observed for 
the~region~\mbox{$10<\pty<50\gevc$}, 
\mbox{$\left|\yy\right|<1.2$}.
Therefore, the efficiencies are calculated under the~assumption of 
unpolarised production of \ups~mesons 
and no corresponding systematic uncertainty is assigned on 
the~cross-section. 
Under~the~assumption of transversely polarised \ups~mesons with \mbox{$\uplambda_{\vartheta}=0.2$} 
in the~\lhcb~kinematic region,\footnote{The CMS~measurements for \YoneS~mesons are consistent 
with small transverse polarisation in the~helicity frame with 
the~central values for the~polarisation parameter~\mbox{$0\lesssim\uplambda_{\vartheta}\lesssim0.2$}~\cite{Chatrchyan:2012woa}.} 
the~total efficiency  
would result in an~decrease of~3\%~\cite{LHCb-PAPER-2015-045}.

\boldmath\section{Kinematic distributions of  $\ups\Charm$~events}\unboldmath 
\label{sec:diff}

The differential distributions are important for 
the~determination of the~production mechanism.
In this section,
the~shapes of  differential distributions for 
$\YoneS\Dz$ and $\YoneS\Dp$~events are studied.  
Assuming that the~production mechanism 
of $\ups\Charm$~events is essentially
the~same at $\sqs=7$~and 8\tev,
both samples are treated together in this~section.

\begin{figure}[tb]
  \setlength{\unitlength}{1mm}
  \centering
  \begin{picture}(155,122)
    %
    \put(  0, 62){ 
      \includegraphics*[width=75mm,height=60mm,%
      ]{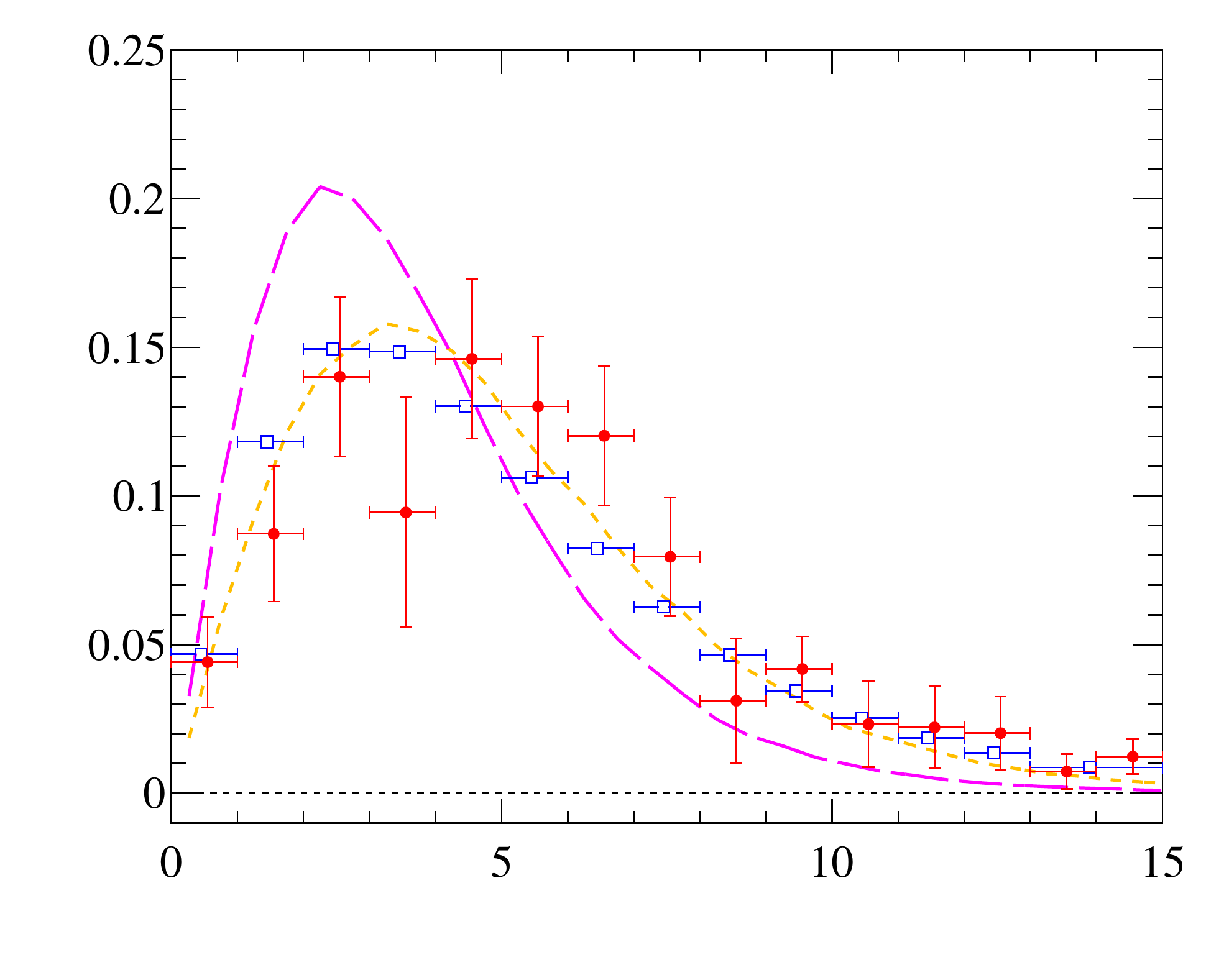}
    }
    \put(  80,62){ 
      \includegraphics*[width=75mm,height=60mm,%
      ]{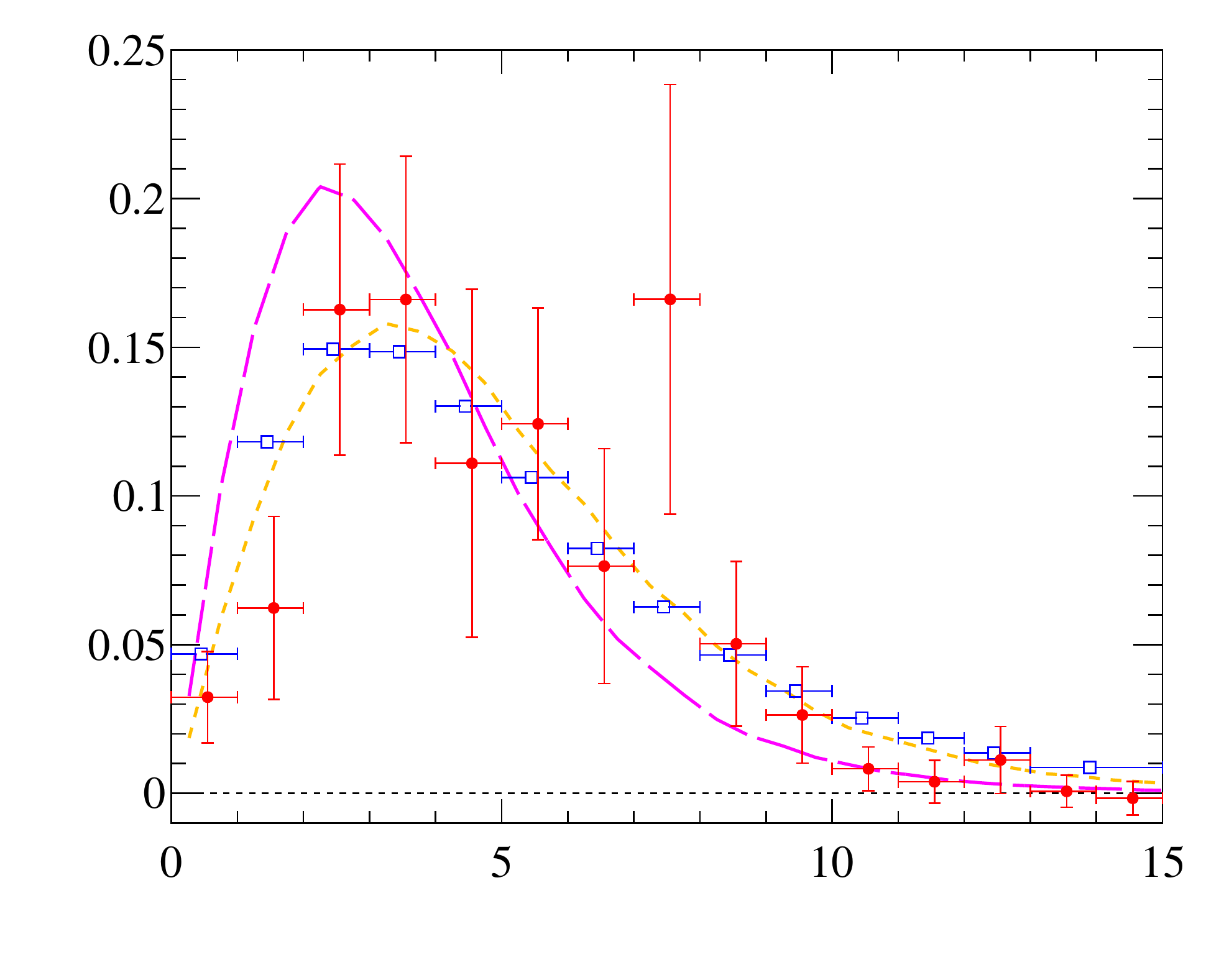}
    }
    \put(  0, 00){ 
      \includegraphics*[width=75mm,height=60mm,%
      ]{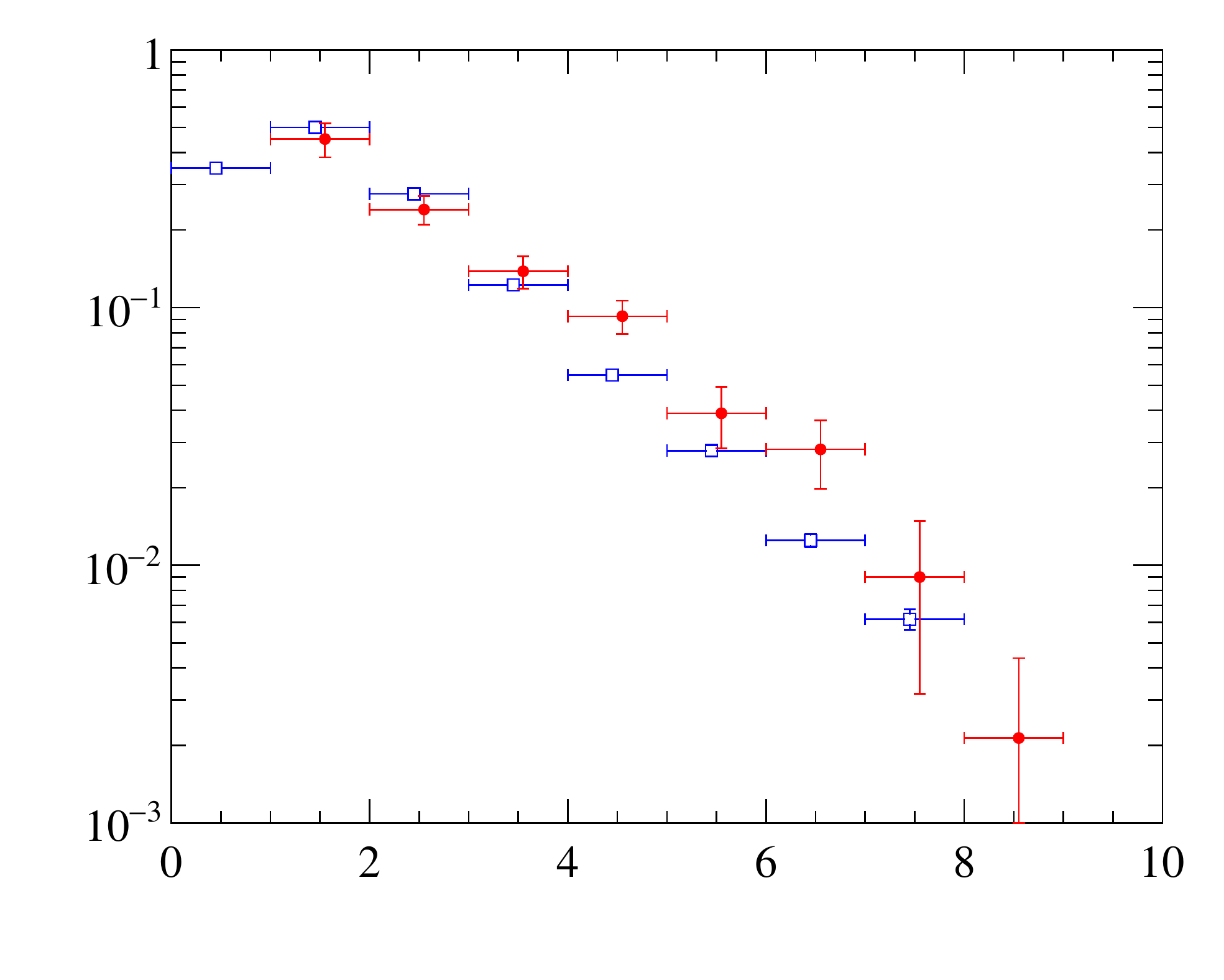}
    }
    \put( 80,  0){ 
      \includegraphics*[width=75mm,height=60mm,%
      ]{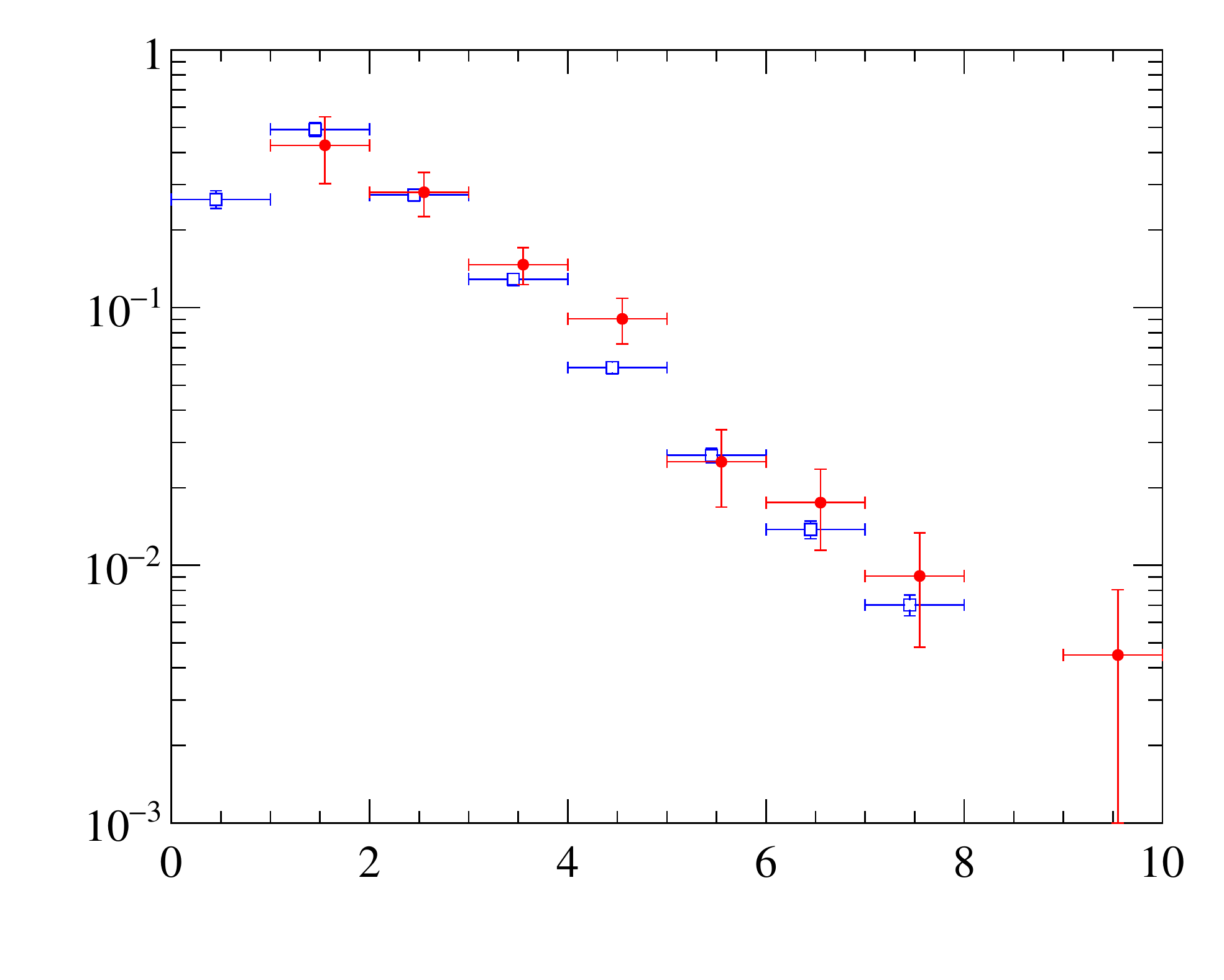}
    }
    \put ( -3, 92) { \begin{sideways}$\tfrac{1}{\upsigma}\tfrac{\deriv\upsigma}{\deriv\pty}~~~\left[\tfrac{1}{1\gevc}\right]$\end{sideways}} 
    \put ( 77, 92) { \begin{sideways}$\tfrac{1}{\upsigma}\tfrac{\deriv\upsigma}{\deriv\pty}~~~\left[\tfrac{1}{1\gevc}\right]$\end{sideways}} 
    \put ( -3, 30) { \begin{sideways}$\tfrac{1}{\upsigma}\tfrac{\deriv\upsigma}{\deriv\ptC}~~~\left[\tfrac{1}{1\gevc}\right]$\end{sideways}} 
    \put ( 77, 30) { \begin{sideways}$\tfrac{1}{\upsigma}\tfrac{\deriv\upsigma}{\deriv\ptC}~~~\left[\tfrac{1}{1\gevc}\right]$\end{sideways}} 
    \put( 35,  63) { $p^{\YoneS}_{\mathrm{T}}$ } \put( 58, 63) { $\left[\!\gevc\right]$ }  
    \put(115,  63) { $p^{\YoneS}_{\mathrm{T}}$ } \put(138, 63) { $\left[\!\gevc\right]$ }  
    \put( 35,   1) { $p^{\Dz}_{\mathrm{T}}$ }  \put(  58, 1) { $\left[\!\gevc\right]$ }  
    \put(115,   1) { $p^{\Dp}_{\mathrm{T}}$ }  \put( 138, 1) { $\left[\!\gevc\right]$ }  
    \put( 45,110) { a)~$\begin{array}{l} \lhcb \\ \YoneS\Dz\end{array}$ }
    \put(125,110) { b)~$\begin{array}{l} \lhcb \\ \YoneS\Dp\end{array}$ }
    \put( 45, 48) { c)~$\begin{array}{l} \lhcb \\ \YoneS\Dz\end{array}$ }
    \put(125, 48) { d)~$\begin{array}{l} \lhcb \\ \YoneS\Dp\end{array}$ }
  \end{picture}
  \caption { \small
    Background-subtracted and 
    efficiency-corrected \pty\,(top) and \ptC\,(bottom)~distributions for 
    $\YoneS\Dz$~events\,(left) and 
    $\YoneS\Dp$~event\,(right).
    The~transverse momentum spectra, 
    derived within the~DPS~mechanism using the~measurements 
    from Refs.~\cite{LHCb-PAPER-2012-041,LHCb-PAPER-2015-045},
    are shown with the~open\,(blue) squares.
    The~SPS predictions~\cite{Baranov:private} for the~\pty~spectra are shown 
    with dashed\,(orange) 
    and 
    long\nobreakdash-dashed\,(magenta) 
    curves for calculations based on 
    the~$\kT$\nobreakdash-factorization and 
    the~collinear approximation, respectively.
    All~distributions are normalized to unity.
  }
  \label{fig:props_pt}
\end{figure}

\begin{figure}[t]
  \setlength{\unitlength}{1mm}
  \centering
  \begin{picture}(155,120)
    %
    \put(  0, 60){ 
      \includegraphics*[width=75mm,height=60mm,%
      ]{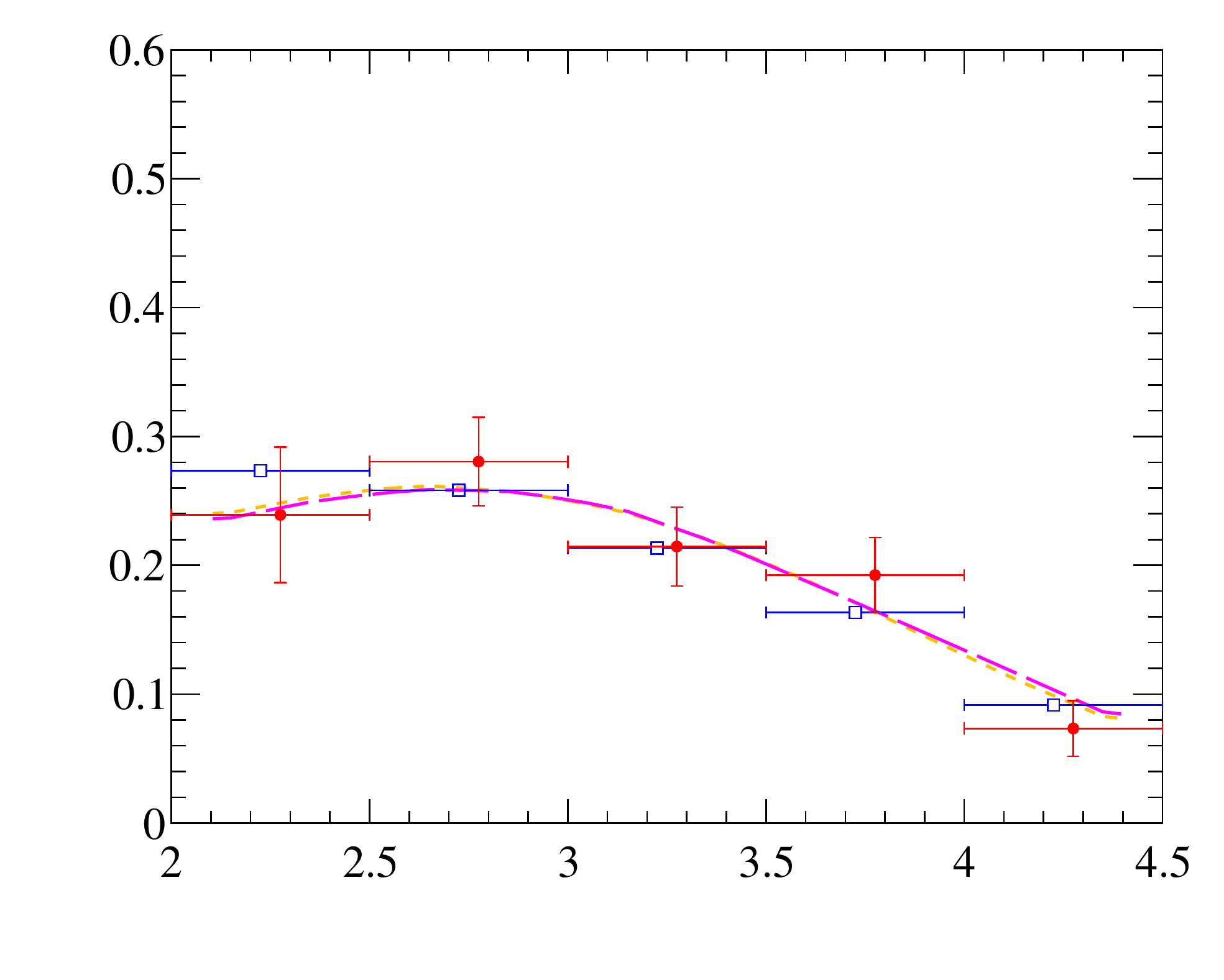}
    }
    \put( 80, 60){ 
      \includegraphics*[width=75mm,height=60mm,%
      ]{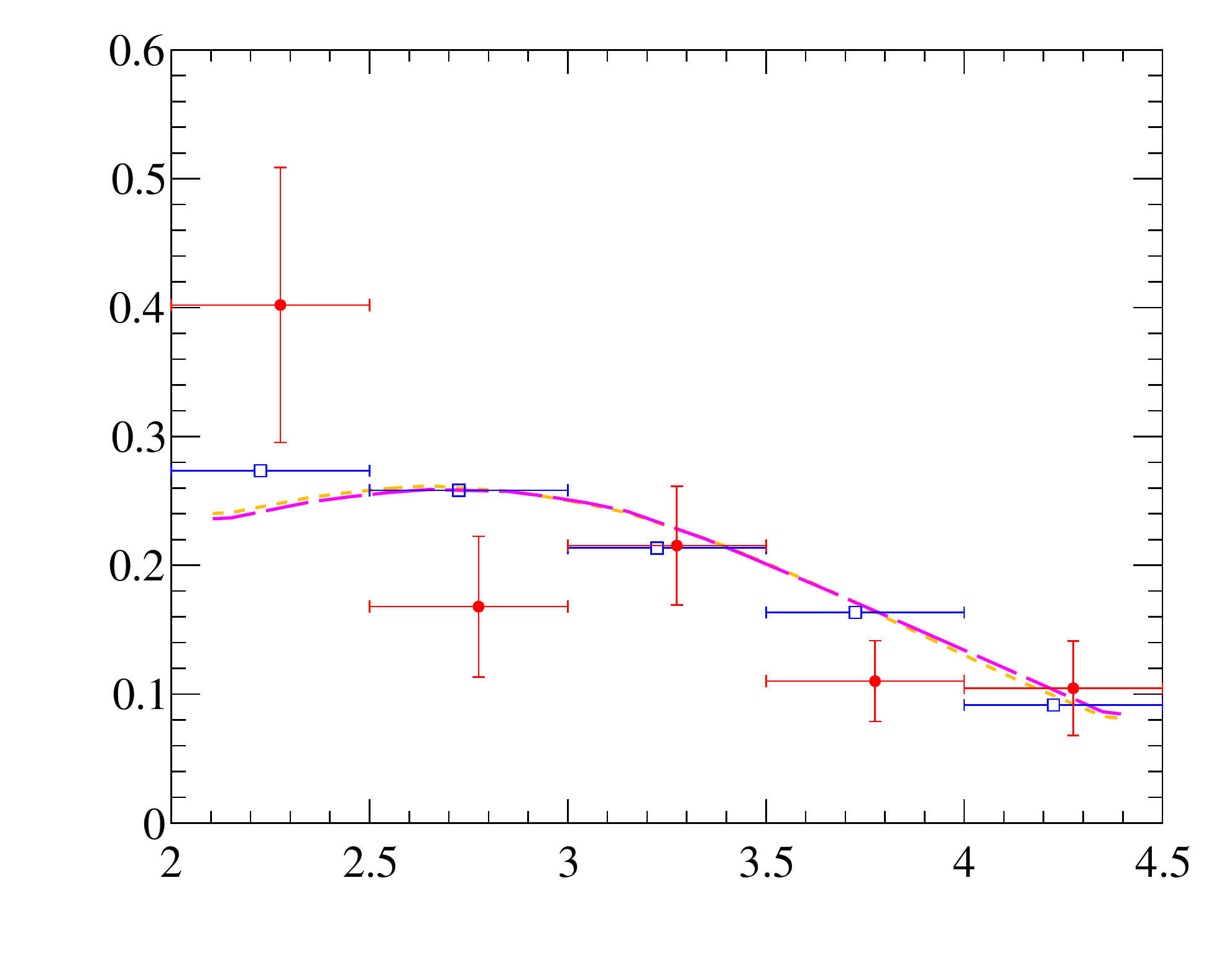}
    }
    \put(  0,  0){ 
      \includegraphics*[width=75mm,height=60mm,%
      ]{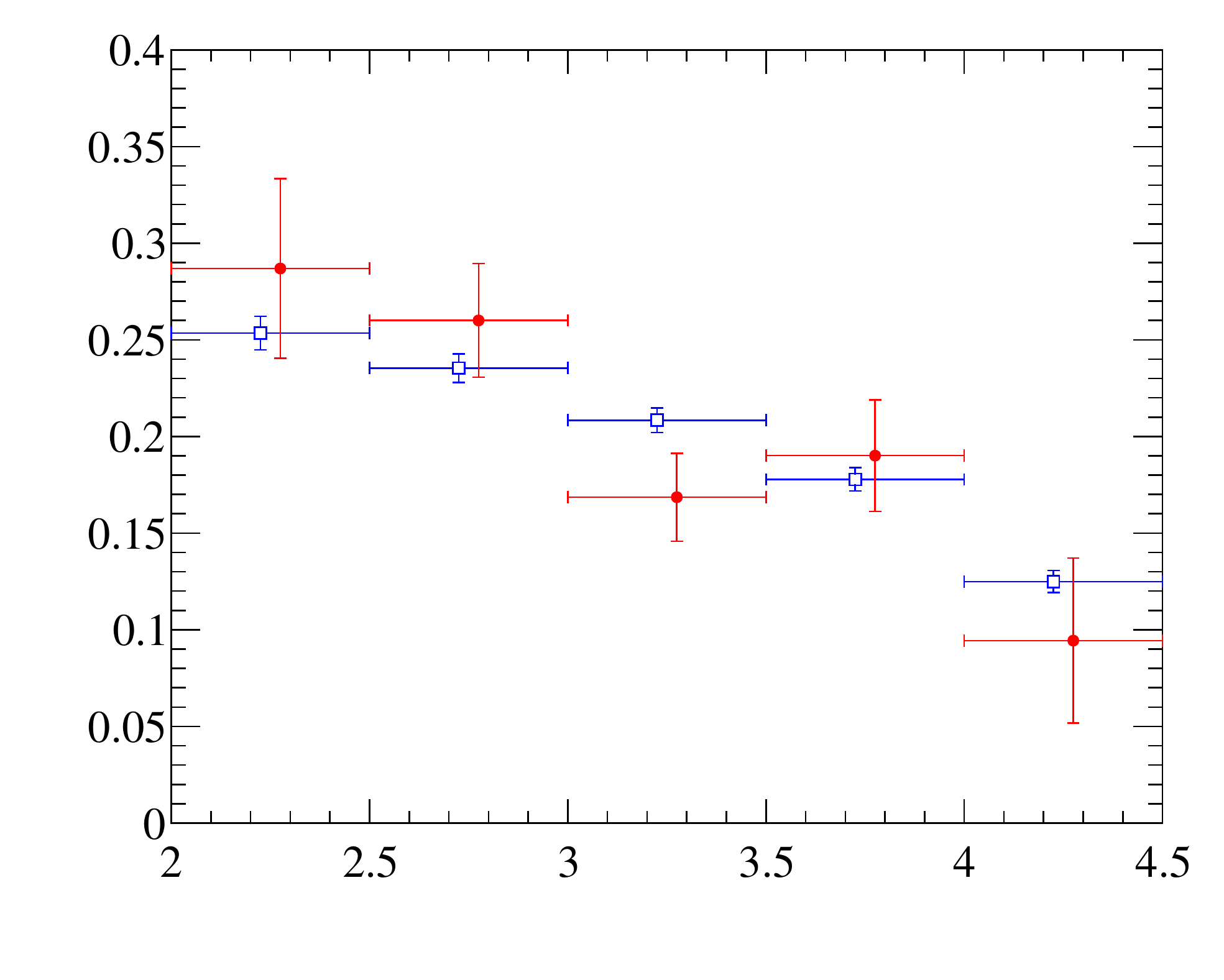}
    }
    \put( 80,  0){ 
      \includegraphics*[width=75mm,height=60mm,%
      ]{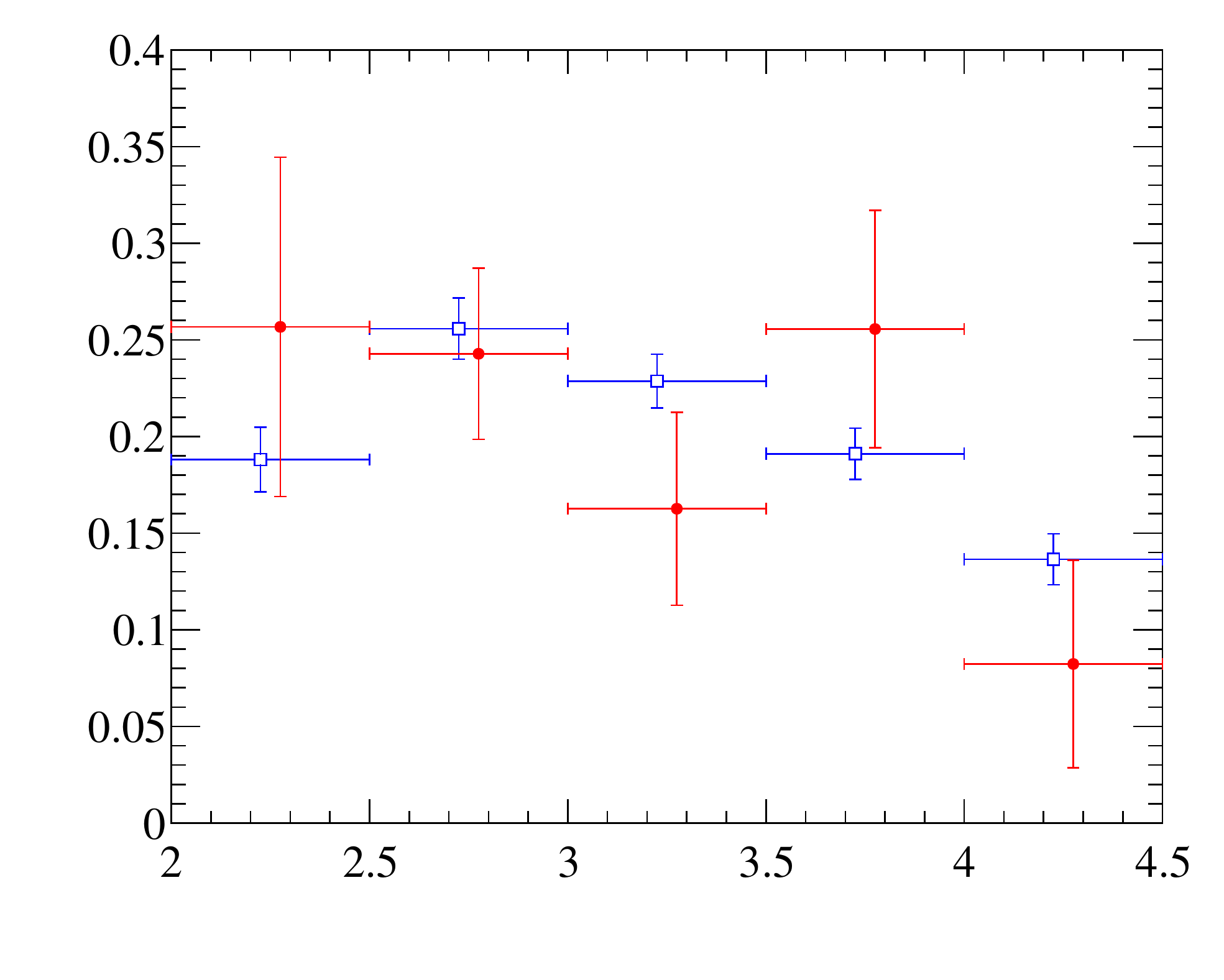}
    }
    \put ( -2, 96) { \begin{sideways}$\tfrac{1}{\upsigma}\tfrac{\deriv\upsigma}{\deriv\yy}~~~\left[\tfrac{1}{0.5}\right]$\end{sideways}} 
    \put ( 78, 96) { \begin{sideways}$\tfrac{1}{\upsigma}\tfrac{\deriv\upsigma}{\deriv\yy}~~~\left[\tfrac{1}{0.5}\right]$\end{sideways}} 
    \put ( -2, 36) { \begin{sideways}$\tfrac{1}{\upsigma}\tfrac{\deriv\upsigma}{\deriv\yC}~~~\left[\tfrac{1}{0.5}\right]$\end{sideways}} 
    \put ( 78, 36) { \begin{sideways}$\tfrac{1}{\upsigma}\tfrac{\deriv\upsigma}{\deriv\yC}~~~\left[\tfrac{1}{0.5}\right]$\end{sideways}} 
    \put( 40,  61) { $y^{\YoneS}$ } 
    \put(120,  61) { $y^{\YoneS}$ } 
    \put( 40,   1) { $y^{\Dz}$  } 
    \put(120,   1) { $y^{\Dp}$  } 
    \put( 45,108) { a)~$\begin{array}{l} \lhcb \\ \YoneS\Dz\end{array}$ }
    \put(125,108) { b)~$\begin{array}{l} \lhcb \\ \YoneS\Dp\end{array}$ }
    \put( 45, 48) { c)~$\begin{array}{l} \lhcb \\ \YoneS\Dz\end{array}$ }
    \put(105, 48) { d)~$\begin{array}{l} \lhcb \\ \YoneS\Dp\end{array}$ }
  \end{picture}
  \caption { \small
    Background-subtracted and 
    efficiency-corrected
    \yy\,(top) and \yC\,(bottom)~distributions for 
    $\YoneS\Dz$\,(left) and 
    $\YoneS\Dp$\,(right)~events.
    The~rapidity spectra, derived within the~DPS~mechanism using 
    the~measurements from Refs.~\cite{LHCb-PAPER-2012-041,LHCb-PAPER-2015-045},
    are shown with the~open\,(blue) squares.
    The~SPS predictions~\cite{Baranov:private} for the~\yy~spectra 
        are shown 
        with dashed\,(orange) 
        and 
        long\nobreakdash-dashed\,(magenta) 
        curves for calculations based on 
        the~$\kT$\nobreakdash-factorization and 
        the~collinear approximation, respectively.
    All~distributions are normalized to unity.
  }
  \label{fig:props_y}
\end{figure}

\begin{figure}[t]
  \setlength{\unitlength}{1mm}
  \centering
  \begin{picture}(155,120)
    %
    \put(  0, 60){ 
      \includegraphics*[width=75mm,height=60mm,%
      ]{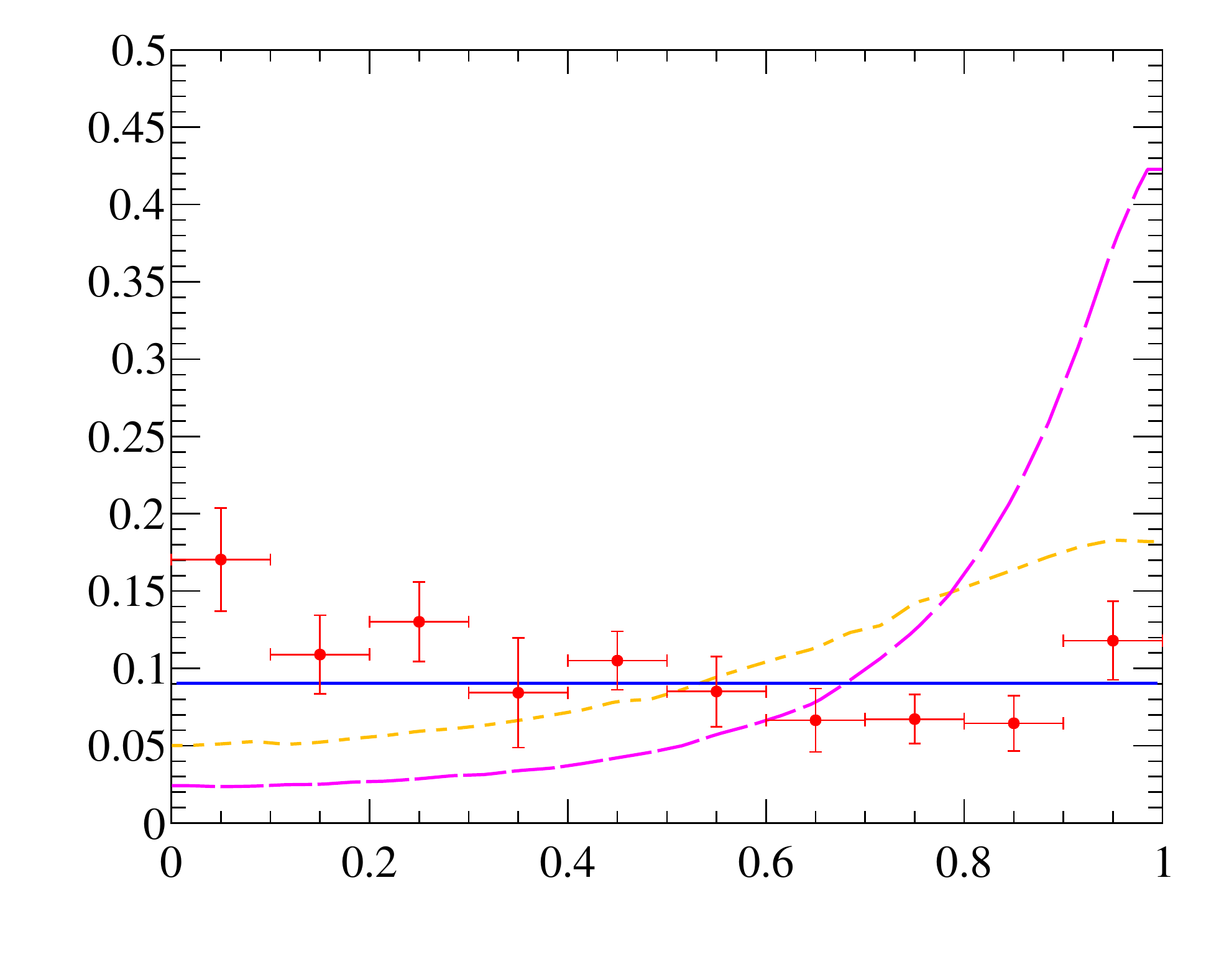}
    }
    \put( 80, 60){ 
      \includegraphics*[width=75mm,height=60mm,%
      ]{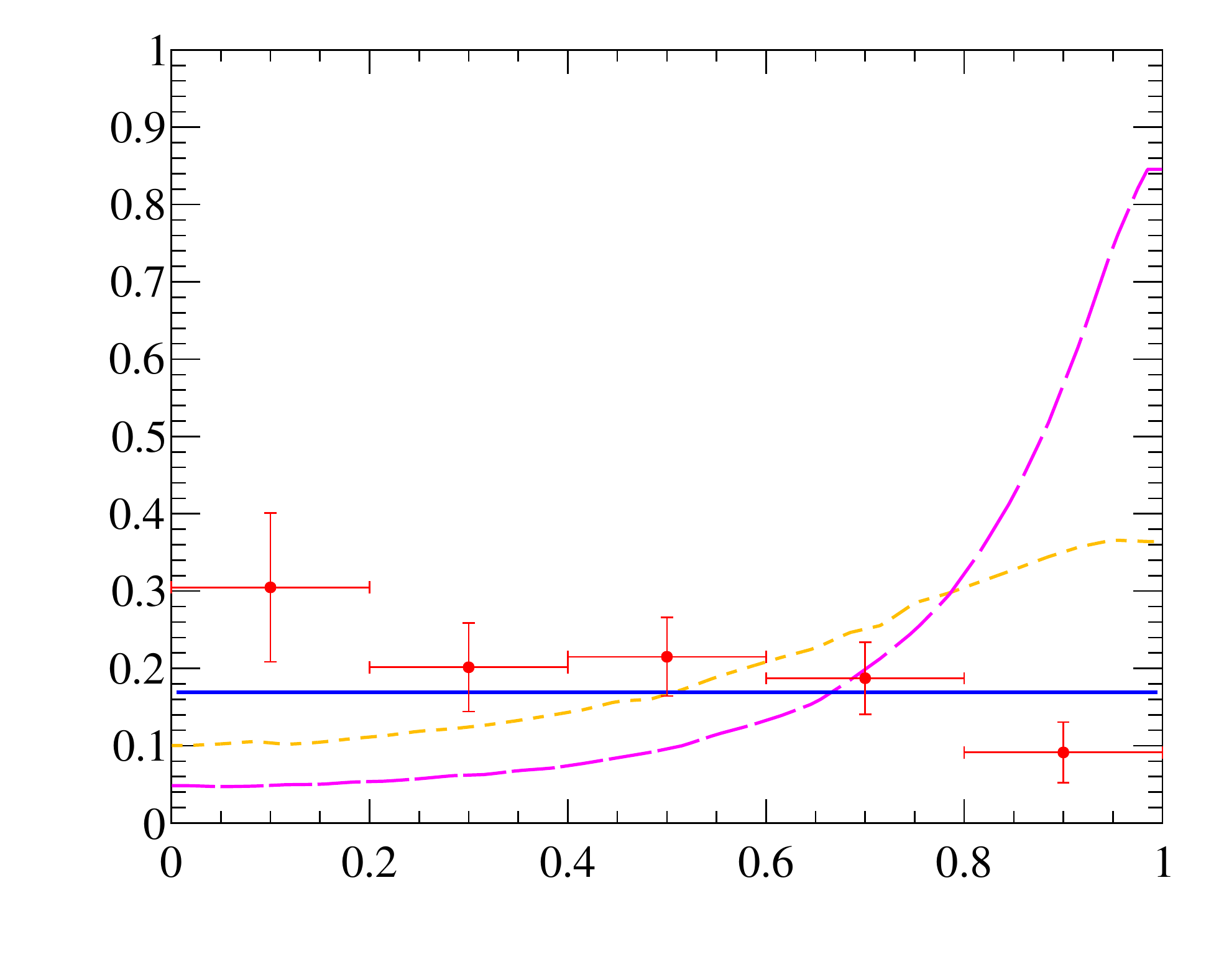}
    }
    \put ( -2, 95) { \begin{sideways}$\tfrac{1}{\upsigma}\tfrac{\deriv\upsigma}{\deriv\left|\Delta\phi\right|}~~~\left[\tfrac{\pi}{0.1}\right]$\end{sideways}} 
    \put ( 78, 95) { \begin{sideways}$\tfrac{1}{\upsigma}\tfrac{\deriv\upsigma}{\deriv\left|\Delta\phi\right|}~~~\left[\tfrac{\pi}{0.2}\right]$\end{sideways}} 
    \put( 35,  61) { $\left|\Delta\phi\right|/\pi$ } 
    \put(115,  61) { $\left|\Delta\phi\right|/\pi$ } 
    \put(  0, 0){ 
      \includegraphics*[width=75mm,height=60mm,%
      ]{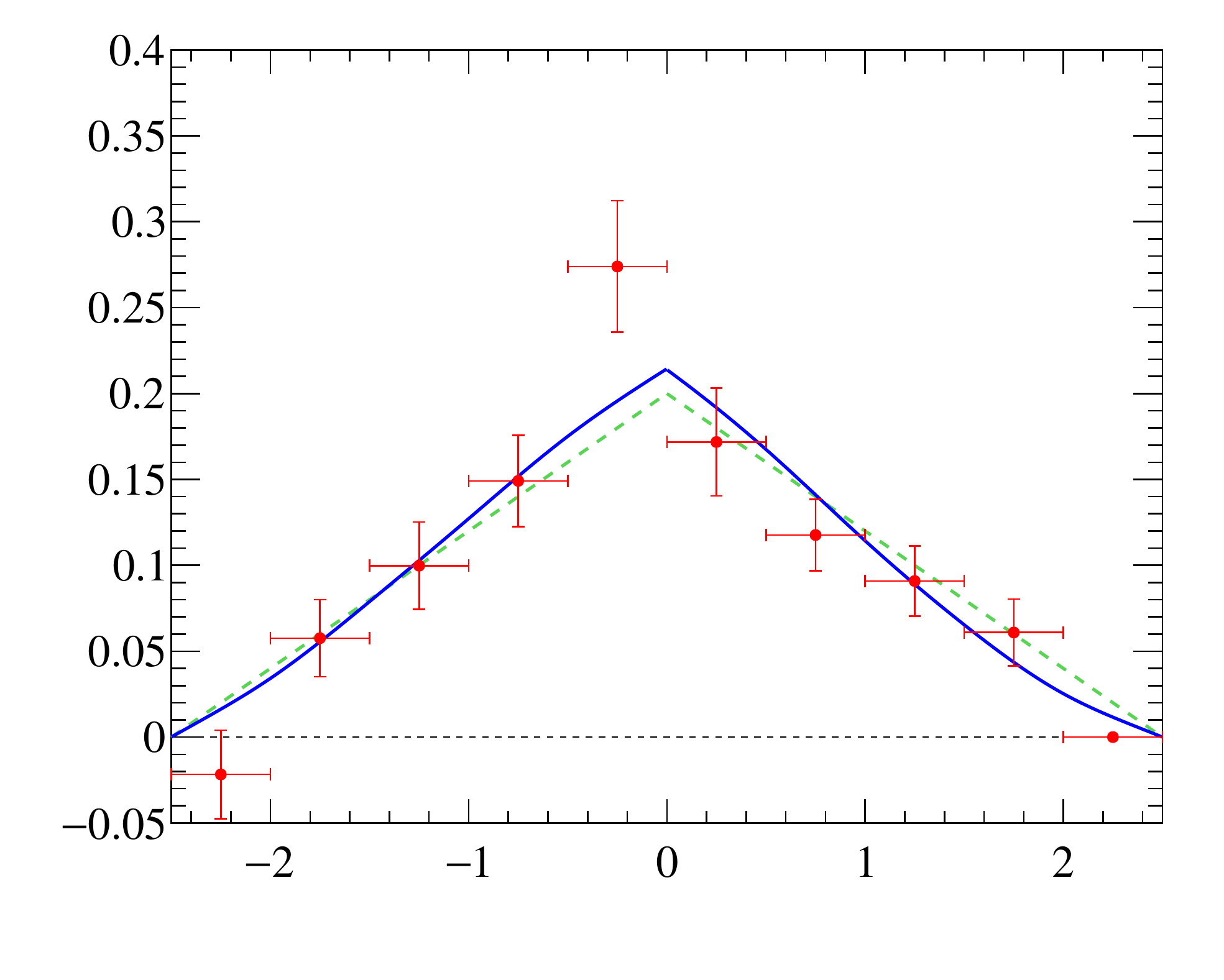}
    }
    \put( 80, 0){ 
      \includegraphics*[width=75mm,height=60mm,%
      ]{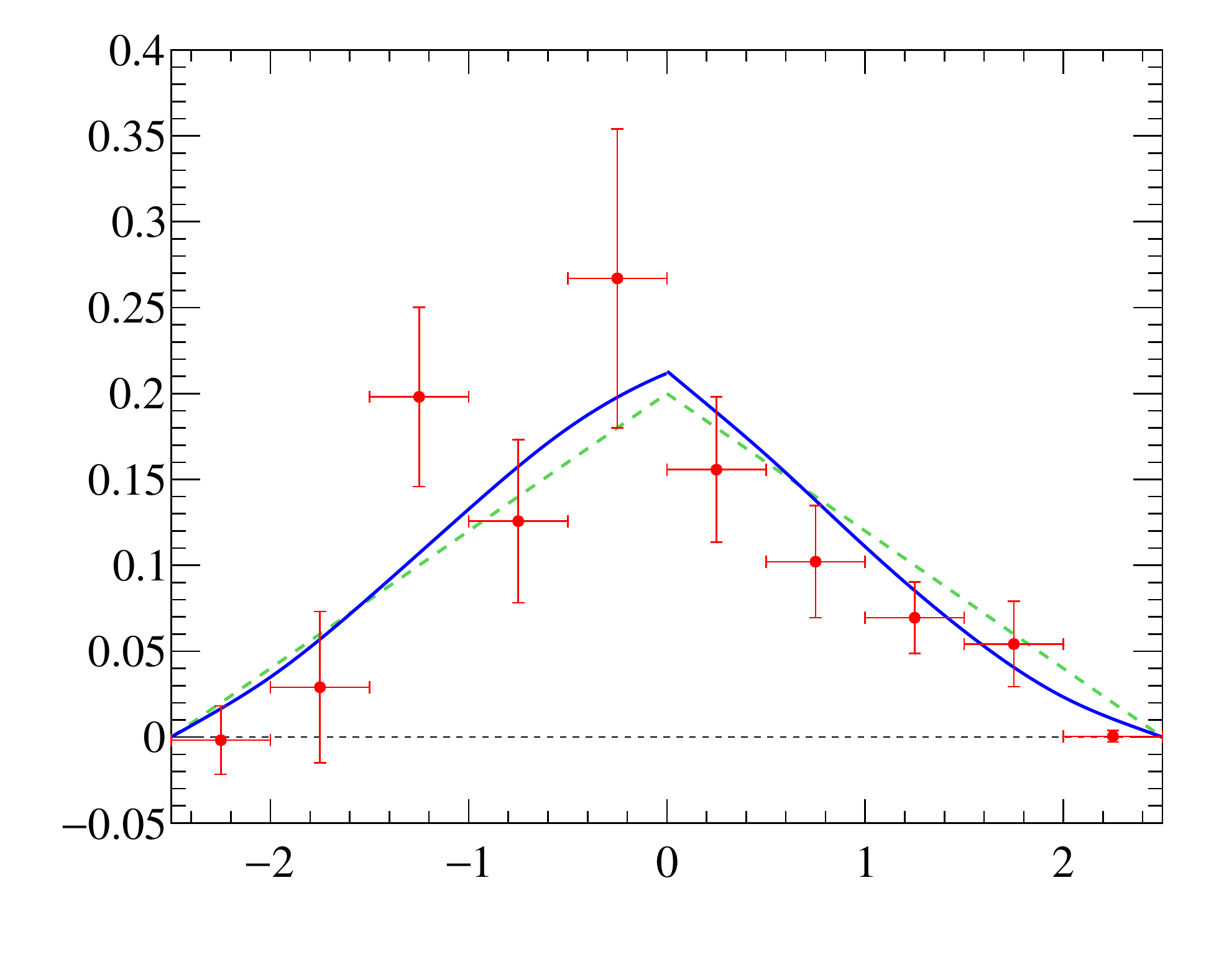}
    }
    \put ( -2, 37) { \begin{sideways}$\tfrac{1}{\upsigma}\tfrac{\deriv\upsigma}{\deriv\Delta y}~~~\left[\tfrac{1}{0.5}\right]$\end{sideways}} 
    \put ( 78, 37) { \begin{sideways}$\tfrac{1}{\upsigma}\tfrac{\deriv\upsigma}{\deriv\Delta y}~~~\left[\tfrac{1}{0.5}\right]$\end{sideways}} 
    \put( 40,   1) { $\Delta y$ } 
    \put(120,   1) { $\Delta y$ }
    \put( 45,108) { a)~$\begin{array}{l} \lhcb \\ \YoneS\Dz\end{array}$ }
    \put(125,108) { b)~$\begin{array}{l} \lhcb \\ \YoneS\Dp\end{array}$ }
    \put( 45, 48) { c)~$\begin{array}{l} \lhcb \\ \YoneS\Dz\end{array}$ }
    \put(125, 48) { d)~$\begin{array}{l} \lhcb \\ \YoneS\Dp\end{array}$ }
  \end{picture}
  \caption { \small
    Background-subtracted and efficiency-corrected
    distributions
    for
    $\left|\Delta\phi\right|/\pi$\,(top) and
    $\Delta y$\,(bottom)
    for 
    $\YoneS\Dz$\,(left) and 
    $\YoneS\Dp$\,(right)~events.
    Straight lines in the~$\left|\Delta\phi\right|/\pi$~plots 
    show the~result of the~fit with a~constant function.
    The~SPS predictions~\cite{Baranov:private} for the~shapes 
        of $\Delta\phi$~distribution 
        are shown 
        with dashed\,(orange) 
        and 
        long\nobreakdash-dashed\,(magenta) 
        curves for calculations based on 
        the~$\kT$\nobreakdash-factorization and 
        the~collinear approximation, respectively.
    The~solid\,(blue) curves in the~$\Delta y$~plots show the~spectra 
    obtained using a~simplified simulation based on data from    
    Refs.~\cite{LHCb-PAPER-2012-041,LHCb-PAPER-2015-045}.
    The~dashed\,(green) lines show the~triangle function expected 
    for totally uncorrelated production of two particles, 
    uniformly distributed in rapidity.
    All~distributions are normalized to unity.
  }
  \label{fig:props_dphidy}
\end{figure}

\begin{figure}[htb]
  \setlength{\unitlength}{1mm}
  \centering
  \begin{picture}(155,122)
    %
    \put(  0,62){ 
      \includegraphics*[width=75mm,height=60mm,%
      ]{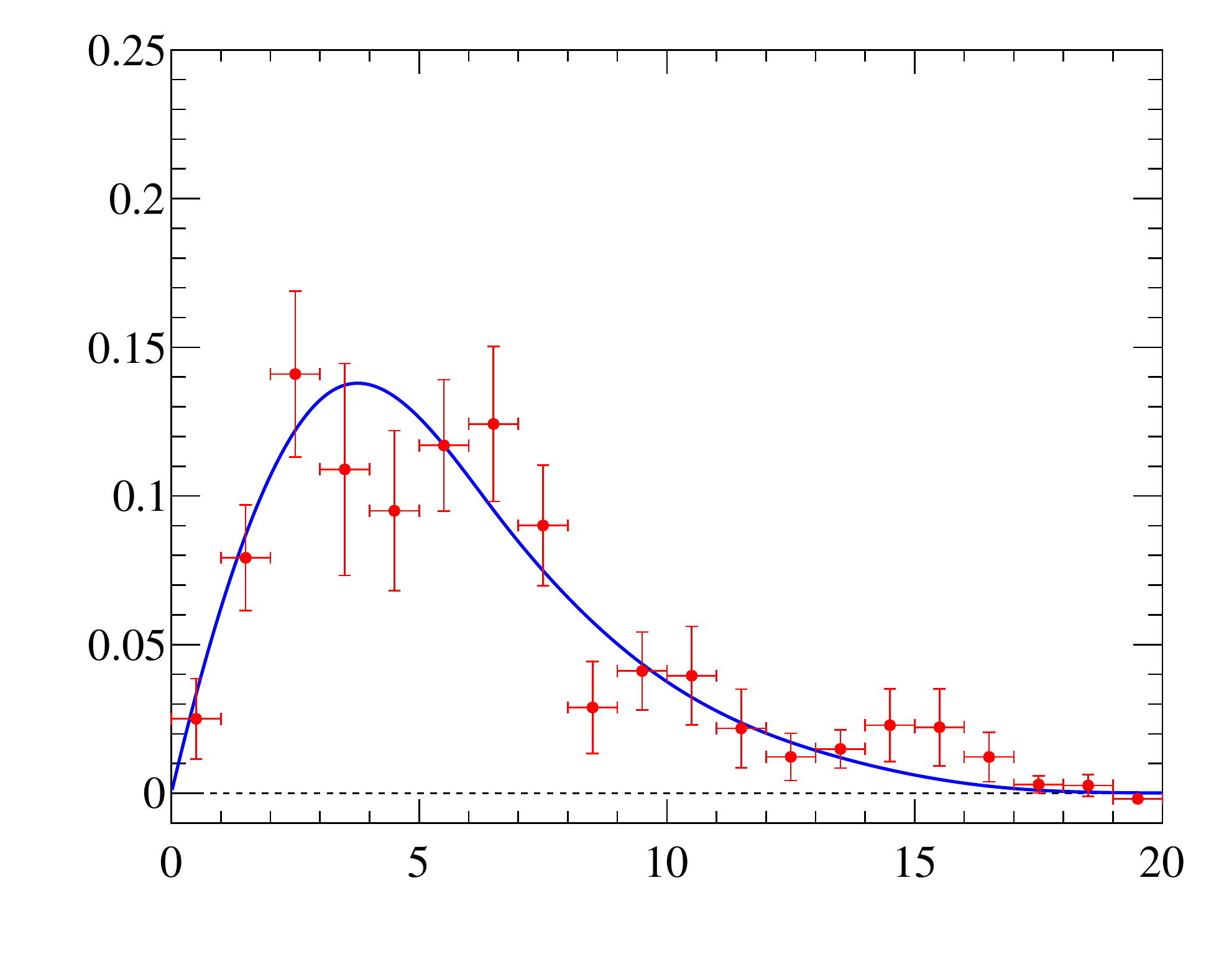}
    }
    \put( 80,62){ 
      \includegraphics*[width=75mm,height=60mm,%
      ]{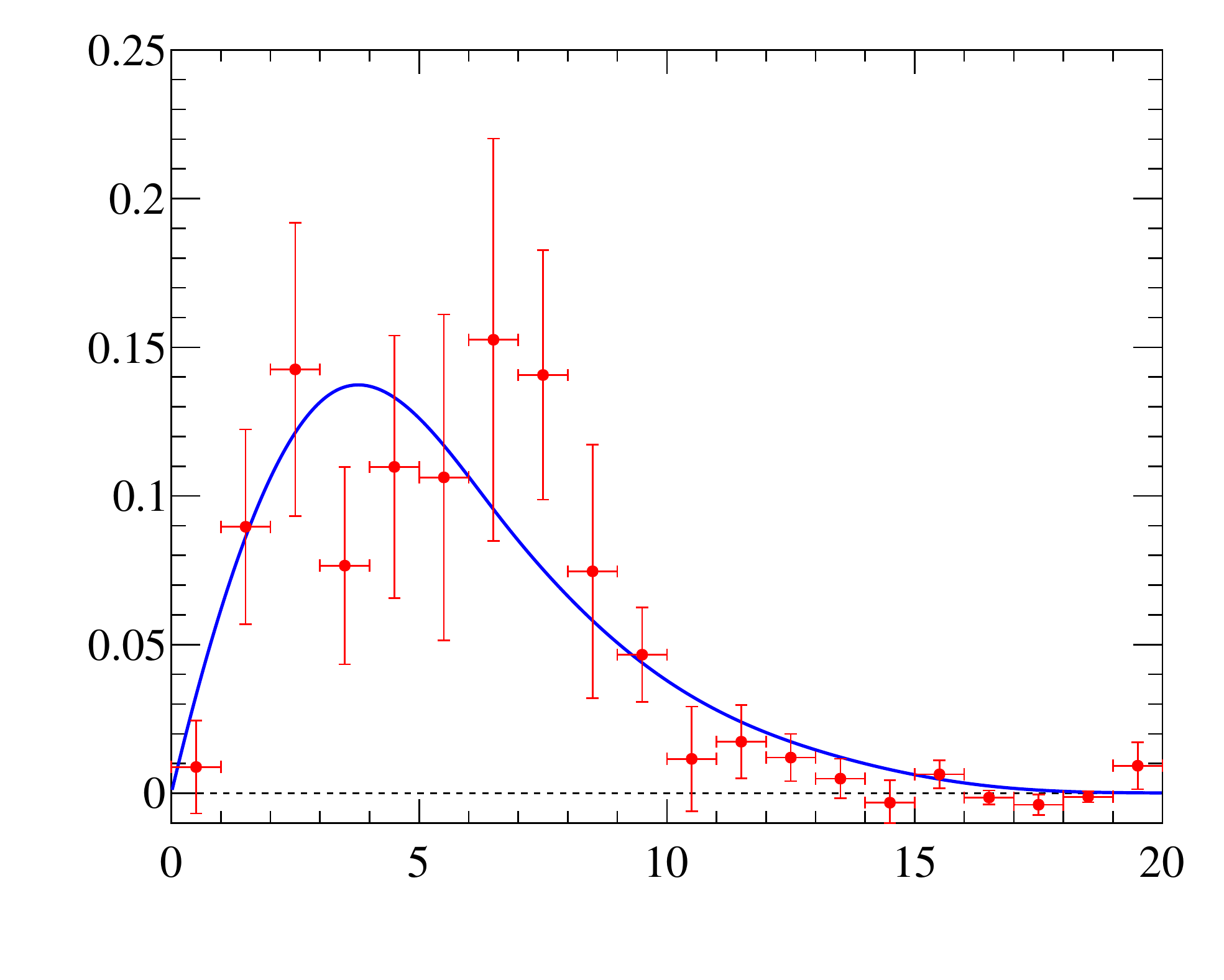}
    }
    \put ( -3, 92) { \begin{sideways}$\tfrac{1}{\upsigma}\tfrac{\deriv\upsigma}{\deriv \pt }~~~\left[\tfrac{1}{1\gevc}\right]$\end{sideways}} 
    \put ( 77, 92) { \begin{sideways}$\tfrac{1}{\upsigma}\tfrac{\deriv\upsigma}{\deriv \pt }~~~\left[\tfrac{1}{1\gevc}\right]$\end{sideways}} 
    \put( 40, 63) { $p^{\YoneS\Dz}_{\mathrm{T}}$ }
    \put(120, 63) { $p^{\YoneS\Dp}_{\mathrm{T}}$ }
    \put( 58, 63) { $\left[\!\gevc\right]$ } 
    \put(138, 63) { $\left[\!\gevc\right]$ } 
    \put(  0, 0){ 
      \includegraphics*[width=75mm,height=60mm,%
      ]{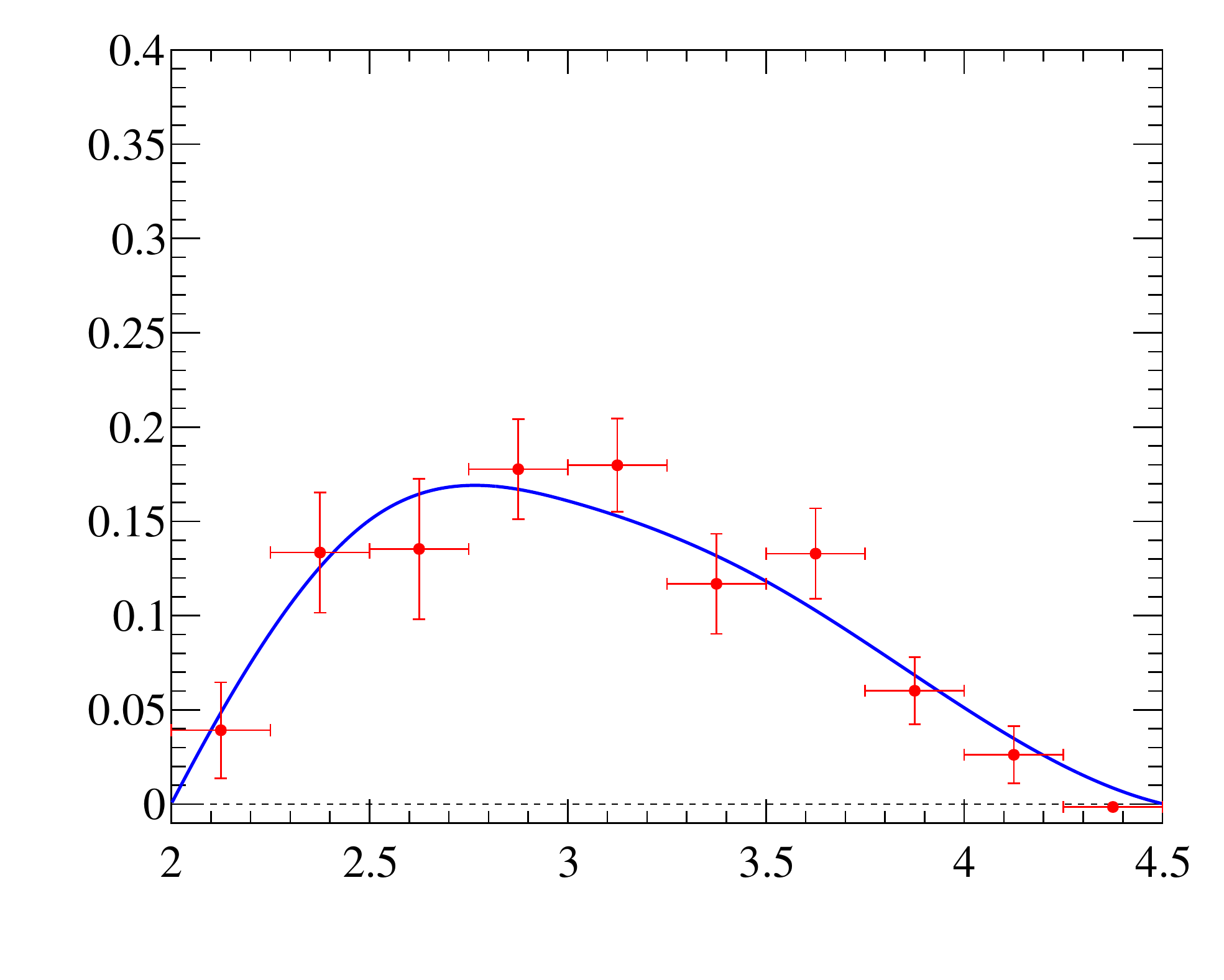}
    }
    \put( 80, 0){ 
      \includegraphics*[width=75mm,height=60mm,%
      ]{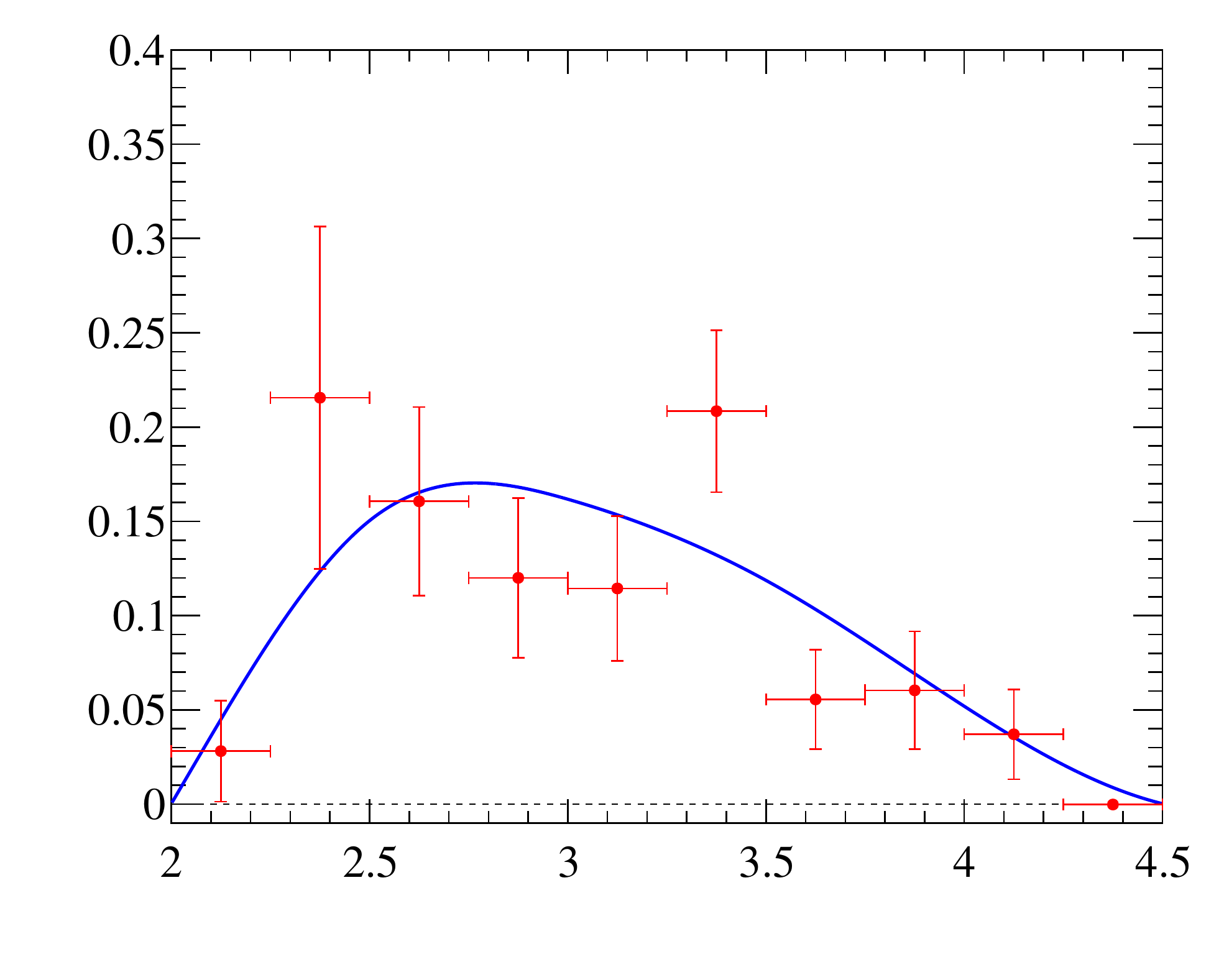}
    }
    \put ( -2, 37) { \begin{sideways}$\tfrac{1}{\upsigma}\tfrac{\deriv\upsigma}{\deriv y}~~~\left[\tfrac{1}{0.25}\right]$\end{sideways}} 
    \put ( 78, 37) { \begin{sideways}$\tfrac{1}{\upsigma}\tfrac{\deriv\upsigma}{\deriv y}~~~\left[\tfrac{1}{0.25}\right]$\end{sideways}} 
    \put( 40,   1) { $y^{\YoneS\Dz}$ } 
    \put(120,   1) { $y^{\YoneS\Dp}$ }
    \put( 45,110) { a)~$\begin{array}{l} \lhcb \\ \YoneS\Dz\end{array}$ }
    \put(125,110) { b)~$\begin{array}{l} \lhcb \\ \YoneS\Dp\end{array}$ }
    \put( 45, 48) { c)~$\begin{array}{l} \lhcb \\ \YoneS\Dz\end{array}$ }
    \put(125, 48) { d)~$\begin{array}{l} \lhcb \\ \YoneS\Dp\end{array}$ }
  \end{picture}
  \caption { \small
    Background-subtracted and efficiency-corrected
    $p_{\mathrm{T}}^{\YoneS\Charm}$\,(top) and  
    $y^{\YoneS\Charm}$\,(bottom)~distributions for 
    $\YoneS\Dz$\,(left) and~$\YoneS\Dp$\,(right)~events.
    The~blue curves show the spectra 
    obtained using a~simplified simulation based on data from 
    Refs.~\cite{LHCb-PAPER-2012-041,LHCb-PAPER-2015-045}.
    All~distributions are normalized to unity.
  }
  \label{fig:props_ptyyc}
\end{figure}

\begin{figure}[htb]
  \setlength{\unitlength}{1mm}
  \centering
  \begin{picture}(155,120)
    %
    \put(  0,60){ 
      \includegraphics*[width=75mm,height=60mm,%
      ]{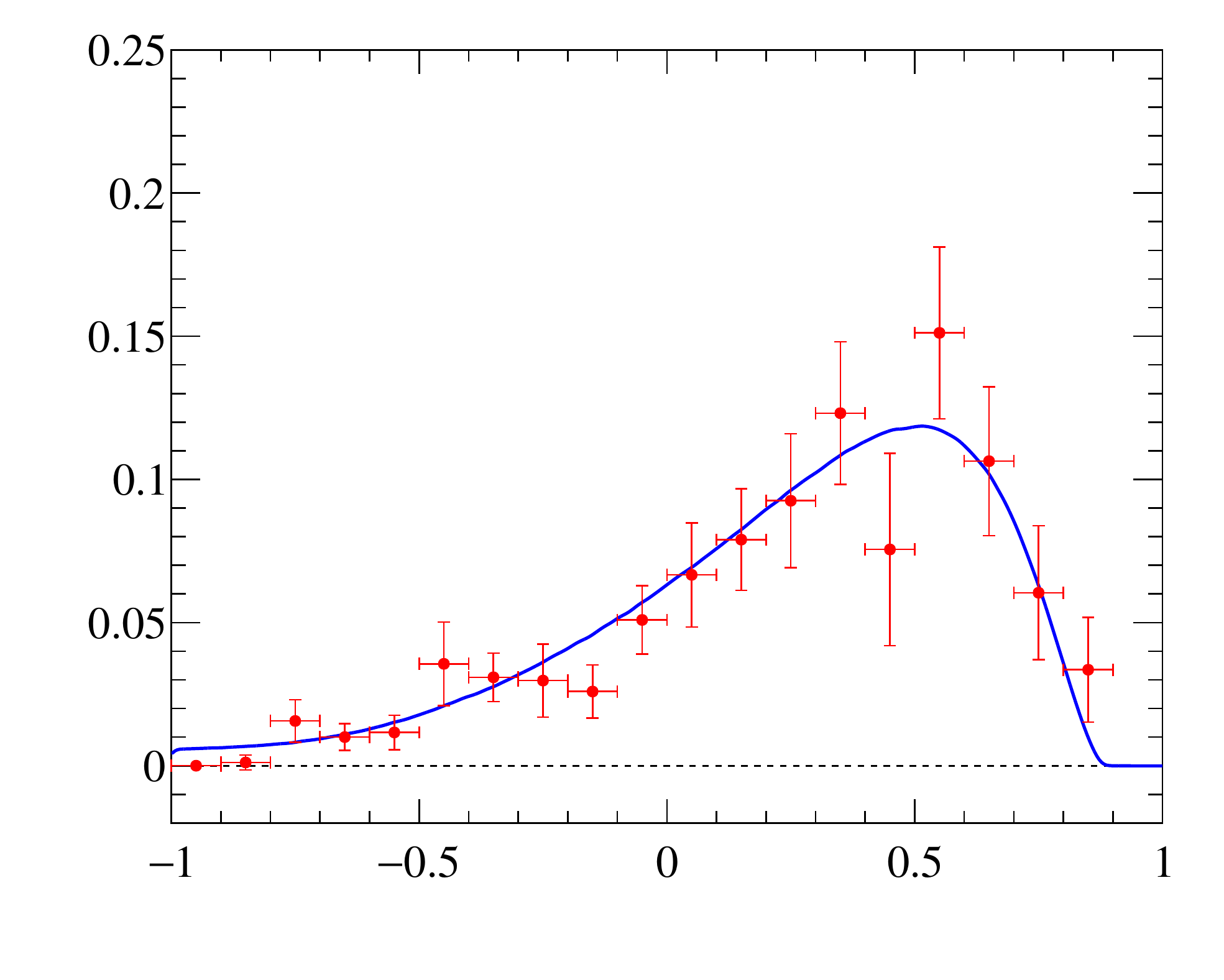}
    }
    \put( 80,60){ 
      \includegraphics*[width=75mm,height=60mm,%
      ]{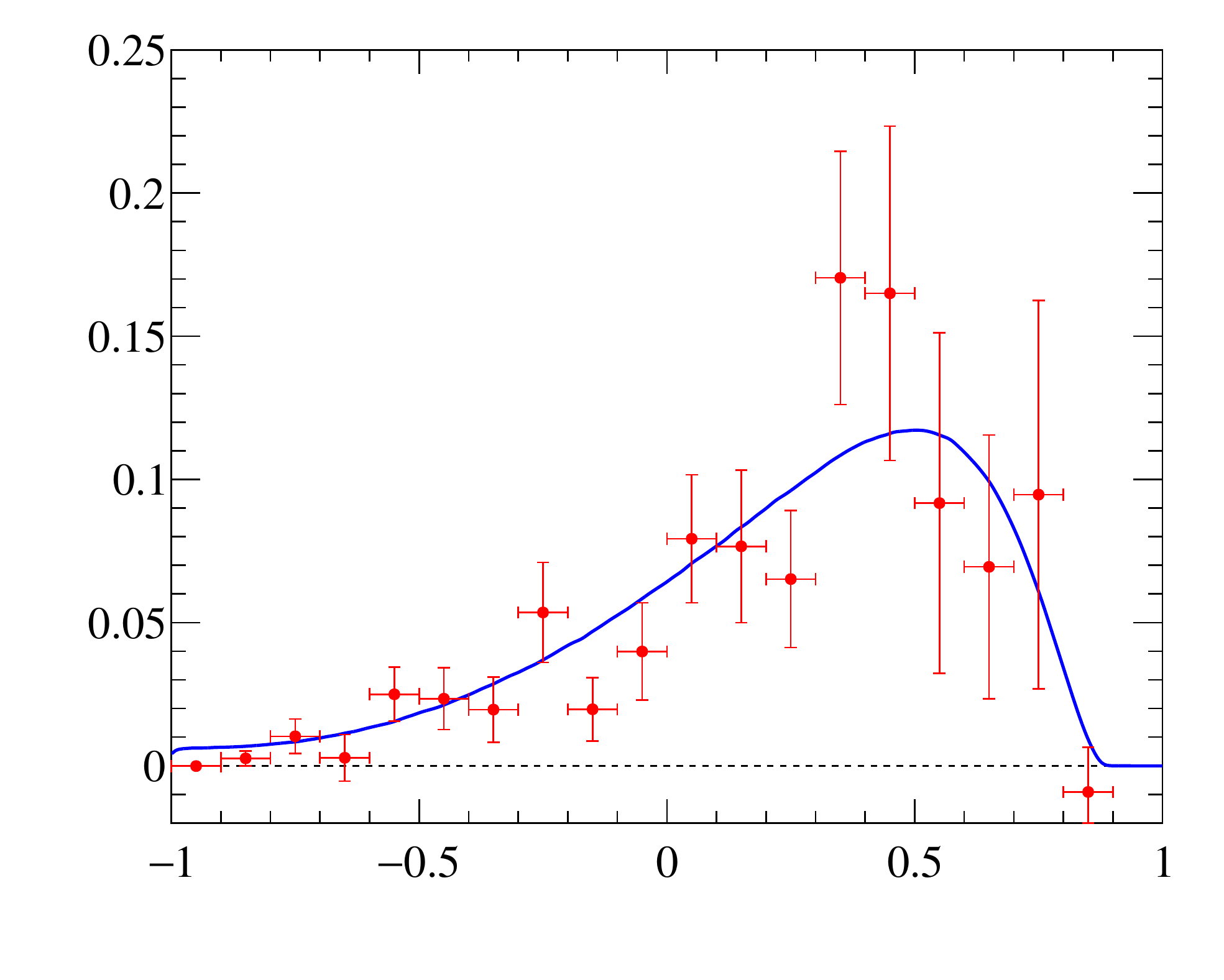}
    }
    \put ( -2, 95) { \begin{sideways}$\tfrac{1}{\upsigma}\tfrac{\deriv\upsigma}{\deriv\mathscr{A}_{\mathrm{T}}}~~~\left[\tfrac{1}{0.1}\right]$\end{sideways}} 
    \put ( 78, 95) { \begin{sideways}$\tfrac{1}{\upsigma}\tfrac{\deriv\upsigma}{\deriv\mathscr{A}_{\mathrm{T}}}~~~\left[\tfrac{1}{0.1}\right]$\end{sideways}} 
    \put( 40,  62) { $\mathscr{A}_{\mathrm{T}}$ } 
    \put(120,  62) { $\mathscr{A}_{\mathrm{T}}$ }
    \put(  0, 0){ 
      \includegraphics*[width=75mm,height=60mm,%
      ]{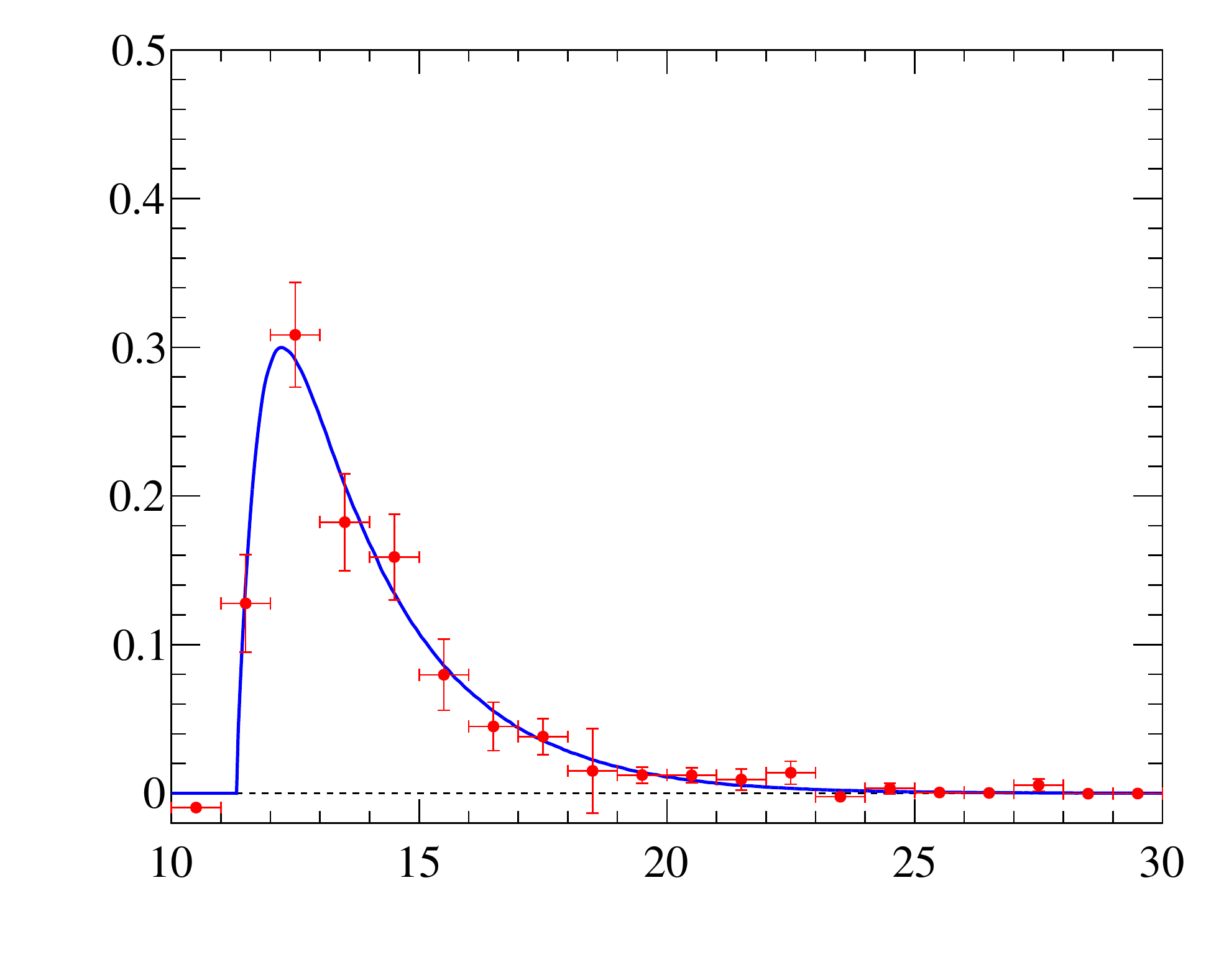}
    }
    \put( 80, 0){ 
      \includegraphics*[width=75mm,height=60mm,%
      ]{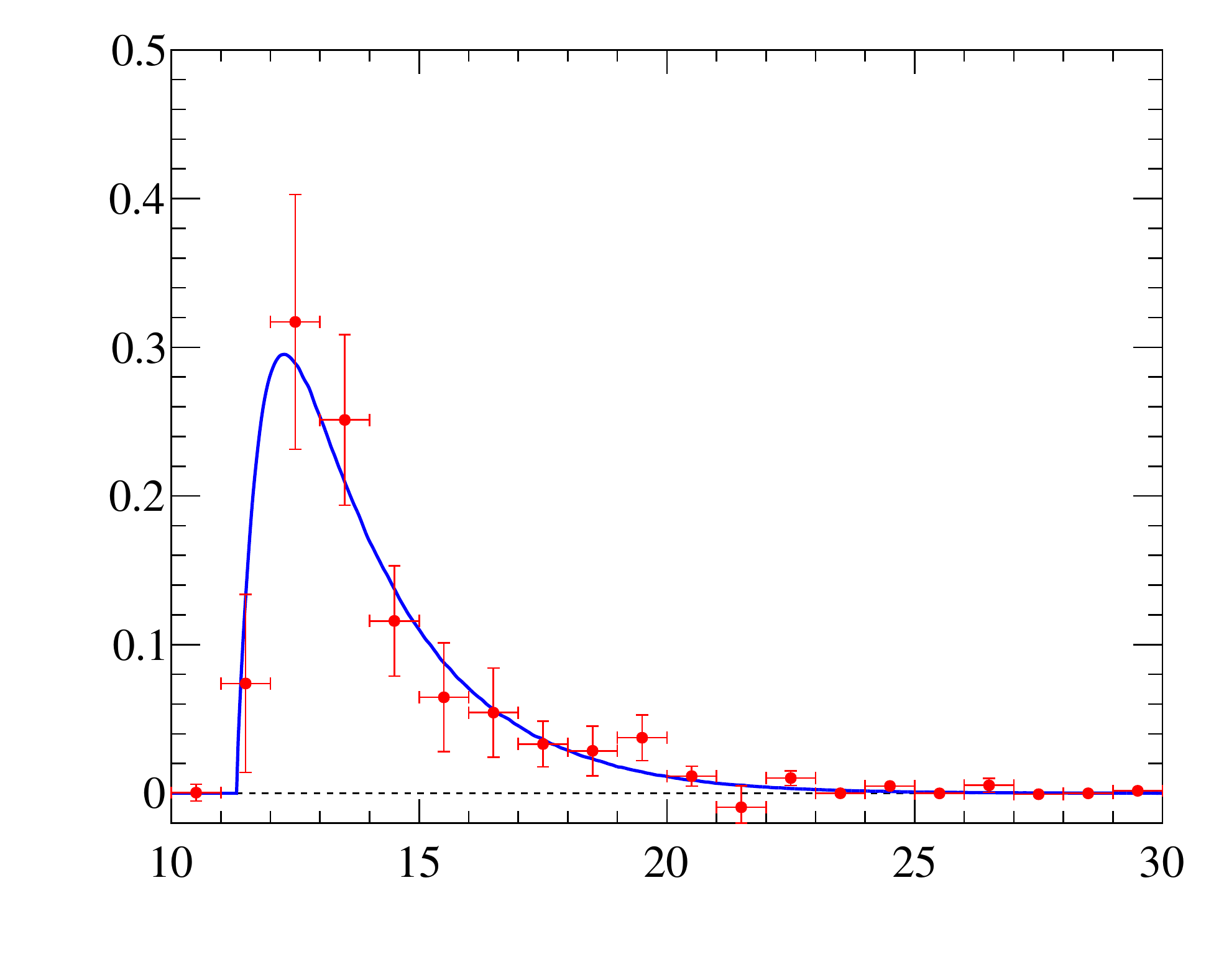}
    }
    \put ( -2, 30) { \begin{sideways}$\tfrac{1}{\upsigma}\tfrac{\deriv\upsigma}{\deriv m }~~~\left[\tfrac{1}{1\gevcc}\right]$\end{sideways}} 
    \put ( 78, 30) { \begin{sideways}$\tfrac{1}{\upsigma}\tfrac{\deriv\upsigma}{\deriv m }~~~\left[\tfrac{1}{1\gevcc}\right]$\end{sideways}} 
    \put( 40,   1) { $m^{\YoneS\Dz}$ }  \put( 58, 1) { $\left[\!\gevcc\right]$ } 
    \put(120,   1) { $m^{\YoneS\Dp}$ }  \put(138, 1) { $\left[\!\gevcc\right]$ } 
    \put( 15,108) { a)~$\begin{array}{l} \lhcb \\ \YoneS\Dz\end{array}$ }
    \put( 95,108) { b)~$\begin{array}{l} \lhcb \\ \YoneS\Dp\end{array}$ }
    \put( 45, 48) { c)~$\begin{array}{l} \lhcb \\ \YoneS\Dz\end{array}$ }
    \put(125, 48) { d)~$\begin{array}{l} \lhcb \\ \YoneS\Dp\end{array}$ }
  \end{picture}
  \caption { \small
    Background-subtracted and efficiency-corrected
    $\mathscr{A}_{\mathrm{T}}$\,(top)
    and $m^{\YoneS\Charm}$\,(bottom) ~distributions for 
    $\YoneS\Dz$\,(left)~and $\YoneS\Dp$\,(right)~events.
    The~blue curves show the spectra 
    obtained using a~simplified simulation based on data from 
    Refs.~\cite{LHCb-PAPER-2012-041,LHCb-PAPER-2015-045}.
    All~distributions are normalized to unity.
  }
  \label{fig:props_aptmass}
\end{figure}

The normalized differential distribution for 
each variable $v$ is calculated as 
\begin{equation} 
\dfrac{1}{\upsigma}
\dfrac{\mathrm{d}\upsigma}{\mathrm{d} v} = 
\dfrac{1}{N_{\mathrm{corr}}^{\ups\Charm}}
\dfrac{ N_{\mathrm{corr},i}^{\ups\Charm} } {\Delta v},  \label{eq:diff}
\end{equation} 
where $N_{\mathrm{corr},i}^{\ups\Charm}$~is the~number of 
efficiency-corrected signal events in bin $i$ of width~$\Delta v$, and 
$N_{\mathrm{corr}}^{\ups\Charm}$ is the~total number of efficiency-corrected 
events. 
The differential distributions are presented for the~following variables 
\begin{itemize}
\item[-] $p_{\mathrm{T}}^{\YoneS}$, the~transverse momentum of the~\YoneS~meson; 
\item[-] $p_{\mathrm{T}}^{\Charm}$, the~transverse momentum of the~$\Dz(\Dp)$~meson;
\item[-] $y^{\YoneS}$, the~rapidity  of the~\YoneS~meson; 
\item[-] $y^{\Charm}$, the~rapidity of the~$\Dz(\Dp)$~meson;
\item[-] $\Delta \phi = \phi^{\YoneS} - \phi^{\Charm}$, the~difference in azimuthal
  angles between  the~\YoneS~and the~\Charm~mesons;
\item[-] $\Delta y    = y^{\YoneS} - y^{\Charm}$, the~difference in rapidity between 
 the~\YoneS~and the~\Charm~mesons;
\item[-] $p_{\mathrm{T}}^{\YoneS\Charm}$, the~transverse momentum of the~$\YoneS\Charm$~system;
\item[-] $y^{\YoneS\Charm}$, the~rapidity of the~$\YoneS\Charm$~system;
\item[-] $\mathscr{A}_{\mathrm{T}} = \dfrac{p^{\YoneS}_{\mathrm{T}} - p^{\Charm}_{\mathrm{T}}}
  {p^{\YoneS}_{\mathrm{T}} + p^{\Charm}_{\mathrm{T}}}$, 
  the~\pt~asymmetry for the~\YoneS~and the~\Charm~mesons;
\item[-] $m^{\YoneS\Charm}$, the~mass of the~$\YoneS\Charm$~system.
\end{itemize} 
The~distributions are shown in 
Figs.~\ref{fig:props_pt}, 
\ref{fig:props_y},
\ref{fig:props_dphidy},
\ref{fig:props_ptyyc}
and~\ref{fig:props_aptmass}.
Only statistical uncertainties are displayed on these figures,
as the~systematic uncertainities 
discussed in Sect.~\ref{sec:syst} are small.
For all variables the width of the~resolution
function is much smaller than the~bin width, \ie the~results
are not affected by bin\nobreakdash-to\nobreakdash-bin migration.

The~shapes of the~measured differential distributions are 
compared with the~SPS and DPS predictions.
The~DPS predictions are deduced from the~measurements given 
in Refs.~\cite{LHCb-PAPER-2012-041,LHCb-PAPER-2015-045},
using a~simplified simulation assuming uncorrelated 
production of the~\YoneS~and charm hadron.
The~agreement between all measured distributions 
and the~DPS predictions is good. 
For~the~SPS mechanism, 
the~predictions~\cite{Baranov:private} based on 
$\kT$\nobreakdash-factorization~\cite{
       Gribov:1984tu,*Levin:1990gg,
       Baranov:2002cf,
       Andersson:2002cf,*Andersen:2003xj,*Andersen:2006pg,
       Baranov:2006dh,Baranov:2006rz,Baranov:2012fb}
using the~transverse momentum dependent gluon density 
from Refs.~\cite{Jung:2012hy,Hautmann:2013tba,Jung:2014vaa}
are used along with 
the~collinear approximation~\cite{Berezhnoy:2015jga}
with the~leading\nobreakdash-order 
gluon density  taken from Ref.~\cite{Gluck:1998xa}.
The~transverse momentum and rapidity distributions 
of  \YoneS~mesons also agree well with
SPS~predictions based on $\kT$\nobreakdash-factorization,
while 
the~shape of the~transverse momentum 
spectra of \ups~mesons disfavours 
the SPS~predictions obtained using 
the~collinear approximation.
The~shapes of the~\yy~distribution have very limited sensitivity 
to the~underlying production mechanism.  

The distribution $\left|\Delta \phi\right|$ is presented
\mbox{in~Fig.~\ref{fig:props_dphidy}(a,b)}.
The~DPS mechanism predicts a~flat distribution 
in $\Delta\phi$, while for SPS  a~prominent enhancement  
at \mbox{$\left|\Delta\phi\right|\sim\pi$}~is 
expected in collinear approximation.
The~enhancement is partly reduced 
    taking into account transverse momenta 
    of collinding partons~\cite{Baranov:2006rz,vanHameren:2015wva}
    and it is expected to be further smeared out 
    at next\nobreakdash-to\nobreakdash-leading order. 
The~measured distributions for \YoneS\Dz~and \YoneS\Dp~events, 
shown \mbox{in~Fig.~\ref{fig:props_dphidy}(a,b)} 
agree with a~flat distribution. The~fit result with a~constant function 
gives a~$p$-value of 6\%\,(12\%) for
the~$\YoneS\Dz\,(\YoneS\Dp)$~case,
indicating
that the~SPS contribution to the~data is small.
The~shape of $\Delta y$ distribution is defined primarily 
by the~acceptance of LHCb experiment $2<y<4.5$ 
and has no sensitivity to the~underlying production 
mechanism, in the~limit of current statistics.

\section{Systematic uncertainties}
\label{sec:syst}

The~systematic uncertainties
related to the measurement of
the~production cross-section for $\ups\Charm$~pairs 
are summarized in Table~\ref{tab:syst} and discussed in detail 
in the~following.

\begin{table}[t]
  \centering
  \caption{ \small 
    Summary of relative systematic uncertainties for $\upsigma^{\ups\Charm}$\,(in \%).
    The total systematic uncertainty does not include 
    the~systematic uncertainty related to the~knowledge 
    of integrated luminosity~\cite{LHCb-PAPER-2014-047}.
    The symbol $\oplus$ denotes the~sum in quadrature.
  } \label{tab:syst}
  \vspace*{3mm}
  \begin{tabular*}{0.99\textwidth}{@{\hspace{2mm}}l@{\extracolsep{\fill}}cc@{\hspace{2mm}}}
    Source                      &   $\upsigma^{\ups\Dz}$& $\upsigma^{\ups\Dp}$ 
    \\[1mm]
    \hline
    \\[-2mm]
    Signal $\ups\Charm$~extraction          &        &  \\ 
    \quad $\ups$ and \Charm~signal shapes   
    &  $0.1\oplus0.3 $    
    &  $0.1\oplus0.5 $  \\
    \quad 2D fit model          &  0.4  & 0.7  \\ 
    \ups~radiative tail         &  1.0  & 1.0  \\ 
    Efficiency corrections      &  0.1  & 1.3  \\ 
    Efficiency calculation      &       &      \\ 
    \quad muon identification 
    & 0.2  &  0.2     \\
    \quad hadron identification 
    & 0.5  &  0.8    \\
    \quad simulated samples size   & 0.2   & 0.2  \\ 
    \quad tracking   
    & $0.4 \oplus4\times0.4 $
    & $0.5 \oplus5\times0.4 $   \\ 
    \quad hadronic interactions 
    & $2\times1.4 $
    & $3\times1.4 $   \\ 
    \quad trigger                  & 2.0   & 2.0  \\ 
    \quad data-simulation 
          agreement                & 1.0   & 1.0  \\  
    $\mathscr{B}_{\Charm}$          & 1.3   & 2.1  
    \\[1mm]
    \hline 
    \\[-2mm]
    Total   &  4.3   &  5.9  \\ 
  \end{tabular*}   
\end{table}

\begin{table}[t]
  \centering
  \caption{ \small 
    Summary of relative systematic uncertainties for $\upsigma_{\mathrm{eff}}$\,(in \%).
    The~reduced uncertainty for \Charm~hadron production cross-section, 
    denoted as $\updelta(\upsigma^{\Charm})$,  
    is recalculated from Ref.~\cite{LHCb-PAPER-2012-041} 
    taking into account the~cancellation of 
    correlated systematic uncertainties.
  } \label{tab:systseff}
  \vspace*{3mm}
  \begin{tabular*}{0.95\textwidth}{@{\hspace{5mm}}l@{\extracolsep{\fill}}cc@{\hspace{5mm}}}
    Source                      
    &   $\left.\upsigma_{\mathrm{eff}}\right|_{\ups\Dz}$
    &   $\left.\upsigma_{\mathrm{eff}}\right|_{\ups\Dp}$
    \\[1mm]
    \hline 
    \\[-2mm]
    Signal $\ups\Charm$~extraction          &        &  \\ 
    \quad $\ups$ and \Charm~signal shapes   
    &  $0.1 \oplus0.3 $    
    &  $0.1 \oplus0.5 $  \\
    \quad  2D~fit model         &  0.4  & 0.7  \\ 
    Efficiency corrections      &  0.1  & 1.3  \\ 
    Efficiency calculation      &       &      \\ 
    \quad hadron identification 
    & 0.5  &  0.8    \\
    \quad simulated samples size       &   0.2     & 0.2  \\ 
    $\updelta(\upsigma^{\Charm})$               & 6.7     & 9.7   \\
    FONLL extrapolation\,($\sqs=8\tev$ only)   & 2.1     & 2.1 
    \\[1mm]
    \hline 
    \\[-2mm]
    Total~$\begin{array}{l}\sqs=7\tev \\ \sqs=8\tev\end{array}$ 
    &  $\begin{array}{c} 6.7 \\  7.0            \end{array}$ 
    &  $\begin{array}{c} 9.9 \\ 10.1\phantom{0} \end{array}$ 
  \end{tabular*}   
\end{table}

\begin{table}[t]
  \centering
  \caption{ \small 
    Summary of relative systematic uncertainties for 
    the~ratios~$R^{\ups\Charm}$ and $R^{\Dz/\Dp}$\,(in \%).
  } \label{tab:systr}
  \vspace*{3mm}
  \begin{tabular*}{0.99\textwidth}{@{\hspace{2mm}}l@{\extracolsep{\fill}}ccc@{\hspace{2mm}}}
    Source                      
    & $R^{\ups\Dz}$ 
    & $R^{\ups\Dp}$  
    & $R^{\Dz/\Dp}$
    \\[1mm]
    \hline 
    \\[-2mm]
    Signal extraction           &        &   &  \\ 
    \quad $\ups$ and \Charm~signal shapes   
    &  $0.1 \oplus0.3 $    
    &  $0.1 \oplus0.5 $  
    &  $0.3 \oplus0.5 $  \\
    \quad 2D~fit model              
    & 0.4  
    & 0.7  
    & $0.4 \oplus 0.7 $ \\ 
    Efficiency corrections      
    & 0.1  
    & 1.3  
    & $0.1 \oplus 1.3 $  \\ 
    Efficiency calculation:     &        &  & \\ 
    \quad hadron identification 
    &  0.5  
    &  0.8  
    &  $0.5 \oplus 0.8 $  \\
    \quad tracking   
    & $0.4 \oplus4\times0.4 $
    & $0.5 \oplus5\times0.4 $  
    & $0.6 \oplus1\times0.4 $ \\ 
    \quad hadronic interactions 
    & $2\times1.4 $
    & $3\times1.4 $    
    & $1\times1.4 $  \\ 
    \quad data-simulation agreement 
    & 1.0   
    & 1.0  
    & $1.0 \oplus 1.0 $ \\  
    \quad simulated samples size    
    & 0.2   
    & 0.2  
    & $0.2 \oplus 0.2 $ \\ 
    $\mathscr{B}_{\Charm}$      
    & 1.3   
    & 2.1  
    & $1.3 \oplus 2.1 $   
    \\[1mm]
    \hline 
    \\[-2mm]
    Total   &  3.4   &  5.3   & $3.8$
  \end{tabular*}   
\end{table}

The signal shapes and parameters are taken from 
fits to large low-background inclusive 
$\ups\to\mumu$ and charm samples.  
The~parameters, signal peak positions and resolutions and 
the~tail parameters for the~double-sided Crystal~Ball 
and the~modified Novosibirsk functions,
are varied within their uncertainties as~determined 
from the~calibration samples. 
The~small difference in parameters between the~data sets obtained 
at $\sqs=7$~and $8\tev$ is also used to assign the~systematic 
uncertainty. For \Dz~and \Dp~signal peaks 
alternative 
fit models have been used, namely a~double-sided asymmetric variant 
of an~Apolonious~function~\cite{Santos:2013gra}
without power-law tail, 
a~double-sided Crystal~Ball function and 
an~asymmetric Student\nobreakdash-$t$~shape. 
The~systematic uncertainty related to the~parameterization of
the~combinatorial 
background is determined by varying
the~order of the~polynomial function 
in Eq.~\eqref{eq:b} between zeroth and second order. 
For the~purely combinatorial background 
component\,(last line in Eq.~\eqref{eq:pdf}), 
a~non\nobreakdash-factorizable function is used 
\begin{equation}
  F^{BB}( m_{\mumu}, m_{\Charm} ) 
  \propto 
  \mathrm{e}^{-\upbeta_1  m_{\mumu} - \upbeta_2 m_{\Charm}} 
  \times 
  \left( \sum^{n}_{i=0} \sum^{k}_{j=0} 
    \upkappa^2_{ij} 
    {P}^i_{n}\left( m_{\mumu} \right)
    {P}^j_{k}\left( m_{\Charm} \right)
  \right), 
\end{equation}
where the~parameters $\upbeta_1$, $\upbeta_2$ and $\upkappa_{i,j}$
are allowed to float in the~fit, and 
\mbox{$P^i_{n}$} and
\mbox{$P^j_{k}$}~are basic Bernstein polynomials,
and the~order of these polynomials, $n$ and~$k$, 
is varied between zero and two. 
The~corresponding variations of $\ups\Charm$~signal yields 
are taken as the~systematic uncertainty related to
the~description of the~signal and background components. 

Other systematic uncertainties
are related to 
the~imperfection of the~\photos~generator~\cite{Golonka:2005pn} 
to describe the~radiative tails
in \mbox{$\ups\to\mumu$}~decays.
This systematic is studied 
in Ref.~\cite{LHCb-PAPER-2011-003} and taken to be~1\%.

The systematic uncertainty related to efficiency correction 
is estimated using an~alternative technique for the~determination 
of $N^{\ups\Charm}_{\mathrm{corr}}$, 
where the~efficiency\nobreakdash-corrected yields are obtained via
\begin{equation}
  N^{\ups\Charm}_{\mathrm{corr}}= \sum_i \dfrac{ w_i } {\varepsilon^{\mathrm{tot}}},
\end{equation}
where $w_i$ is the~signal event weight, obtained with
the~\sPlot 
technique~\cite{Pivk:2004ty} using fits
to the~efficiency\nobreakdash-uncorrected data sets,
and $\varepsilon^{\mathrm{tot}}$~is a~total efficiency for the~given event, defined 
with  Eq.~\eqref{eq:effic}. The~difference
in~the~efficiency\nobreakdash-corrected yields with respect to
the~baseline approach 
of $0.1\,(1.3)\%$ for $\ups\Dz\,(\ups\Dp)$, 
is assigned as the~corresponding systematic uncertainty.

The systematic uncertainty related to
the~particle identification
is estimated to be~0.2\% for muons
and $0.5\,(0.8)\%$ for hadrons 
for the~\mbox{$\YoneS\Dz\,(\YoneS\Dp)$}~case
and is obtained 
from the~uncertainties for 
the~single particle identification efficiencies
using an~error propagation technique
with a~large number of pseudoexperiments.
The~same approach is used to~propagate the~uncertainties in 
$\varepsilon^{\mathrm{acc}}$,
$\varepsilon^{\mathrm{rec}}$ and 
$\varepsilon^{\mathrm{trg}}$ related to the~limited simulation 
sample~size.

The~efficiency is corrected using data-driven
techniques to account for small differences in 
the~tracking efficiency between data and 
simulation~\mbox{\cite{LHCb-DP-2013-002,LHCb-DP-2013-001}}.
The~uncertainty in the~correction factor is propagated 
to the cross-section measurement using pseudoexperiments 
resulting in a~global 0.4\,(0.5)\%~systematic uncertainty 
for the~$\ups\Dz\,(\ups\Dp)$~cases plus 
an~additional uncertainty of 0.4\%~per track. 
The~knowledge of the~hadronic interaction length 
of~the~detector results in an~uncertainty
of 1.4\%~per final\nobreakdash-state hadron~\cite{LHCb-DP-2013-002}.

The systematic uncertainty associated with the~trigger requirements 
is assessed by studying the~performance of the~dimuon trigger
for \YoneS~events selected using the~single muon 
high-\pt~trigger~\cite{LHCb-DP-2012-004}
in data and simulation. The~comparison is performed 
in bins of the \YoneS~meson  transverse momentum and rapidity 
and the~largest observed difference of 2.0\% is assigned 
as the~systematic uncertainty associated with 
the~imperfection of trigger simulation~\cite{LHCb-PAPER-2015-045}.

Using large samples of  low-background  inclusive 
\mbox{$\ups\to\mumu$}, 
\mbox{$\Dz\to\Km\pip$} and 
\mbox{$\Dp\to\Km\pip\pip$}~events,
good agreement between data and simulation 
is observed for the~selection variables used in this analysis,
in particular for dimuon and charm vertex quality and 
$\chisq_{\mathrm{fit}}(\ups)/\mathrm{ndf}$.
The~small differences seen would affect the~efficiencies 
by less than~1.0\%, which is conservatively taken as 
a~systematic uncertainty accounting for 
the~disagreement between data and simulation.

The systematic uncertainty related to
the~uncertainties of the~branching fractions 
of \Dz~and \Dp~mesons is~1.3\% and 2.1\%~\cite{PDG2014}.
The integrated luminosity is measured using 
a~beam-gas imaging method~\cite{FerroLuzzi:2005em,LHCb-PAPER-2011-015}.
The absolute luminosity scale is determined 
with \mbox{$1.7\,(1.2)\%$}~uncertainty 
for the~sample collected at~\mbox{$\sqs=7\,(8)\tev$},
dominated by the~vertex resolution 
for beam-gas interactions, 
the~spread of the~measurements
and the~detector alignment~\cite{LHCb-PAPER-2014-047,LHCb-PAPER-2011-015,Barschel:1693671}.

The selection criteria 
    favour the~selection of charm hadrons produced promptly 
    at the~$\proton\proton$~collision vertex and 
    significantly suppress the~feed down from charm hadrons produced 
    in decays of beauty hadrons.  
The~remaining feed down is estimated separately for 
    DPS and SPS processes with 
    the~simultaneous production of  
    an~\ups~meson and a~$\bquark\bquarkbar$-pair.
    The former is estimated using simulation, 
    normalized to the~measured $\bquark\bquarkbar$~and 
    $\cquark\cquarkbar$~ production cross-sections~\cite{LHCb-PAPER-2010-002,LHCb-PAPER-2012-041}
    and validated using a~data\nobreakdash-driven technique.
    It~is found to be smaller than 1.5\%
    of the~observed signal
    and is neglected.
The~contribution from SPS processes with the~associated
    production of \ups~meson and $\bquark\bquarkbar$~pairs 
    is estimated using the~prediction for the~ratio of 
    production cross\nobreakdash-sections,
    \begin{equation}
      \dfrac
      { \upsigma^{\ups\bquark\bquarkbar}} 
      { \upsigma^{\ups\cquark\cquarkbar}} = (2-5)\%\,, \label{eq:rbbcc}
    \end{equation}
    obtained using the~$\kT$\nobreakdash-factorization approach
    with the~transverse momentum dependent gluon density 
    taken from Refs.~\cite{Jung:2012hy,Hautmann:2013tba,Jung:2014vaa}.
    The~uncertainty reflects the variation of scale and 
    the~difference with results obtained using 
    the~collinear approximation with the~gluon density  
    from Ref.~\cite{Gluck:1998xa}.
    Combining the estimates from 
    Eqs.~\eqref{eq:rbbcc}, \eqref{eq:r} and~\eqref{eq:rkt}
    with the~probability for a~charm hadron from the~decay of 
    beauty hadron to pass the~selection criteria, 
    this feed down is found to be totally negligible.

    The~effect of possible extreme polarization scenarios
    for \ups~mesons from SPS processes is proportional
    to the~SPS contamination, $\upalpha_{\mathrm{SPS}}$,
    and could lead to  $+0.08\,(-0.16)\alpha_{\mathrm{SPS}}$ correction~\cite{LHCb-PAPER-2011-036}
    to the~cross\nobreakdash-sections $\upsigma^{\ups\Charm}$
    and the~ratios  $R^{\ups\Charm}$
    for totally transverse\,(longitudinal) polarizations
    of \ups~mesons in centre-of-mass helicity frame.
    It~is very small for small SPS contamination.
    The~corresponding corrections to ratios
    $R^{\Dz/\Dp}$ are non-zero only
    if SPS has different contributions
    to $\ups\Dz$ and $\ups\Dp$~production processes
    and accouts for $+0.08\,(-0.16)\Delta\alpha_{\mathrm{SPS}}$,
    where $\Delta\alpha_{\mathrm{SPS}}$ is the~difference in SPS contaminations
    to the considered processes.
    The~same estimate is valid also for
    the~ratios $R^{\YoneS/\YtwoS}$.

A~large part of the~systematic uncertainties cancels in the~ratio
$R^{\ups\Charm}$ and  in the variable~$\upsigma_{\mathrm{eff}}$.
The~systematic uncertainties for 
$\upsigma_{\mathrm{eff}}$, 
$R^{\ups\Charm}$ and 
$R^{\Dz/\Dp}$ 
are summarized 
in Tables~\ref{tab:systseff} and 
\ref{tab:systr}. 
For~the~production cross\nobreakdash-section of charm mesons 
at $\sqs=8\tev$ the~measured cross\nobreakdash-section at $\sqs=7\tev$
is extrapolated using FONLL~calculations~\cite{Cacciari:1998it,Cacciari:2001td,Cacciari:2012ny}.
The~uncertainty related to the~imperfection of the~extrapolation is 
estimated from the~comparison of the~measured ratio 
$\upsigma^{\Charm}_{\sqs=13\tev}/\upsigma^{\Charm}_{\sqs=7\tev}$~\cite{LHCb-PAPER-2012-041,LHCb-PAPER-2015-041}
and the~corresponding FONLL estimate. 
As a~result of this comparison the~\Charm~hadron production cross-section 
is scaled up by 2.7\% and 
a~systematic uncertainty of 2.1\% is assigned.
The~systematic uncertainty for the~ratios $R_{\Charm}^{\YtwoS/\YoneS}$
is  small compared to the~statistical uncertainty and is neglected.

\section{Results and discussion}
\label{sec:results}

The~associated production of \ups~and charm mesons is studied.
Pair~production of 
$\YoneS\Dz$, 
$\YtwoS\Dz$, 
$\YoneS\Dp$, 
$\YtwoS\Dp$ and 
$\YoneS\Ds$~states 
is observed 
with significances exceeding five~standard deviations.
The~production cross-sections in the~fiducial region 
$2.0<\yy<4.5$, $\pty<15\gevc$, 
$2.0<\yC<4.5$ and $1<\ptC<20\gevc$
are measured for  $\YoneS\Dz$ and $\YoneS\Dp$~final states at 
$\sqs=7$~and 8\tev as:  
\begin{eqnarray*}
  \mathscr{B}_{\mumu}\times\upsigma^{\YoneS\Dz}_{\sqs=7\tev}  
  & =  &           155   \pm 21  \stat \pm \phantom{1}7\syst \pb\,,  \\ 
  \mathscr{B}_{\mumu}\times \upsigma^{\YoneS\Dp}_{\sqs=7\tev}  
  & =  & \phantom{0}82   \pm 19  \stat \pm \phantom{1}5 \syst \pb\,,  \\ 
  \mathscr{B}_{\mumu}\times \upsigma^{\YoneS\Dz}_{\sqs=8\tev}  
  & =  &           250   \pm 28  \stat \pm           11\syst \pb\,,    \\
  \mathscr{B}_{\mumu}\times \upsigma^{\YoneS\Dp}_{\sqs=8\tev} 
  & =  & \phantom{0}80   \pm 16  \stat \pm \phantom{1}5  \syst \pb\,,  
\end{eqnarray*}
where the~first uncertainty is statistical, and the~second 
is the~systematic uncertainty from Table~\ref{tab:syst}, 
combined with the~uncertainty related to the~knowledge of 
the~luminosity.
All~these measurements are statistically limited.
The measured cross-sections 
are in agreement with the~DPS~expectations from Eq.~\eqref{eq:dpsth}, 
and significantly exceed the~expectations 
from the~SPS~mechanism in Eqs.~\eqref{eq:r} and~\eqref{eq:rkt}.
Differential kinematic distributions are studied for $\ups\Dz$ and $\ups\Dp$~final states.
All~of them are in good agreement with DPS~expectations as the~main production mechanism.

The~ratios of the~cross-sections for  $\YoneS\Dz$~and $\YoneS\Dp$ are 
\begin{eqnarray*}
  R^{\Dz/\Dp}_{\sqs=7\tev}  = \dfrac {\upsigma^{\YoneS\Dz}_{\sqs=7\tev}}
  {\upsigma^{\YoneS\Dp}_{\sqs=7\tev}} 
  & = &  1.9 \pm  0.5 \stat \pm 0.1 \syst\,,  \\ 
  R^{\Dz/\Dp}_{\sqs=8\tev} = \dfrac {\upsigma^{\YoneS\Dz}_{\sqs=8\tev} }
  {\upsigma^{\YoneS\Dp}_{\sqs=8\tev}} 
  & = &  3.1  \pm 0.7 \stat \pm 0.1 \syst\,, \\ 
\end{eqnarray*} 
where the~systematic uncertainty is discussed in detail in Sect.~\ref{sec:syst}.
The~results are compatible  with 
the~DPS expectation of $2.41\pm0.18$ from Eq.~\eqref{eq:rdd}.

The~cross-section ratios $R^{\ups\Charm}$ are measured to be
\begin{eqnarray*}
  R^{\YoneS\Dz}_{\sqs=7\tev} 
  = \left.\dfrac{\upsigma^{\YoneS\Dz}}{\upsigma^{\YoneS}}\right|_{\sqs=7\tev}
  & = &  \left( 6.3  \pm 0.8 \stat \pm 0.2 \syst \right)\%\,,  \\
  R^{\YoneS\Dp}_{\sqs=7\tev} 
  = \left.\dfrac{\upsigma^{\YoneS\Dp}}{\upsigma^{\YoneS}}\right|_{\sqs=7\tev}
  & = &  \left( 3.4  \pm 0.8 \stat \pm 0.2 \syst \right)\%\,,  \\
  R^{\YoneS\Dz}_{\sqs=8\tev}
  = \left. \dfrac{\upsigma^{\YoneS\Dz}}{\upsigma^{\YoneS}}\right|_{\sqs=8\tev}
  & = &  \left( 7.8  \pm 0.9 \stat \pm 0.3 \syst \right)\%\,,  \\
  R^{\YoneS\Dp}_{\sqs=8\tev} 
  = \left. \dfrac{\upsigma^{\YoneS\Dp}}{\upsigma^{\YoneS}} \right|_{\sqs=8\tev}
  & = &  \left( 2.5  \pm 0.5 \stat \pm 0.1 \syst \right)\%\,.  
\end{eqnarray*}
Extrapolating the~ratios $R^{\ups\Charm}$ down to \mbox{$\ptC=0$} using
the~measured transverse momentum spectra of \Dz~and \Dp~mesons from Ref.~\cite{LHCb-PAPER-2012-041},
and using the~fragmentation fractions  
\mbox{$f(\cquark\to\Dz)=0.565\pm0.032$} and 
\mbox{$f(\cquark\to\Dp)=0.246\pm0.020$}, measured at \epem~colliders 
operating at a~centre-of-mass energy close to the \YfourS~resonance~\cite{Amsler:2008zzb:Frag}, 
the~ratios $R^{\ups\cquark\cquarkbar}$ are calculated to be   
\begin{eqnarray*}
  R^{\YoneS\cquark\cquarkbar}_{\sqs=7\tev} 
  = \left.\dfrac{\upsigma^{\YoneS\cquark\cquarkbar}}{\upsigma^{\YoneS}}\right|_{\sqs=7\tev}
  & = &  \left( 7.7  \pm 1.0 \right)\%\,,  \\
  R^{\YoneS\cquark\cquarkbar}_{\sqs=8\tev} 
  = \left.\dfrac{\upsigma^{\YoneS\cquark\cquarkbar}}{\upsigma^{\YoneS}}\right|_{\sqs=8\tev}
  & = &  \left( 8.0 \pm  0.9 \right)\%\,,  
\end{eqnarray*}
which significantly exceed SPS expectations from 
Eqs.~\eqref{eq:r} and~\eqref{eq:rkt}.

The large statistical uncertainty for the~other $\ups\Charm$~modes 
does not allow to obtain a~numerical model\nobreakdash-independent 
measurement, but, assuming  similar kinematics for~$\YtwoS$~and charm
mesons to the~prompt production, the~following ratios are measured
\begin{eqnarray*}
  R_{\Dz}^{\YtwoS/\YoneS}   =  
  \mathscr{B}_{2/1} \times \dfrac
  {  \upsigma^{\YtwoS \Dz}_{\sqs=7\tev\phantom{\&8}}}
  {  \upsigma^{\YoneS \Dz}_{\sqs=7\tev\phantom{\&8}}} 
  & = &  (13  \pm5  )\%\,, 
  \\
  R_{\Dz}^{\YtwoS/\YoneS}   =  
  \mathscr{B}_{2/1} \times \dfrac
  {  \upsigma^{\YtwoS \Dz}_{\sqs=8\tev\phantom{\&7}}}
  {  \upsigma^{\YoneS \Dz}_{\sqs=8\tev\phantom{\&7}}} 
  & = &  (20  \pm4  )\%\,, 
\end{eqnarray*} 
where  $\mathscr{B}_{2/1}$~is the~ratio of dimuon branching fractions
of \YtwoS and \YoneS~mesons and 
where the~systematic uncertainties are negligible compared 
to statistical uncertainties. 
These~values are smaller than, but compatible with
the~DPS~expectations from Eq.~\eqref{eq:r21}.
For~the~$\ups\Dp$~production one obtains 
\begin{eqnarray*}
  R_{\Dp}^{\YtwoS/\YoneS}   =  
  \mathscr{B}_{2/1} \times \dfrac
  {  \upsigma^{\YtwoS \Dp}_{\sqs=7\tev\phantom{\&8}}}
  {  \upsigma^{\YoneS \Dp}_{\sqs=7\tev\phantom{\&8}}} 
  & = &  (22 \pm 7  )\%\,, 
  \\
  R_{\Dp}^{\YtwoS/\YoneS}   =  
  \mathscr{B}_{2/1} \times \dfrac
  {  \upsigma^{\YtwoS \Dp}_{\sqs=8\tev\phantom{\&7}}}
  {  \upsigma^{\YoneS \Dp}_{\sqs=8\tev\phantom{\&7}}} 
  & = &  (22  \pm 6)\%\,, 
\end{eqnarray*}
where again the~systematic uncertainties are negligible 
with respect to the~statistical ones and are ignored.
These~values are compatible with 
the~DPS expectation of 25\% from Eq.~\eqref{eq:r21}.

Neglecting the contributions from SPS mechanism,
the~effective cross-section $\upsigma_{\mathrm{eff}}$~is determined
using Eq.~\eqref{eq:sdps} for the~$\sqs=7\tev$~data as 
\begin{eqnarray*}
  \left.\upsigma_{\mathrm{eff}}\right|_{\YoneS\Dz} & = & 19.4\pm 2.6\stat \pm 1.3\syst \mbarn\,, \\ 
  \left.\upsigma_{\mathrm{eff}}\right|_{\YoneS\Dp} & = & 15.2\pm 3.6\stat \pm 1.5\syst \mbarn\,.
\end{eqnarray*} 
The~central values of  
    $\upsigma_{\mathrm{eff}}$ increase by up to 10\% 
    if  SPS~contribution exceeds 
    by a factor of two the~central value from Eq.~\eqref{eq:r}.
Both values are consistent with previous measurements of 
$\upsigma_{\mathrm{eff}}$~\cite{
  Abe:1993rv,
  Abe:1997xk,
  Abazov:2009gc,
  LHCb-PAPER-2012-003,
  Aad:2013bjm,
  Chatrchyan:2013xxa,
  Abazov:2014fha,
  Bansal:2014paa,
  Astalos:2015ivw},
and their~average~is 
\begin{equation*}
  \left.\upsigma_{\mathrm{eff}}\right|_{\YoneS\D^{0,+},\sqs=7\tev} 
  = 18.0 \pm 2.1\stat \pm 1.2\syst
  = 18.0\pm 2.4\mbarn\,.  
\end{equation*}
For the~$\sqs=8\tev$~data
the~effective cross\nobreakdash-section 
$\upsigma_{\mathrm{eff}}$~is estimated using
the~measured \YoneS~cross\nobreakdash-section 
at $\sqs=8\tev$~\cite{LHCb-PAPER-2015-045}
combined
with $\upsigma^{\Charm}$, extrapolated from
\mbox{$\sqs=7\tev$}~\cite{LHCb-PAPER-2012-041} to
\mbox{$\sqs=8\tev$} using 
FONLL~calculations~\cite{Cacciari:1998it,Cacciari:2001td,Cacciari:2012ny}.
The~obtained effective DPS~cross-sections are:  
\begin{eqnarray*}
  \left.\upsigma_{\mathrm{eff}}\right|_{\YoneS\Dz} & = & 17.2\pm 1.9\stat \pm 1.2\syst \mbarn\,, \\ 
  \left.\upsigma_{\mathrm{eff}}\right|_{\YoneS\Dp} & = & 22.3\pm 4.4\stat \pm 2.2\syst \mbarn\,, 
\end{eqnarray*}
The~mean value of 
\begin{equation}
  \left.\upsigma_{\mathrm{eff}}\right|_{\YoneS\D^{0,+},\sqs=8\tev} 
  = 17.9 \pm 1.8\stat \pm 1.2 \syst
  = 17.9\pm 2.1\mbarn\,,  
\end{equation}
is in good agreement with those obtained for
$\sqs=7\tev$~data.
Averaging  these~values,
$\upsigma_{\mathrm{eff}}$
is found to be 
\begin{equation*}
  \left.\upsigma_{\mathrm{eff}}\right|_{\YoneS\D^{0,+}} 
  = 18.0 \pm 1.3\stat \pm 1.2\syst
  = 18.0 \pm 1.8\mbarn\,.
\end{equation*}

The~large value of the~cross-section for the~associated production 
of \ups~and open charm hadrons, compatible with the~DPS estimate
of Eq.~\eqref{eq:dpsr},  has important consequences
for the~interpretation of heavy-flavor production measurements,
especially inclusive measurements and possibly for~$\bquark$\nobreakdash-flavor 
tagging~\cite{LHCb-PAPER-2011-027,LHCb-CONF-2012-033,LHCb-CONF-2012-026,LHCb-PAPER-2015-027}, 
where the~production of uncorrelated charm~hadrons
    could affect the~right assignment of the~initial
    flavour of the~studied beauty hadron.

\section{Summary}\label{sec:summary}

The~associated production of \ups~mesons with open charm hadrons 
is observed in $\proton\proton$~collisions at
centre-of-mass energies of~$7$ and~$8\tev$
using~data samples corresponding to integrated luminosities 
of~\mbox{$1\invfb$}~and  \mbox{$2\invfb$}
respectively, collected with the~LHCb detector. 
The~production of 
$\YoneS\Dz$, 
$\YtwoS\Dz$, 
$\YoneS\Dp$, 
$\YtwoS\Dp$ and 
$\YoneS\Ds$~pairs  
is observed  with significances larger 
than 5~standard deviations.
The~production cross-sections in the~fiducial region 
$2.0<\yy<4.5$, $\pty<15\gevc$, 
$2.0<\yC<4.5$ and $1<\ptC<20\gevc$
are measured for  $\YoneS\Dz$ and $\YoneS\Dp$~final states at 
$\sqs=7$~and 8\tev.
The measured cross-sections 
are in agreement with DPS~expectations 
and significantly exceed the~expectations 
from the~SPS~mechanism.
The~differential kinematic distributions 
for $\ups\Dz$ and $\ups\Dp$ are studied
and  all are found to be in good agreement with
the~DPS~expectations as the~main production mechanism.
The~measured effective cross-section $\upsigma_{\mathrm{eff}}$~is
in agreement with most previous measurements.%

\bigskip
\section*{Acknowledgements}

 
\noindent 
We thank 
J.~R.~Gaunt, 
P.~Gunnellini,
M.~Diehl, 
A.~K.~Likhoded, 
A.~V.~Luchinsky, 
R.~Maciu\l{}a,
S.~Poslavsky
and 
A.~Szczurek
for interesting and 
stimulating discussions on the~SPS and DPS mechanisms.
We~are greatly indebted to  S.~P.~Baranov for 
providing us with predictions Eqs.~\eqref{eq:rkt}~and~\eqref{eq:rbbcc}
and the~differential kinematic distributions 
for $\ups+\cquark\cquarkbar$ SPS~process.
We~express our gratitude to our colleagues in the~CERN
accelerator departments for the excellent performance of the~LHC. 
We~thank the technical and administrative staff at the~LHCb
institutes. 
We~acknowledge support from CERN and from the~national
agencies: CAPES, CNPq, FAPERJ and FINEP\,(Brazil); 
NSFC\,(China);
CNRS/IN2P3\,(France); 
BMBF, DFG and MPG\,(Germany); 
INFN\,(Italy); 
FOM and NWO\,(The~Netherlands); 
MNiSW and NCN\,(Poland); 
MEN/IFA\,(Romania); 
MinES and FANO\,(Russia); 
MinECo\,(Spain); 
SNSF and SER\,(Switzerland); 
NASU\,(Ukraine); 
STFC\,(United Kingdom); 
NSF\,(USA).
We~acknowledge the~computing resources that are provided by CERN, 
IN2P3\,(France), 
KIT and DESY\,(Germany), 
INFN\,(Italy), 
SURF\,(The~Netherlands), 
PIC\,(Spain), 
GridPP\,(United Kingdom), 
RRCKI\,(Russia), 
CSCS\,(Switzerland), 
IFIN\nobreakdash-HH\,(Romania), 
CBPF\,(Brazil), 
PL\nobreakdash-GRID\,(Poland) and 
OSC\,(USA). 
We~are indebted to the communities behind the~multiple open 
source software packages on which we depend. 
We~are also thankful for the~computing resources and 
the~access to software R\&D tools provided by Yandex LLC\,(Russia).
Individual groups or members have received support from 
AvH Foundation\,(Germany),
EPLANET, Marie Sk\l{}odowska\nobreakdash-Curie Actions and ERC\,(European Union), 
Conseil G\'{e}n\'{e}ral de Haute\nobreakdash-Savoie, Labex ENIGMASS and OCEVU, 
R\'{e}gion Auvergne\,(France), 
RFBR\,(Russia), 
GVA, XuntaGal and GENCAT\,(Spain), The~Royal Society 
and Royal Commission for the~Exhibition of 1851\,(United Kingdom).


\addcontentsline{toc}{section}{References}
\setboolean{inbibliography}{true}
\bibliographystyle{LHCb}
\bibliography{local,main,LHCb-PAPER,LHCb-CONF,LHCb-DP,LHCb-TDR}

\newpage


\centerline{\large\bf LHCb collaboration}
\begin{flushleft}
\small
R.~Aaij$^{38}$, 
C.~Abell\'{a}n~Beteta$^{40}$, 
B.~Adeva$^{37}$, 
M.~Adinolfi$^{46}$, 
A.~Affolder$^{52}$, 
Z.~Ajaltouni$^{5}$, 
S.~Akar$^{6}$, 
J.~Albrecht$^{9}$, 
F.~Alessio$^{38}$, 
M.~Alexander$^{51}$
% ...

{\footnotesize \it
% ...
$ ^{5}$Clermont Universit\'{e}, Universit\'{e} Blaise Pascal, CNRS/IN2P3, LPC, Clermont-Ferrand, France\\
$ ^{6}$CPPM, Aix-Marseille Universit\'{e}, CNRS/IN2P3, Marseille, France\\
% ...
$ ^{9}$Fakult\"{a}t Physik, Technische Universit\"{a}t Dortmund, Dortmund, Germany\\
% ...
$ ^{37}$Universidad de Santiago de Compostela, Santiago de Compostela, Spain\\
$ ^{38}$European Organization for Nuclear Research (CERN), Geneva, Switzerland\\
% ...
$ ^{40}$Physik-Institut, Universit\"{a}t Z\"{u}rich, Z\"{u}rich, Switzerland\\
% ...
$ ^{46}$H.H. Wills Physics Laboratory, University of Bristol, Bristol, United Kingdom\\
% ...
$ ^{51}$School of Physics and Astronomy, University of Glasgow, Glasgow, United Kingdom\\
$ ^{52}$Oliver Lodge Laboratory, University of Liverpool, Liverpool, United Kingdom\\
% ...
}
\end{flushleft}

\end{document}